\documentclass[preprint,pre,aps,showpacs,amsfonts,amssymb,floatfix,onecolumn,superscriptaddress]{revtex4-2}
\usepackage[T1]{fontenc}
\usepackage[utf8]{inputenc}
\usepackage{enumerate}
\usepackage[toc,page]{appendix}
\usepackage{graphicx}
\usepackage{amsmath}
\usepackage{amsfonts}
\usepackage{amssymb}
\usepackage{mathtools}
\usepackage{multirow}
\usepackage{bm}
\usepackage{bbm}
\usepackage{pifont}
\usepackage{color}
\usepackage{relsize}
\usepackage[update,prepend]{epstopdf} 
\usepackage{tikz}
\usepackage{lmodern} 
\usepackage{hyperref,cleveref}
\usepackage[activate={true,nocompatibility},final,tracking=true,kerning=true,spacing=true,factor=1100,stretch=10,shrink=10]{microtype}

\usetikzlibrary{shapes}
\usetikzlibrary{plotmarks}

\newcommand*{\diff}{\mathop{\!\mathrm{d}\!}}
\newcount\hour \newcount\minute
\hour=\time \divide \hour by 60 \minute=\time \count99=\hour
\multiply \count99 by -60 \advance\minute by \count99 \hour=\time
\divide \hour by 60
\date{Draft, \today, \number\hour:\number\minute}

\let\vec\mathbf
\newcommand{\ii}{\mathrm{i}}
\newcommand{\Sc}{\operatorname{Sc}}
\let\epsilon\varepsilon

\makeatother 

\begin{document}
\title{Asymptotically exact scattering theory of the Kuramoto-Vicsek model} 
\author{Thomas Ihle}
\affiliation{Institute for Physics,
Greifswald University, Felix-Hausdorff-Str. 6, 17489 Greifswald, Germany}
\author{Horst-Holger Boltz}
\affiliation{Institute for Physics,
Greifswald University, Felix-Hausdorff-Str. 6, 17489 Greifswald, Germany}
\author{R{\"u}diger K{\"u}rsten}
\affiliation{Institute for Physics,
Greifswald University, Felix-Hausdorff-Str. 6, 17489 Greifswald, Germany}
\affiliation{Internationales Universitätszentrum, TU Bergakademie Freiberg, Freiberg, Germany}
\author{Benjamin Lindner}
\affiliation{Institute for Physics,
Humboldt University, Berlin, Germany}

\begin{abstract}
We consider the Kuramoto-Vicsek model of non-Brownian self-propelled particles with velocity-alignment interactions. 
Starting from the exact $N$-particle Liouville equation, 
a kinetic equation for the 
one-particle distribution function is obtained in a self-consistent manner without additional input.
We show that the usual mean-field assumption of molecular chaos  which involves a simple factorization of
the N-particle probability leads to qualitatively wrong predictions such
as an infinite coefficient of self-diffusion. Going beyond mean-field and applying the refined assumption of \emph{one-sided molecular chaos} where the two-particle-correlations during binary interactions are explicitly taken into account, we analytically calculate the scattering of particles in the limit of low density and obtain 
explicit expressions for the dynamical noise of
an effective one-particle Langevin-equation and the corresponding self-diffusion. The theory is developed in detail for anti-aligning couplings, where collision times are finite and exact analytical results are obtainable.
In this calculation, the so-called superposition principle of traditional kinetic theory is modified to
handle a system with non-Hamiltonian dynamics involving phase-space compression.
The predicted theoretical expressions for the relaxation of hydrodynamic modes and the self-diffusion coefficient
are in excellent, quantitative agreement with agent-based
simulations at small density and coupling.
At large particle densities, a given particle is constantly approached and abandoned by different collision partners.
Modeling this switching of partners by a random telegraph process and  exactly solving a self-consistent integral equation, 
we obtain explicit expressions for
the noise correlations of the effective one-particle Langevin-equation.
Comparison with agent-based simulation shows very good agreement. 

We also consider the effect of frozen disorder in the particle speeds and show how the limit of vanishing disorder can be used to calculate the exact Boltzmann collision operator for positive alignment strengths overcoming secular terms introduced by the unphysical binary collision approximation in this regime.
\end{abstract}


\maketitle


\section{Introduction}
\label{sec:intro}

Ensembles of interacting, self-propelled particles (SPPs) provide the most common realization of active matter and have been attracting 
much attention from the statistical physics and soft matter communities. 
During the last 30 years, SPPs have been extensively studied as minimal representations of many living and synthetic systems from insect swarms and bird flocks to 
pedestrians and robots \cite{vicsek_12,marchetti_13,menzel_15,chate_20,liebchen_22}. 
A wealth of intriguing collective states, including wave formation, swirling, laning and mesoscale turbulence can be obtained by 
surprisingly 
simple microscopic models for the particles \cite{bechinger_16}.  
One prominent class of such models is characterized by a velocity-alignment rule among neighboring particles and goes back
to the famous Vicsek-model (VM)~\cite{vicsek_95,czirok_97,nagy_07,chate_08,ihle_13,ginelli_16,kuersten_20a} 
which favors parallel alignment of propulsion directions. 
Another well-established model is the Active Brownian Particle model 
\cite{lindner_08,romanczuk_12,fily_12,bialke_12,redner_13,speck_14,digregorio_18,caporusso_20,pinto2025} where
self-propelled particles interact via isotropic repulsion due to excluded volume. 
Both model classes contain explicit noise sources that mimic, for example, environmental disturbances or alignment errors. 
The interesting features of these models such as pattern formation and collective motion are a result of the interplay of self-propulsion, noise, 
alignment and/or steric avoidance.

Since a global theoretical framework comparable to equilibrium statistical thermodynamics is still missing for such far-from-equilibrium systems with 
many interacting objects, researchers mostly rely on agent-based computer simulations, e.g. \cite{chate_08,digregorio_18,chate_20,kuersten_20a}, and hydrodynamic theories which are constructed by 
symmetry arguments \cite{toner_95,toner_98,toner_12,toner} or 
are derived from the microscopic rules by means of mean-field assumptions 
\cite{bertin_06,baskaran_08b,peruani_08,bertin_09,ihle_11,peshkov_12,grossmann_13,degond2013,chepizhko_14,bonilla2023,lier2026}. 
Efforts to go beyond mean-field in a self-consistent way \cite{chou_15,patelli_21,kuersten_21b} are sparse as they are usually prohibitively difficult or 
only work for specific systems in certain ranges of parameter 
space.

SPPs with alignment interactions form large networks of rotators, where links between rotators are defined 
if they are within each other's interaction range. The connectivity of the network
evolves in time and depends on the history of
the directions of the rotators. 
Networks of interacting rotators are studied in connection with spiking nerve cells in the brain, pacemaker cells in the heart, 
or the interacting cells in developing tissue. The equations of motion of these rotators are almost identical to the equations of SPPs 
with chirality where the chirality is given by the oscillation frequency of a particular rotator.  
The main difference is the absence of evolution for the rotator positions, hence, 
the network configuration is typically assumed as frozen. 
Similar to active matter, research in this area often focuses on collective phenomena 
like synchronization, global oscillations, and waves. 
However, the emergence of asynchronous irregular activity instead of some form of macroscopic order is actually more typical, e.g., in the 
awake behaving animal \cite{poulet_08,harris_11,vanvreeswijk_96}. 
A full understanding of the rich temporal structure of the asynchronous state is still an open challenge. 
In this state, units behave quasistochastically because they are driven by a large number of other likewise quasistochastic units.
The statistics of the driving amount to an effective, dynamical network noise whose correlations 
depend in a non-trivial way on both the oscillator and network properties.
Recently, progress was made for a system of permanently but randomly coupled rotators in the 
asynchronous state: within 
a stochastic mean-field approximation an effective Langevin equation for the rotators with temporally correlated 
noise sources was established and the noise correlations were calculated self-consistently \cite{vanmeegen_18,ranft_22a,ranft_22b,kati2024}.

In this paper\footnote{A shorter version of this work has been presented in Ref. \cite{ihle_23a}.}, we show the details on how the temporal correlations of the network noise 
can be analytically determined in a history-dependent temporal network of Non-Brownian 
self-propelled 
particles in
the asynchronous state.
To this end and similar to ref.~\citenum{vanmeegen_18},
we pursue the main idea of Brownian motion and
assume that the effects of the surrounding
rotators
on a focal rotator can be mapped to a Gaussian network noise term $\xi(t)$, leading to an effective, one-particle
Langevin-equation
for the angular change of the focal rotator.
Since the particles are mobile, the network noise manifests itself in the self-diffusion of the particles, which is one of the
predicted
quantities of our theory.
At large particle densities, this is achieved by means of a self-consistent mapping of the network dynamics to a birth-death process, whereas
at small densities, we develop a quantitative scattering theory beyond mean-field by using a first-principle, non-local closure of the 
BBGKY-hierarchy.
We give a particular example of a system of self-propelled particles, where the usual mean-field factorization 
of the N-particle probability distributions, 
often called molecular chaos assumption, 
leads to unphysical results whereas the non-local closure gives quantitatively correct predictions for the dynamics of the system, 
even far from the stationary state.
Our theory opens a route for quantitative treatment and derivations of hydrodynamic equations beyond mean-field for other, 
more complex systems of self-propelled particles with, for example, chiral \cite{liebchen_17,levis_19,kuersten_23a,wagner2026}, nematic \cite{ginelli_10,peruani_10}, 
bounded-confidence \cite{lorenz_07,romensky_14}, vision-cone  \cite{barberis_16,negi_22} or other 
non-reciprocal \cite{fruchart_21,kreienkamp_22}  interactions. The theory has already been extended to binary 
mixtures of active particles \cite{kuersten_2025}
where it quantitatively reproduces the effect of the self-propulsion 
speed on the order/disorder transition. 
The theory has also been generalized to models with very small external noise \cite{kuersten_2025} and non-reciprocal couplings~\cite{mihatsch2025}.

We consider a minimalist version of the already bare-bones model of SPPs with Kuramoto-like alignment 
\cite{degond2008,peruani_08,farrell_12,chepizhko_13,chepizhko_21,zhao_21,chen_23} without any noise term and without chirality. The deterministic nature of the model enables an exact treatment of the two-particle scattering problem. For anti-aligning couplings ($\Gamma<0$), collision times are always finite, which permits a clean analytical development; we show in section~\ref{sec:disorder} that the results extend to the aligning case by a physical regularization.
We develop a scattering theory which starts at the N-particle Liouville equation and the corresponding BBGKY-hierarchy of evolution equations for reduced
probability densities.
The simplicity of the anti-alignment interactions allows us to analytically determine the cross section of the SPPs and to 
explicitly solve the evolution equation of the two-particle probability density for two 
interacting particles in the low density limit. Reinserting this solution in the first BBGKY-equation amounts to a non-local closure of this equation, leading to correction terms absent in the usual mean-field closure.  

There are only a few model systems of many interacting particles for which it is possible to analytically derive a Langevin-equation 
by explicitly integrating over the irrelevant degrees of
freedom. 
Some examples are described in the text book by Zwanzig~\cite{zwanzig_book}.
More recent examples are given, e.g., in Refs. \citenum{vanmeegen_18} and \citenum{netz_18}.
Here, we provide another example where this is possible and where
the approach is asymptotically exact in the limit of vanishing density and interaction strength.

\subsection{The model}
\label{sec:model}

We consider $N$ point-particles with constant speed $v_0$ in two dimensions and periodic boundary conditions.
The positions $\vec{r}_i(t)=(x_i(t),y_i(t))$ and the flying directions $\theta_i(t)$ of the particles are updated by the following rules,
\begin{subequations}
\begin{eqnarray}
\label{POS_EQ}
& & {\diff \vec{r}_i \over \diff t}=v_0\,\hat{\vec n}_i \\
\label{ANGLE_EQ}
& & {\diff\theta_i \over \diff t}={\Gamma\over N_i^{\beta}}\sum_{j\epsilon\Omega_i}
\sin(\theta_j-\theta_i)\,. 
\end{eqnarray}
\end{subequations}
Here, $\hat{\vec n}_i=\hat{\vec n}(\theta_i)=(\cos{\theta_i},\sin{\theta_i})$ 
is a unit vector which points in the flying direction of particle $i$, and $\Gamma$
is the interaction strength, which can take either sign.

In regular Vicsek-like models with polar order \cite{vicsek_95,czirok_97,nagy_07,peruani_08,farrell_12,chepizhko_13,chepizhko_21,zhao_21}, 
$\Gamma$ is positive and supports ``ferromagnetic'' alignment.
For $\Gamma<0$, i.e. ``anti-ferromagnetic alignment'' which mimics social distancing
of particles, all binary collision times are finite, which makes the anti-aligning case the natural starting point for the analytical development in sections~\ref{sec:vlasovkinet}–\ref{sec:scatter}.  
Similar anti-alignment rules have been studied before, e.g. in Refs. \cite{grossmann_14,grossmann_15,kreienkamp_22,chatterjee_23, escaff2024, das2024, escaff2024b, escaff2025, boltz2025}.

The sum in eq. (\ref{ANGLE_EQ}) goes over the $N_i$ particles (including particle $i$) that are inside a circle of radius $R$ 
around particle $i$ and form the set $\Omega_i = \{ j\, |\, \lvert \vec r_i -\vec r_j\rvert \leq R\}$.
The exponent $\beta$ is usually chosen to be zero or one and has been shown to significantly impact the formation
of density waves \cite{stroteich_thesis,kuersten_21b} in Vicsek-like models. An important dimensionless parameter of the system
is the scaled density $M$, also called partner number, $M=\pi R^2 \rho$, where $\rho=N/L^2$ is the number density
of the particles. The parameter $M$ describes the average number of interaction partners. 
At small $M\ll 1$, interactions that involve more than two particles are very rare.

In contrast to the ``work horses'' of active matter, such as the active Brownian particle model 
\cite{romanczuk_12,lindner_08,caporusso_20}, the
standard Vicsek-model (VM) \cite{vicsek_95,vicsek_12} or run-and-tumble models for bacterial motion \cite{tailleur_08}, our microscopic model is deterministic and does not contain 
any external noise.
For anti-aligning couplings,
the system is \emph{self-mixing}: the effect of the surrounding particles on a given, focal particle, can be described 
by an effective dynamical noise.

As shown further below, the absence of an external noise term allows us to find an exact solution for the scattering of two particles,
which dominates the dynamics at low densities. We are also able to explicitly calculate the effect of
phase-space compression in the corresponding kinetic theory, something that is rarely, if ever, done. In the case of finite noise, the leading order in the noise strength would be identical given the trivial diffusive terms are included into the then Fokker--Planck equation. Significant deviations are to be expected if the external noise is large enough to be relevant on collisional timescales as it will then affect the collision integrals.

\section{Vlasov-like kinetic theory}
\label{sec:vlasovkinet}

\subsection{The molecular chaos approximation}
\label{subsec:intro_vlasov}

In this section we will first focus on the simplest kinetic approach, a Vlasov-like theory,
where only the first member of the BBGKY-hierarchy is used by simply factorizing the
N-particle distribution function \cite{vlasov_38}.
This type of approach has been very useful in Plasma physics where particles interact with many others
due to long-ranged Coulomb interactions \cite{plasma_vlasov,plasma_vlasov1} and in the theory of dilute electrolytes by 
Debye and H{\"u}ckel \cite{debye_23}.
In the system considered here, one would naively expect it to be useful at large particle number densities and/or large interaction range, where
many particles are within the collision circle of the focal particle, i.e. where $M\gg 1$. 
As we will show further below, this expectation is incorrect for our system which has a continuous time dynamics but no external noise.
Note that the approach of factorizing the
N-particle distribution function, also called \emph{molecular chaos}, can be much better justified in systems with a discrete time step $\Delta t$ such as in the standard VM~\cite{ihle_11,ihle_13,bonilla_18,bonilla_19}. 
This is because there is an additional usually small parameter, the ratio of the interaction radius $R$ to
the mean free path $\Delta t\,v_0$. If this ratio is sufficiently small, two particles that just collided have a very small probability to collide again in the next time step, and thus, particles are mostly uncorrelated before the
next collision. This is not the case in models with continuous time: during the small but finite encounter of
collision partners, these particles undergo correlated collisions.

\subsection{Deriving a one-particle Liouville description}
\label{subsec:fokker}

We define the 3N-dimensional vector 
$\vec{Z}=(\vec{r}_1,\theta_1,\vec{r}_2,\theta_2,\ldots,\vec{r}_N,\theta_N)=(\vec 1,\vec 2,\vec 3\ldots \vec N)$ which describes the microscopic state 
of the system and where we abbreviated the phase of particle $1$, that is $(\vec{r}_1,\theta_1)$ by $\vec 1$ and so on. 
The model equations (\ref{POS_EQ}) and (\ref{ANGLE_EQ}) can now be rewritten as an equation for $\vec{Z}$.
Standard kinetic theory, allows us to see that the N-particle
probability density $P_N=P_N(\vec{Z},t)$ is described by the Liouville-equation:
\begin{eqnarray}
\partial_t P_N=&-&\sum_{i=1}^N\Big\{v_0\,(\hat{\vec n}_i\cdot\vec{\nabla}_i)\, P_N
+\partial_{\theta_i}\,\Big(\sum_{j=1}^N \Big[{\Gamma\over N_i^{\beta}} a_{ji}\, \sin(\theta_j-\theta_i) 
\Big]P_N \Big) \Big\} \label{N_FOKK}
\end{eqnarray}
with $\vec{\nabla}_i \equiv (\partial_{x_i}, \partial_{y_i})$. The  matrix element $a_{ji}$ depends on the positions of the particles and is given
by $a_{ji}=0$ for $|\vec{r}_j-\vec{r}_i|>R$ and 
$a_{ji}=1$ for $|\vec{r}_j-\vec{r}_i|\leq R$.   We are not considering noise here, but standard theory of stochastic systems, see for example \cite{gardiner-book,risken-book,vanKampen-book}, establishes that this holds in the stochastic case as well if diffusive (second derivative) terms are included for the noise, turning this Liouville equation into a Fokker--Planck equation

The exact equation \eqref{N_FOKK} contains too much information and is intractable.
To simplify, we first
factorize the probability distribution on the right hand side of the equation,
$P_N(\vec 1,\vec 2,\ldots,\vec N)=\prod_{j=1}^N P_1(\vec j)$. This neglects correlations among the particles and 
amounts to the mean-field approximation of molecular chaos. 
This approximation 
is widely used
in active particle systems 
\cite{bertin_06,peruani_08,ihle_11,romanczuk_12,grossmann_13,bussemaker_97,romanczuk_12b,reinken_18,benvegnen_22}. 

Next, we multiply \cref{N_FOKK} by the one-point phase space density
\begin{equation}
\Psi_1=\sum_{j=1}^N\delta(\vec{r}-\vec{r}_j)\,\delta(\theta-\theta_j)
\end{equation}
and integrate over all particle positions and angles\footnote{Alternatively, we could just have integrated over all particles, except one, without using $\Psi_1$. However, the use of
phase space density has conceptional and practical advantages, in particular, if one extends the description to two- or
more-point correlations.}.
Here, $(\vec{r}_j,\theta_j)$ is the phase of particle $j$, whereas $(\vec{r},\theta)$ is a field point in phase space.
For more details on the integration procedure, see Refs. \citenum{ihle_16} and \citenum{kuersten_21b}.
Finally, one obtains a kinetic equation, a non-linear one-particle Fokker--Planck-equation without diffusive terms, for the distribution function $f(\vec{r},\theta,t)=N\,P_1(\vec{r},\theta,t)$,
\begin{equation}
\label{ONE_P_FP}
\partial_t f=-v_0\hat{\vec n}(\theta)\cdot\vec{\nabla}f
-\partial_{\theta}[\Gamma\,F\,f]
\end{equation}
with the mean-field force,
\begin{subequations}
\begin{equation}
\label{MEAN_FORCE}
F(\vec{r},\theta)\equiv A\,G_{\beta}(M(\vec{r}))\int_0^{2\pi}\!\!\!\diff\theta_2\,\sin(\theta_2-\theta)\,\bar{f}(\vec{r},\theta_2)
\end{equation}
(where the time-dependence has been omitted for brevity)
and the function $G_{\beta}(\vec{r})$,
\begin{equation}
\label{G_SUM}
G_{\beta}=\sum_{n=2}^{\infty}{\rm e}^{-M} {M^{n-2} \over n^{\beta} (n-2)!}  \text{,}
\end{equation}
which depends on the local partner number $M(\vec{r})$,
\begin{equation}
\label{M_DEF}
M(\vec r)=\int_{\odot}\diff\vec{r}_2 \rho (\vec{r}_2)\text{.}
\end{equation}
Here $\int_{\odot}$ denotes an integral over the collision circle centered at $\vec r$. The quantity $\bar{f}$ in eq. (\ref{MEAN_FORCE}) is the
average of the distribution function over the collision circle, 
\begin{equation}
\bar{f}(\vec{r},\theta)\equiv {1\over A}\int_{\odot}\diff\vec{r}_2 f(\vec{r}_2,\theta)
\end{equation}\end{subequations}
where $A=\pi R^2$ is the area of the collision circle.
For $\beta=0$ and $\beta=1$ the sum in eq. (\ref{G_SUM}) can be evaluated exactly to yield
\begin{subequations}
\begin{align}G_0=1\end{align} and 
\begin{equation}
\label{G1_DEF}
G_1={1\over M}\left[ 1-{1\over M}\left(1-{\rm e}^{-M}\right)\right] \text{.}
\end{equation}	
\end{subequations}
For $M\ll 1$ one finds $G_1=1/2$, and in the opposite limit $M\gg 1$ one obtains $G_1=1/M$.

\subsection{Angular mode equations}
\label{sec:mode}
Defining the angular Fourier-transformation, 
\begin{align}
 \begin{split}
\hat{f}_n(\vec{r},t) = & {1\over 2\pi}\int_0^{2\pi} {\rm e}^{-\ii n\theta}f(\vec{r},\theta,t)\,\diff\theta \\
\label{DEF_ANG_FOUR}
f(\vec{r},\theta,t)=& \sum_{n=-\infty}^{\infty} \hat{f}_n(\vec{r},t)\, {\rm e}^{\ii n\theta}\,\text{,}
 \end{split}
\end{align}
the kinetic equation, \cref{ONE_P_FP}, is transformed into a hierarchy of evolution equations for
the angular modes $\hat{f}_n$:
\begin{equation}
\label{KINETIC-VLASOV1}
\partial_t\hat{f}_n+{v_0\over 2}\left[\nabla^*\hat{f}_{n-1}+\nabla \hat{f}_{n+1}\right]=
-A\,n\,\pi \Gamma G_{\beta}\left[
\hat{f}_{n+1}\hat{f}_{-1}
-\hat{f}_{n-1}\hat{f}_1
\right]
\end{equation}
where $\nabla$ and $\nabla^*$ are the complex nabla operator and its complex conjugate, respectively,
\begin{align}
\begin{split}
\nabla\equiv& \partial_x+\ii\, \partial_y \\
\label{COMPLEX_NABLA}
\nabla^*\equiv& \partial_x-\ii\, \partial_y 
\end{split}
\end{align}
Note that for $|n|\neq 1$,  due to the absence of external angular and positional noise there are no damping terms on
the right hand side of \cref{KINETIC-VLASOV1} of the type $\sim -\hat{f}_n$ or $\sim \nabla\nabla^*\hat{f}_n$.

The first five hierarchy equations for $n=0,1,\ldots 4$ are: 
\begin{align}
\begin{split}
\partial_t\hat{f}_0+{v_0\over 2}\left[\nabla^*\hat{f}_{-1}+\nabla \hat{f}_{1}\right]=&
0\\
\partial_t\hat{f}_1+{v_0\over 2}\left[\nabla^*\hat{f}_{0\,\,\,\,}+\nabla \hat{f}_{2}\right]=&
-A\,\pi \Gamma G_{\beta}\left[
\hat{f}_{2}\hat{f}_{-1}-\hat{f}_0\hat{f}_1
\right]\\
\partial_t\hat{f}_2+{v_0\over 2}\left[\nabla^*\hat{f}_{1\,\,\,\,}+\nabla \hat{f}_{3}\right]=&
-2 A\,\pi \Gamma G_{\beta}\left[
\hat{f}_{3}\hat{f}_{-1}
-\hat{f}_{1}\hat{f}_1
\right]\\
\partial_t\hat{f}_3+{v_0\over 2}\left[\nabla^*\hat{f}_{2\,\,\,\,}+\nabla \hat{f}_{4}\right]=&
-3 A\,\pi \Gamma G_{\beta}\left[
\hat{f}_{4}\hat{f}_{-1}
-\hat{f}_{2}\hat{f}_1
\right]\\
\label{VLASOV_HIERA}
\partial_t\hat{f}_4+{v_0\over 2}\left[\nabla^*\hat{f}_{3\,\,\,\,}+\nabla \hat{f}_{5}\right]=&
-4 A\,\pi \Gamma G_{\beta}\left[
\hat{f}_{5}\hat{f}_{-1}
-\hat{f}_{3}\hat{f}_1
\right]
\end{split}
\end{align}
Here, an important property of the zeroth mode is apparent. Since $\hat{f}_0 = \frac{\rho}{2\pi}$ is the density, its material derivative has to vanish. This conservation of matter constitutes a valuable sanity check to any theory Because the distribution function $f$  is a real function, the negative modes
are given by complex conjugated modes, 
\begin{equation}
\hat{f}_{-n}=
\hat{f}_{n}^* \text{.}
\end{equation}

\section{Scattering theory for small densities}
\label{sec:scatter}

\subsection{Failure of the molecular chaos approximation}

In section \ref{sec:numer}, agent-based simulations of eqs. \eqref{POS_EQ} and \eqref{ANGLE_EQ} are presented.
They show that if the system is initialized in a non-stationary state with strong polar and higher order 
(all modes $\hat{f}_n$, defined in \cref{DEF_ANG_FOUR}, are non-zero), it always relaxes
towards a disordered state, where all modes except $\hat{f}_0$
become zero.
This is in qualitative disagreement with the prediction of the hierarchy equations from Vlasov-like kinetic theory \eqref{VLASOV_HIERA} where
the final, stationary state is not disordered but rather depends on initial conditions, see section \ref{sec:numer}. 
As shown later, this behavior is related to the incorrect prediction of an {\em infinite} coefficient of self-diffusion. 
Since this coefficient is related to the noise strength of an effective one-particle Langevin equation with
dynamical noise, one of the main goals of the work, the derivation of this Langevin-equation,
cannot be achieved by a Vlasov-like mean-field theory.

Therefore, in the current section we construct an improved kinetic theory which goes beyond the simple molecular chaos Ansatz 
and leads to additional dissipative terms in the equations for the angular modes. 
As a result, quantitative agreement for the relaxation of the angular modes and the self-diffusion coefficient is achieved. 
This improved kinetic theory is restricted to small densities; a different theory for very large densities is presented in section~\ref{sec:large_dens}.

At small densities, $M\ll 1$, and in the absence of clustering (as expected in the disordered phase)
most interactions are binary: two particles interact continuously for a duration time $T_{\text{dur}}$ after their first encounter,
and the likelihood for a third particle to join, is negligible.
The time between subsequent encounters, the mean-free-flight time $T_{\text{free}}$, is assumed to be much larger than the duration of 
such a binary collision, $T_{\text{free}}\gg T_{\text{dur}}$.
If the same particles meet again at a later time, they will have lost most of the memory of their interaction due to collisions 
with other particles in the mean time.
Because of this, it is reasonable to assume that the two particles are approximately uncorrelated {\em before} their encounter.
However, directly after their encounter, when their distance becomes larger than $R$ again, they will be correlated.
This means, we can factorize the two-particle probability $P_2$ before the encounter but not directly afterwards.
This approximation has been named {\em one-sided molecular chaos}  in the context of standard kinetic theory \cite{kreuzer_81}. 
Due to the absence of momentum conservation, 
there will be no long-time tails \cite{alder_70,dorfman_70,ernst_70,kawasaki_71,zwanzig_book}
and the assumption of one-sided molecular chaos becomes asymptotically exact for $M\to 0$ if the underlying physics is analytic in $M$.

In 1872, Boltzmann proposed his equation using powerful intuitive arguments. However, mathematical rigorous ways to derive the Boltzmann equation from the microscopic dynamics were 
published only much later~\cite{bogolioubov,kirkwood,born,grad_58}.
Here, we generalize the derivation from Kreuzer~\cite{kreuzer_81} 
to active matter, see also Ref.~\citenum{green_52} for an earlier presentation and Ref.~\citenum{waldmann_58} 
for the derivation of the scattering cross section in regular gases. 
We show in the following that there are several crucial differences between the Boltzmann-equation of a dilute classical gas
and the one for the continuous-time VM.
In particular, in contrast to Hamiltonian dynamics, the phase-space compression factor \cite{evans_08} is nonzero, that is, the total 
time derivative of $P_N$ for the VM does not vanish. This leads to an additional non-trivial factor in the collision integral. 
Furthermore, the interactions are velocity-dependent, and 
the diverging dynamics of the social-distancing interactions for $\Gamma<0$ leads to {\em forbidden} pairs of angles $\theta_1$, $\theta_2$
at interaction distances $|\vec{r}_1-\vec{r}_2|\le R$. That is, there are points in the 6-dimensional phase space of two particles, which have 
zero probability,
$P_2(\theta_1,\vec{r}_1,\theta_2,\vec{r}_2)=0$. These points form the ``forbidden zone'' that is calculated in section \ref{sec:one-sided}.
As a result, the collision integral in this Boltzmann scattering theory is more difficult to evaluate than the one for a regular gas.
Defining the dimensionless interaction strength
\begin{equation}
\label{SC_DEF}
\Sc\equiv \frac{|\Gamma| R}{v_0}
\end{equation}
we will evaluate the Boltzmann equation in the limits of weak coupling, $\Sc\ll 1$.
The quantity $\Sc$ is a measure for the change of the flying direction over the time duration of a binary interaction.
The theory could also be evaluated perturbatively for very strong coupling, $\Sc\gg 1$. However, this  
will be left
for future studies. 

Note that there is a fundamental difference between the scattering theory presented here and the 
``Boltzmann-Ginzburg-Landau approach'' by Peshkov et al. \cite{bertin_06,peshkov_12,peshkov_12b,peshkov_14,thuroff2013,murphy2024}. 
We present a bottom-up approach based on the exact Liouville-equation of a 
particular microscopic model, perform explicit coarse-graining and 
derive
asymptotically exact cross sections in the Boltzmann equation. 
Peshkov et al. already start with a Boltzmann-like kinetic equation that models generic features of systems with alignment interactions.
At this level of modeling, the question about the difference between simple molecular chaos and one-sided molecular chaos is moot and 
does not come up.
However, in many cases, their proposed kinetic equations agree with the Vlasov-like mean-field equations of a particular microscopic model and effectively rely on similar assumptions but
understandably miss the rather non-intuitive couplings between angular Fourier-modes 
(partly due to phase-space compression and the existence of a forbidden zone) needed for a description of that model beyond mean field. The short-comings of such approaches are particularly apparent in variations of the Kuramoto-Vicsek model where additional rotational symmetry forces the mean-field forces to vanish, such as cohesion or turn-away type torques~\cite{das2024} or vision-cone interactions~\cite{negi_22} which both depend on the direction of the displacement vector as well.

\subsection{The first two members of the BBGKY-hierarchy}
\label{sec:firsttwo}

In the following, we focus on the case $\beta=0$ in the microscopic model, \cref{ANGLE_EQ}.
Integrating the N-particle Liouville equation, \cref{N_FOKK}, over all phases, except one,
yields the first member of the BBGKY-hierarchy,
\begin{align}
\begin{split}
\partial_t P_1=&-v_0\hat{\vec n}(\theta)\cdot\vec{\nabla} P_1\\
\label{FIRST_MEM}
 &-(N-1)\,\Gamma\,\partial_{\theta}\int \diff \theta_2\int d\vec{r}_2\,
a(|\vec{r}_2-\vec{r}|)\,\sin(\theta_2-\theta)\,P_2(\vec{r},\theta,\vec{r}_2,\theta_2,t) 
\end{split}
\end{align}
for the one-particle probability density $P_1\equiv P_1(\vec{r},\theta,t)$, and where we introduced
the indicator function $a(r)=1$ for $r\le R$ and $a(r)=0$ for $r>R$.

Next, we multiply \cref{N_FOKK} by the two-point phase space density
\begin{equation}
\Psi_2=\sum_{j=1}^N \sum_{k\neq j}^N \delta(\vec{r}-\vec{r}_j)\,\delta(\theta-\theta_j)
\,\delta(\vec{z}-\vec{r}_k)\,\delta(\beta-\theta_k)
\end{equation}
and integrate over all particle positions and angles.
Here, $(\vec{r}_j,\theta_j)$ is the phase of particle $j$, 
$(\vec{r}_k,\theta_k)$ is the phase of another particle $k$  with $k\neq j$
whereas $(\vec{r},\theta,\vec{z},\beta)$ is a field point in the product phase space of two particles.
This results in the second member of the BBGKY-hierarchy,
\begin{align}
\begin{split}
	 \partial_t P_2=&-v_0\left[\hat{\vec n}(\theta)\cdot\partial_{\vec{r}}+\hat{\vec n}(\beta)\cdot\partial_{\vec{z}} \right] P_2
-\Gamma\,\partial_{\theta}\big[a(|\vec{r}-\vec{z}|)\,\sin(\beta  -\theta)\,P_2\big] \\
 &-\Gamma\,\partial_{\beta }\big[a(|\vec{r}-\vec{z}|)\,\sin(\theta -\beta) \,P_2\big] \\
 & -(N-2)\,\Gamma\,\partial_{\theta}\bigg[
\int \diff \theta_3\int d \vec{r}_3 \,a(|\vec{r}_3-\vec{r}|)\,\sin(\theta_3  -\theta)\,
P_3(\vec{r},\theta,\vec{z},\beta,\vec{r}_3,\theta_3,t) 
\bigg] \\
& -(N-2)\,\Gamma\,\partial_{\beta}\bigg[
\int \diff \theta_3\int d \vec{r}_3 \,a(|\vec{r}_3-\vec{z}|)\,\sin(\theta_3  -\beta)\,
P_3(\vec{r},\theta,\vec{z},\beta,\vec{r}_3,\theta_3,t) 
\bigg] 
\label{SECOND_MEM}
\end{split}
\end{align}
for the two-particle probability density $P_2\equiv P_2(\vec{r},\theta,\vec{z},\beta,t)$. 

\subsection{Derivation of the Boltzmann equation}
\label{sec:boltz1}

To close the first hierarchy equation, \cref{FIRST_MEM},
it suffices to merely know the two-particle probability density $P_2$ {\em inside}
the collision circle, i.e. for $|\vec{r}_2-\vec{r}|\le R$. 
This is because of the finite interaction range, represented by the presence of the indicator function $a(r)$. 
With this restriction, we look at the second hierarchy equation and realize that the three-particle probability density
$P_3$ only contributes if its spatial coordinates are not further apart than $2R$ 
from each other. Thus, terms containing $P_3$ in \cref{SECOND_MEM} refer to the probability of simultaneously observing three particles at such close distances.

At small normalized particle densities, $M=\pi R^2\rho_0\ll 1$, this probability is negligible, that is, we can neglect
three-particle collisions and formally set $P_3=0$.
This \emph{binary-collision approximation} closes the BBGKY-hierarchy and reduces it to just two equations.

However, one can exploit the binary collision approximation even further, ultimately leading to just one kinetic equation.
Following Ref. \citenum{kreuzer_81}
we first drop the time-derivative $\partial_t$ in \cref{SECOND_MEM} which accounts for the overall evolution of the particles over times 
$T_{\text{free}}$. In the collision integral we follow $P_2$, however, only over the much shorter time $T_{\text{dur}}$ of the duration of the two-particle encounter.
The second BBGKY-equation with $P_3=0$ is nothing else than the Liouville equation for a two-particle system. 
By solving this first order partial differential equation by the method of characteristics \cite{ihle_23a}, 
it can be shown that including the term $\partial_t P_2$
just leads to a correction of higher order in the density and becomes negligible at $M\ll 1$.
Hence, for simplicity we set $\partial_t P_2=0$, and \cref{SECOND_MEM} reduces to,
\begin{eqnarray}
& & {-}\Gamma\partial_{\theta}\!\big[a(|\vec{r}{-}\vec{z}|)\!\sin(\beta  {-}\theta)P_2\big] 
\label{P2_SOL}
\!\approx\!\Gamma\partial_{\beta }\!\big[a(|\vec{r}{-}\vec{z}|)\!\sin(\theta {-}\beta) P_2\big] 
\!+\!v_0\!\left[\hat{\vec n}(\theta)\cdot\partial_{\vec{r}}\!+\!\hat{\vec n}(\beta)\cdot\partial_{\vec{z}} \right]\! P_2 \text{.}
\end{eqnarray}
Since we are only interested in $P_2$ inside the collision circle, $|\vec{r}{-}\vec{z}|\le R$, we have $a=1$ and the 
left hand side of \cref{P2_SOL} is substituted into the collision integral of the first hierarchy equation, \cref{FIRST_MEM} after setting $\vec{z}
=\vec{r}_2$, $\beta=\theta_2$.
The collision integral then becomes equal to
\begin{equation}
\label{JCOLL1}
J^{(\text{coll})}= (N-1)\,\int \diff\theta_2\int_{\odot} \diff\vec{r}_2\,\bigg[
\Gamma\,\partial_{\theta_2}\big[\sin(\theta -\theta_2) \,P_2\big]
+v_0\left[\hat{\vec n}(\theta)\cdot\partial_{\vec{r}}+\hat{\vec n}(\theta_2)\cdot\partial_{\vec{r}_2} \right] P_2
\bigg] \text{,}
\end{equation}
where $\int_{\odot}$ denotes an integral over the collision circle, centered at position $\vec{r}$.
The first term does not contribute, as one can show by partial integration with respect to $\theta_2$.
Finally, one arrives at the following kinetic equation
\begin{equation}
\partial_t P_1+v_0\hat{\vec n}(\theta)\cdot\vec{\nabla} P_1=J^{(\text{coll})} \label{eq:def_jcoll}
\end{equation}
with the collision integral
\begin{equation}
J^{(\text{coll})}=
\label{JCOLL2}
 (N-1)\,\int \diff \theta_2\int_{\odot} \diff \vec{r}_2\,\bigg[
v_0\left[\hat{\vec n}(\theta)\cdot\partial_{\vec{r}}+\hat{\vec n}(\theta_2)\cdot\partial_{\vec{r}_2} \right] P_2(\vec{r},\theta,\vec{r}_2,\theta_2,t)
\bigg]\,,
\end{equation}
where the spatial integration goes over the area of the collision circle.

\begin{figure}
\begin{center}
\vspace{-1.0cm}
\includegraphics[width=4.1in,angle=0]{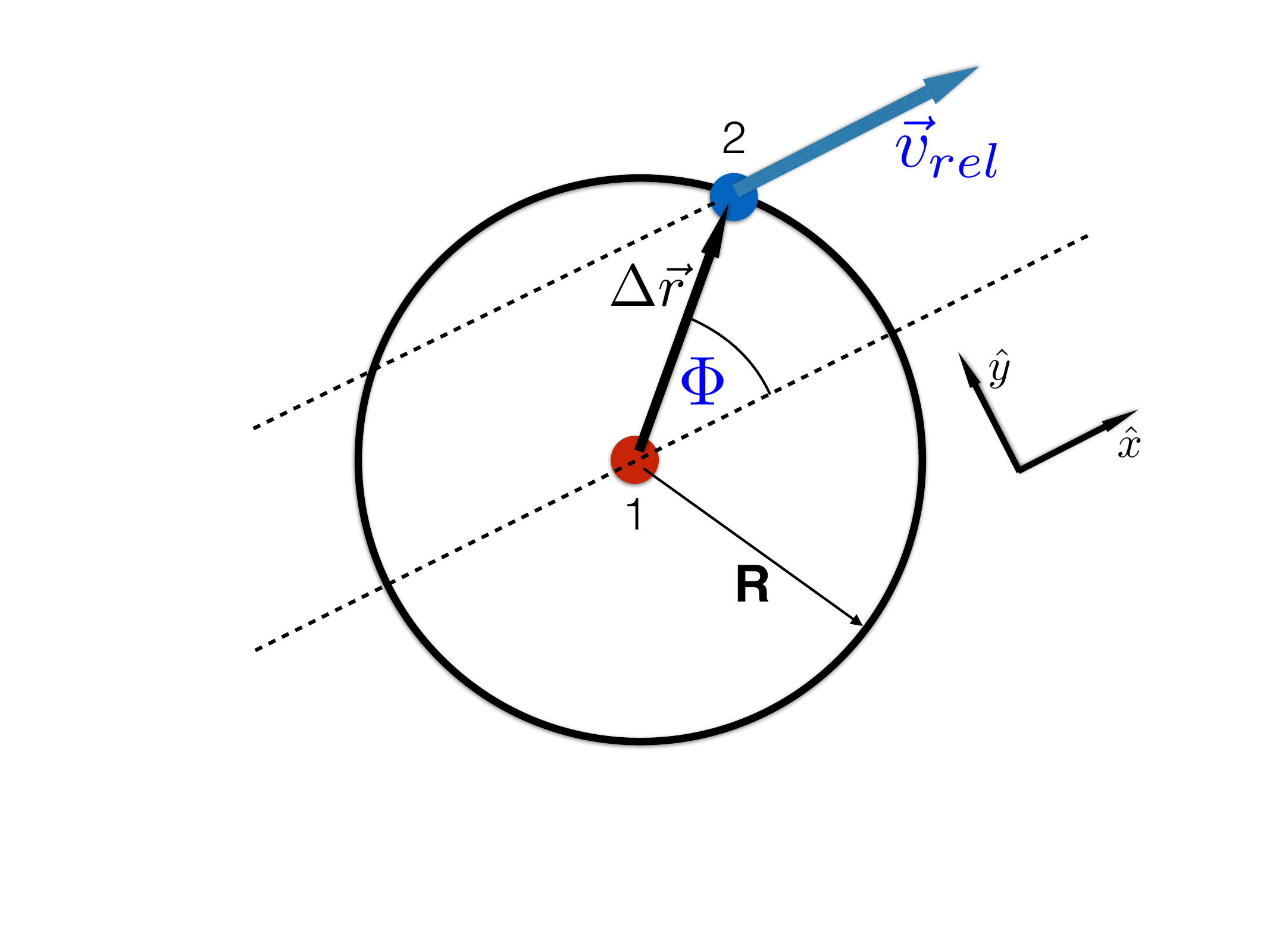}
\vspace{-2.0cm}
\caption{Schematic picture of two colliding particles in the frame of particle 1, where particle 2 has the relative velocity
$\vec{v}_{\text{rel}}=\vec{v}_2-\vec{v}_1$ and particle 1 is at rest. 
The x-axis is chosen parallel to $\vec{v}_{\text{rel}}$ to define the polar angle $\Phi$.
Here, particle 2 is {\em receding}, i.e. is about to leave the interaction circle of the focal particle because $\Phi\in [-\pi/2,\pi/2]$.
}
\label{FIG_COLLCIRCLE1}
\vspace{-0.3cm}
\end{center}
\end{figure}

Another consequence of the neglect of triple collisions is that $P_2$ can depend on $\vec{r}_2$ only through
the difference $\Delta \vec{r}=\vec{r}_2-\vec{r}$ which allows us to write $\partial_{\vec{r}}=-\partial_{\Delta \vec{r}}$
and $\partial_{\vec{r}_2}=\partial_{\Delta \vec{r}}$.
The two-dimensional Gauss-theorem for a vector field $\vec{A}$,
\begin{equation}
\int \diff \vec{r}\, \operatorname{div}\!\vec{A}=\oint \vec{A}\cdot \hat{\vec r}\,\diff s
\end{equation}
allows us to express the integral over the area of the collision circle as an integral over the edge of this circle.
Using the relative velocity of particle $2$ in the co-moving frame of the focal particle,  $\vec{v}_{\text{rel}}=\vec{v}_2-\vec{v}=v_0(\hat{\vec n}(\theta_2)-\hat{\vec n}(\theta))$, as new 
$x$-axis, we define the polar angle $\phi$ as the angle between $\vec{v}_{\text{rel}}$ and $\Delta \vec{r}$, see Fig.~\ref{FIG_COLLCIRCLE1}.
Substituting the arc length $\diff s=R\,\diff \phi$ and the radial vector $\hat{\vec r}=R(\cos\phi, \sin\phi)$ we arrive at
\begin{equation}
\label{JCOLL3}
J^{(\text{coll})}=(N-1)R \,\int_0^{2\pi} \diff\theta_2\int_0^{2\pi} \diff\phi\; 
v_{\text{rel}}\,\cos\phi\, P_2
\end{equation}
The advantages of this representation are that we only need to specify $P_2$ on the edge of the interaction circle and that the number
of integrations is reduced by one.
It is interesting to note that neither the interaction strength $\Gamma$ nor the interaction kernel $\sin(\theta_2-\theta)$
occur anymore in the collision integral. 
However, there has to be an implicit dependence on the coupling strength $\Gamma$. 
Therefore, as a consistence test, let us check the limit $\Gamma\to 0$ where particles evolve independently of each other and the collision integral is expected to vanish.
For a non-interacting system the two-particle density $P_2$ factorises exactly, that is
$P_2=P_1(\vec{r},\theta)\,P_1(\vec{r}_2,\theta_2)$. 
Since there is no dependence on the vector connecting both particles, and thus no dependence on $\phi$,
the integral over the polar angle gives zero. Thus, as expected, the collision integral vanishes at $\Gamma=0$.

This observation also means that a naive factorization of $P_2$ in \cref{JCOLL3} similar to the one in the derivation 
of the Vlasov equation cannot capture the effect of the interactions.
This is were the concept of {\em one-sided molecular chaos} comes into play, where the factorisation is only used when the two particles
start their interaction.
In Fig. \ref{FIG_COLLCIRCLE1} we see that if the angle
$\phi$ is between $\pi/2$ and $3\pi/2$, the two particles reduce their mutual distance from an initial distance $R$, 
i.e. are {\em approaching}.
If $\phi$ is between $-\pi/2$ and $+\pi/2$, the distance between them will be increasing, away from the initial value of $R$. 
In this case we have {\em receding} particles.
We split the integral over $\phi$ into two domains, one for approaching particles and one for receding particles,
and for a periodic integrand we have
\begin{equation}
\int_0^{2\pi}\diff\phi\ldots=
\int_{-\pi/2}^{\pi/2} \diff\phi\ldots+
\int_{\pi/2}^{3\pi/2}\diff\phi\ldots\quad\text{.}
\end{equation}
In the approaching domain, we assume molecular chaos at impact, that is 
we factorize $P_2=P_1(\vec{r},\theta)\,P_1(\vec{r}_2,\theta_2)$ and introduce the new variable $\hat{\phi}=\phi-\pi$, 
leading to $\cos\phi=-\cos\hat{\phi}$ and the same integration boundaries as in the receding domain. Relabeling $\hat{\phi}$ by $\phi$ again, we obtain:
\begin{equation}
\label{JCOLL4}
J^{(\text{coll})}=(N-1)R \,\int_0^{2\pi} \diff\theta_2\, v_{\text{rel}} \int_{-\pi/2}^{\pi/2} \diff\phi\;
\cos\phi\, \big[P_2|_{\text{rec}}-P_1(\vec{r},\theta,t)\,P_1(\vec{r}_2,\theta_2,t)\big]\bigg|_{\Delta r=R} \text{.}
\end{equation}
In the receding part of the integral, $P_2$ 
cannot be factorized because until this time $t$ (when the two particles are about to leave their encounter) 
they had continuously interacted over the time period $T_{\text{dur}}$ and thus, are correlated.
However, (i) we exactly know how they have interacted during that time, and (ii) we assume that they were uncorrelated at the earlier ``entrance'' 
time $t_0=t-T_{\text{dur}}$ when they 
first came in contact with
each other. 
Since we know the positions and angles of the particles when they emerge from their encounter at time $t$ 
(because this is given by the integration variables in the receding part of the collision integral) , and since the dynamics
is deterministic, we can integrate the microscopic evolution equations {\em backwards in time}  until 
the entrance time $t_0$ where the mutual distance is again at the value $R$. 
It turns out that this backtracing from the exit time $t$ to time $t_0<t$ can be done exactly for the noise-free VM,
see Appendix \ref{app:A} where the dynamics of two interacting particles is calculated explicitly. 

For a fluid with Hamiltonian dynamics, the so-called super position principle \cite{kreuzer_81} relates probability densities of particles
at different times. It therefore allows the factorization of $P_2$ using $P_1$ at earlier times, and hence incorporates the
 aforementioned 
assumption (ii) in a simple way. 
However, the superposition principle relies on the fact that the phase space compression factor \cite{evans_08} is zero  
in Hamiltonian systems, which is not the case here.
Therefore, in the following section we develop a modified superposition principle for the noise-free VM.

\subsection{Phase space compression and superposition principle}
\label{sec:compress}

From the Liouville theory of classical mechanics it is well-known that
the infinitesimal volume of phase space does not change along a trajectory and that the total time derivative
of the phase space density is zero.
However, if the equations of motion are not generated by a Hamiltonian, this is not necessarily the case.
Instead, a phase space compression factor $\Lambda_N$ \cite{evans_08} determines the total time derivative for a N-particle system, 
\begin{equation}
{\diff \over \diff t} P_N=-\Lambda_N\, P_N
\end{equation}
where $\Lambda_N$ depends on the phases of the system. For $N=2$ interacting particles,
the total time derivative reads,
\begin{equation}
{\diff \over \diff t} P_2(\vec{r}(t),\theta(t),\vec{z}(t),\beta(t),t)=
 \bigg[\partial_t+\dot{\vec{r}}\cdot\partial_{\vec{r}}+\dot{\theta}\,\partial_{\theta}
+      \dot{\vec{z}}\cdot\partial_{\vec{z}}+\dot{\beta}\, \partial_{\beta}
\bigg]\,P_2 \text{.}
\end{equation}
Inserting
the microscopic rules of the noise- and field-free VM, eqs.\eqref{POS_EQ} and \eqref{ANGLE_EQ}, 
at close range, results in
\begin{equation}
\label{TOTAL_DERIV}
{\diff \over \diff t} P_2=
 \bigg[\partial_t+v_0\hat{\vec n}(\theta)\cdot\partial_{\vec{r}}+\Gamma\sin(\beta-\theta)\,\partial_{\theta}
+      v_0\hat{\vec n}(\beta)\cdot\partial_{\vec{z}}+\Gamma\sin(\theta-\beta)\, \partial_{\beta}
\bigg]\,P_2\text{.}
\end{equation}
Using the product rule in the differentiations in the second hierarchy equation, eq. (\ref{SECOND_MEM}),
at zero noise and without external field, gives
\begin{equation}
\label{SECOND_VAR}
\big[\partial_t+v_0\hat{\vec n}(\theta)\cdot\partial_{\vec{r}}+\Gamma\sin(\beta-\theta)\,\partial_{\theta}
+      v_0\hat{\vec n}(\beta)\cdot\partial_{\vec{z}}+\Gamma\sin(\theta-\beta)\, \partial_{\beta}
\big]\,P_2=2\Gamma \cos(\beta-\theta)\,P_2 \text{.}
\end{equation}
Terms containing $P_3$ vanish exactly because only two particles exist in this case.
Replacing the right side of \cref{TOTAL_DERIV} by eq.~\eqref{SECOND_VAR} yields
\begin{equation}
\label{LAMBDA_EQ}
{\diff \over \diff t} P_2= 2\Gamma \,\cos(\beta-\theta)\,P_2
\end{equation}
where we can read off the phase space compression factor for the two-particle VM,
\begin{equation}
\label{LAMBDA_RES}
\Lambda_2=-2\Gamma \,\cos(\beta-\theta) \text{.}
\end{equation}
Taking time as the only independent variable, \cref{LAMBDA_EQ} can be integrated from time $t_0$ to $t$.
Changing variables $\vec{z}=\vec{r}_2$, $\beta=\theta_2$ one obtains:
\begin{align}
\begin{split}
	 &P_2(\vec{r}(t),\theta(t),\vec{r}_2(t),\theta_2(t),t)= \\
	\label{FACTOR_REC0}
	 &P_2(\vec{r}(t_0),\theta(t_0),\vec{r}_2(t_0),\theta_2(t_0),t_0)
	\;\mathrm{exp}\Big[2 \Gamma\int_{t_0}^t \diff \tilde{t}\, \cos\{\theta_2(\tilde{t})-\theta(\tilde{t})\}\Big] \text{.}
\end{split}
\end{align}
This is the superposition principle for the VM. It reduces to the common principle of a classical gas in the limit of vanishing
alignment, $\Gamma=0$. 
Note that the extended superposition principle depends on the microscopic details of the model.
As pointed out in ref.~\citenum{ihle_23a} the same result \eqref{FACTOR_REC0} can be obtained when 
the Liouville-equation of a two-particle system (within interaction range $R$)
is solved exactly by the method of characteristics, where the characteristics are given by the actual trajectories of the particles.

\subsection{One-sided molecular chaos and the forbidden zone}
\label{sec:one-sided}

\begin{figure}
\begin{center}
\vspace{-0.2cm}
\includegraphics[width=4.1in,angle=0]{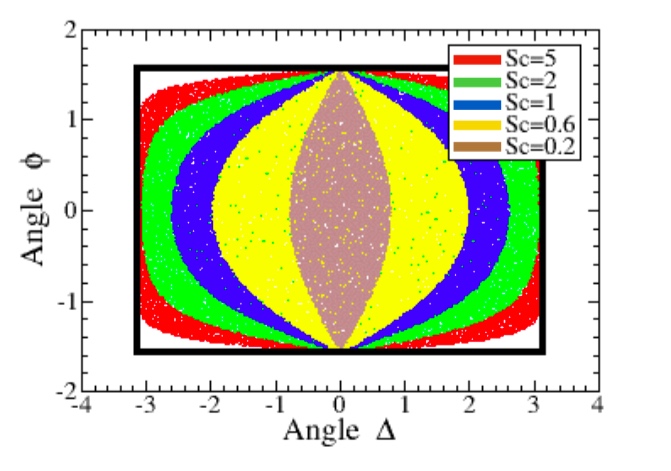}
\vspace{-0.5cm}
\caption{Forbidden zone in the angular plane of angles $\phi$ and $\Delta=\theta_2-\theta$.
The black box denotes the original integration domain, $\Delta \in [-\pi,\pi]$, $\phi \in [-\pi/2,\pi/2]$.
For $\Sc=0.2$, exit states in the inner seed-shaped brown area are forbidden, i.e. do not occur in reality and, thus, 
have zero probability. 
For $\Sc=0.6$ states in the yellow and brown areas
are forbidden, and so on.
The size of the forbidden area increases with anti-alignment strength $\Sc$. At large $\Sc\gg 1$, only states in the four 
corners of the original
integration area are ``allowed'', i.e. contribute to the collision integral.
}
\label{FIG_FORBIDDEN}
\vspace{0.3cm}
\end{center}
\end{figure}

We are now in position to complete the treatment of {\em one-sided molecular chaos} from section \ref{sec:boltz1}:
The quantity $P_2$ of receding particles that end their encounter at time $t$ 
is expressed as a functional of $P_1$ by means of the superposition principle under the assumption 
that the two particles were statistically independent at the earlier time
$t_0$, at the start of their encounter,
\begin{align}
\begin{split}
&P_2(\vec{r}(t),\theta(t),\vec{r}_2(t),\theta_2(t),t)\approx \\
\label{FACTOR_REC1}
& P_1(\vec{r}(t_0),\theta(t_0),t_0)\;P_1(\vec{r}_2(t_0),\theta_2(t_0),t_0)
\,\mathrm{exp}\Bigg[2 \Gamma\int_{t_0}^t \diff \tilde{t}\, \cos\{\theta_2(\tilde{t})-\theta(\tilde{t})\}\Bigg]
\end{split}
\end{align}
However, this factorization at an earlier time for receding particles fails for cases where the dynamics of the model
does not allow particular two-particle states $(\vec{r}(t),\theta(t),\vec{r}_2(t),\theta_2(t))$ at all. 
For $M\to 0$ these cases do not occur in reality but in the mathematical evaluation of the collision integral, and have to be properly taken into account.
As explained further below in the evaluation of \cref{TDUR1}, for anti-alignment interactions, that is for $\Gamma<0$, this often occurs for 
small
differences between the two angles at the exit time $t$. 
In those cases, we simply set $P_2(\vec{r}(t),\theta(t),\vec{r}_2(t),\theta_2(t),t)=0$ in the collision integral.
This amounts to a reduction of the integration domain in the collision integral for receding particles at exit time $t$.
The removed section of the integration domain will be called {\em forbidden zone} and is pictured 
in Fig.~\ref{FIG_FORBIDDEN}.
While there are also quite improbable two-particle states for approaching particles, this is always ignored
on the level of a Boltzmann-like description and is subject of further research. Within Boltzmann approaches,
the incoming particles are always considered as completely independent, something we know since 1970 from 
the work by Alder and Wainright \cite{alder_70} 
is not true for regular, classical fluids where momentum is conserved during collisions.

The ``price'' for a Boltzmann-like description, that is, for the simplicity of having an 
equation for $P_1$ alone,
is to sacrifice the complete knowledge of the time evolution of the system.
Our aim here is to obtain a description that is only valid on length scales of the mean free path 
$\lambda_{\text{free}}$ and on time
scales of the mean free time and beyond.
The binary collision approximation applied earlier only makes sense if the duration of the encounters $T_{\text{dur}}$ is much smaller
than $T_{\text{free}}$ and consequently, the interaction range $R$ must also be much smaller than $\lambda_{\text{free}}$.
This means, in a coarse grained description on the scales of $\lambda_{\text{free}}$ and $T_{\text{free}}$ we can assume
$t\approx t_0$ and
$\vec{r}(t)\approx \vec{r}_2(t)\approx \vec{r}(t_0)\approx \vec{r}_2(t_0)$ in the arguments of $P_1$ and $P_2$ in eq. (\ref{FACTOR_REC1}).
However, the angular changes, for example, $\theta(t)-\theta(t_0)$ during the collision, as well as the time difference $t-t_0$
in the phase space compression integral 
can be significant and have to be treated in detail. 
This coarse-graining leads to a simplification of eq. (\ref{FACTOR_REC1}):
\begin{align}\begin{split}
 P_2(\vec{r},\theta,\vec{r}_2,\theta_2,t)\approx 
\label{FACTOR_REC2}
&P_1(\vec{r},\theta_{\text{ent}},t)\,P_1(\vec{r},\theta_{2,\text{ent}},t)
\,\mathrm{exp}\Bigg[2 \Gamma\int_{t_0}^t \diff \tilde{t}\, \cos\{\theta_2(\tilde{t})-\theta(\tilde{t})\}\Bigg]
\end{split}\end{align}
where $\theta_{\text{ent}}$ and $\theta_{2,\text{ent}}$ are the angles of the two particles at the entrance time $t_0$ (when they first
started their encounter) that leads to their departure at the later time $t$ with angles $\theta$ and $\theta_2$.
Thus, these entrance angles are functions of the exit angles $\theta$ and $\theta_2$ as well as of the direction of the 
vector $\Delta \vec{r}$ which connects the particles at the exit time $t$.
These dependencies, such as $\theta_{\text{ent}}=\theta_{\text{ent}}(\theta,\theta_2,\Delta \vec{r})$ and $T_{\text{dur}}=T_{\text{dur}}(\theta,\theta_2,\Delta \vec{r})$
are analytically derived in section \ref{sec:change} using the results from Appendix \ref{app:A} for the dynamics of two particles.
For the noise-free VM, the calculations can be done exactly with the following results:
\begin{figure}
\begin{center}
\vspace{-1.0cm}
\includegraphics[width=4.1in,angle=0]{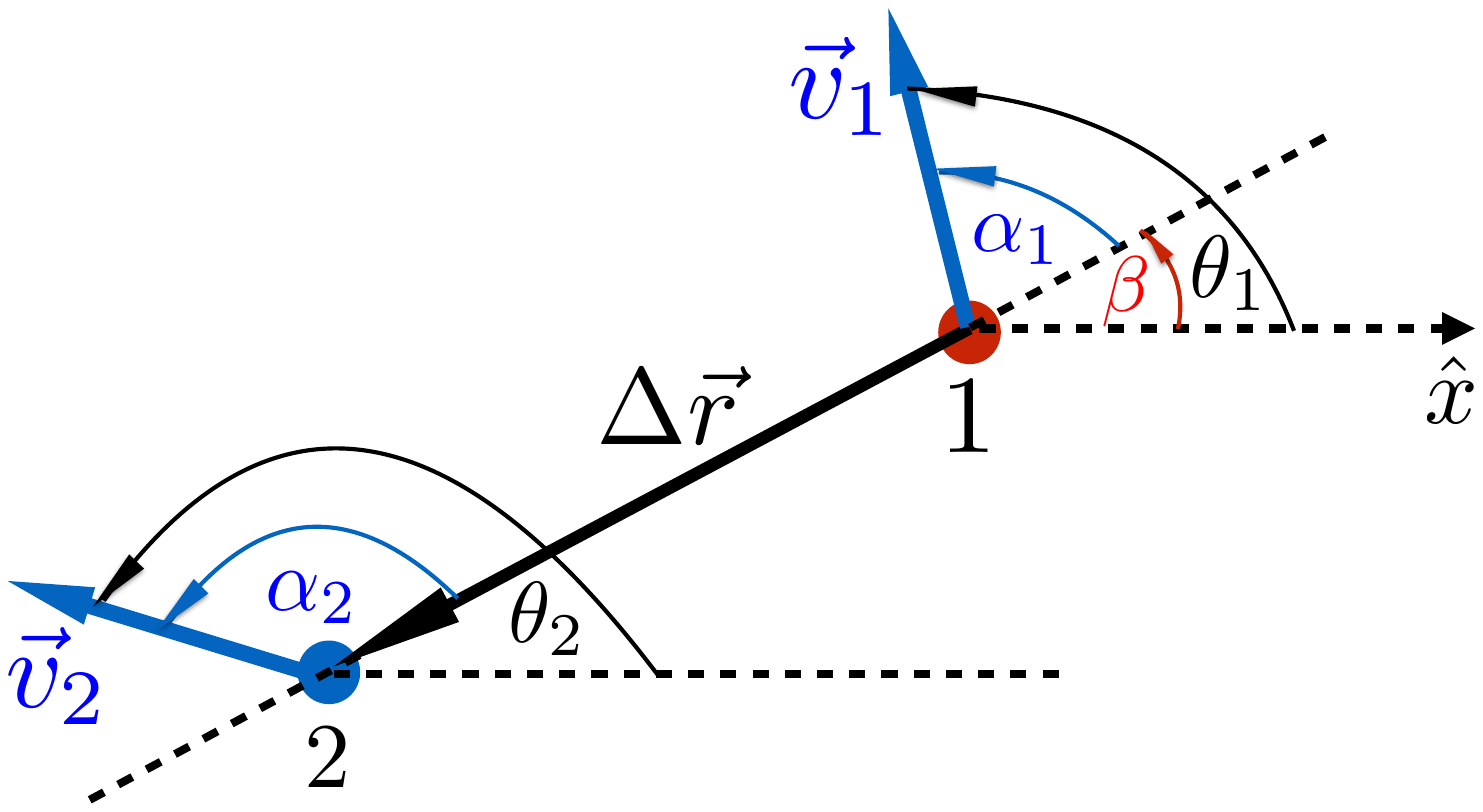}
\vspace{-2.0cm}
\caption{Schematic picture of particles 1 and 2 with the definition of the connecting vector $\Delta\vec{r}$, and the angles
$\alpha_1$, $\alpha_2$, $\beta$, $\theta_1$ and $\theta_2$.
}
\label{FIG_ALPHADEF}
\vspace{-0.3cm}
\end{center}
\end{figure}

In a two-particle interaction, the sum of the angles of the two involved particles
$\tilde{c}=\theta_2+\theta$ is conserved during the time evolution, see Appendix \ref{app:A}.
However,
the angular difference $\Delta= \theta_2-\theta$ changes during the backtracing from $t$ to $\tilde{t}$ 
with $t_0\le \tilde{t}\le t$
according to
\begin{equation}
\label{ANG_DIFF_RES}
\tan\Big[{\Delta(\tilde{t})\over 2}\Big]=
\tan\Big[{\Delta(t  )\over 2}\Big]\,
\mathrm{exp}\big[2\Gamma (t-\tilde{t})\big]
\end{equation}
For $\Gamma<0$ this means that the difference {\em decreases} in the backwards evolution.
The distance between the two particles evolves during backwards evolution in the following way, 
\begin{equation}
\label{DR_EVOLVE}
\Delta r(\tilde{t})=\sqrt{R^2+4v_0^2 G\Big(G+{R\over v_0}\, \sin{c\over 2}\Big)}
\end{equation}
where $G$ is a time-dependent function,
\begin{equation}
\label{GDEF}
G(t-\tilde{t})={1\over 2\,\Gamma}\Bigg[{\rm asinh}\mu-
{\rm asinh}\big\{\mu\, \mathrm{exp}(2\Gamma (t-\tilde{t}))\big\}
\Bigg]\,,
\end{equation}
with the abbreviation $\mu\equiv\tan[\Delta/2]$ where $\Delta\equiv \Delta(t)=\alpha_2(t)-\alpha_1(t)=\theta_2-\theta$.
The quantity $c$ in eq. (\ref{DR_EVOLVE}) is the sum of the flying directions $\alpha_1$ and $\alpha_2$ 
with respect to the connecting vector of the particles, see Fig.~\ref{FIG_ALPHADEF},
and
is taken at the (fixed) exit time $t$, $c\equiv\alpha_1(t)+\alpha_2(t)$.
According to eq. (\ref{GDEF}), at the exit time $\tilde{t}=t$, $G$ is equal to zero, and eq. (\ref{DR_EVOLVE})
delivers the required result: $\Delta r(t)=R$.
Tracing time backwards with $\tilde{t}<t$ (for particles that fulfill the receding condition), 
one sees that independent of the sign of the coupling constant $\Gamma$,
the particle's distance becomes initially smaller than $R$, i.e. the particles engage in the alignment interaction.
Typically, after a certain time, the distance starts increasing again.
At the particular time $\tilde{t}=t_0$ when the bracket in eq. (\ref{DR_EVOLVE}) becomes zero, 
the distance has reached the value $R$ 
again,
which defines the entrance time $t_0$, 
\begin{equation}
\label{ENTRANCE_COND}
G(t-t_0)+{R\over v_0}\, \sin{c\over 2}=0\text{.}
\end{equation}
Substituting $G$ from eq. (\ref{GDEF}) gives the following expression for the
duration $T_{\text{dur}}=t-t_0$ of the interaction,
\begin{equation}
\label{TDUR1}
T_{\text{dur}}={1\over 2\,\Gamma}\,{\rm ln}\Bigg\{
{{\rm sinh}\big[ {\rm asinh}(\mu)+{2\,\Gamma R\over v_0}\,\sin{c\over 2}\big]\over \mu}
\Bigg\}\,.
\end{equation}

The angular difference $\Delta(t_0)$ at the start of an interaction can be determined by inserting the calculated
duration of the interaction, eq. (\ref{TDUR1}), into eq. (\ref{ANG_DIFF_RES}), resulting into
\begin{equation}
\label{FINAL_DIFF1}
{\rm asinh}\Bigg[\tan \bigg({\Delta(t_0)\over 2}\bigg)\Bigg]
={\rm asinh}\Bigg[\tan \bigg({\Delta(t)\over 2}\bigg)\Bigg]+{2\,\Gamma R\over v_0}\,\sin{c\over 2}\text{.}
\end{equation}
One sees that, as expected, for $\Gamma=0$ there is no change of the angles, and for negative $\Gamma$ the
initial angular difference has a smaller absolute value than at the exit point.

If the argument of the logarithm in eq. (\ref{TDUR1}) is negative, there is no real solution for the time duration.
This indicates that the point $(\vec{r},\theta,\vec{r}_2,\theta_2)$ cannot be reached by the dynamics of the system, i.e.
amounts to a point in an inaccessible, or {\em forbidden} part of the 2-particle phase space.
Whenever two particles start an interaction they will never be able to leave their encounter 
with angles from that part of phase space. This is because a non-zero, negative 
interaction strength $\Gamma$ ensures that their
angular difference can not be too small at the exit point.
For positive $\Gamma$ and for receding collisions with $-\pi/2\leq \phi \leq \pi/2$, eq. (\ref{TDUR1}) always has a real 
solution for the duration time $T_{\text{dur}}$, hence, there is no forbidden zone.
Setting the argument of the logarithm to zero gives us criteria for the forbidden zones.
For $\Gamma<0$, a two-particle state is in the forbidden zone if
\begin{equation}
|\mu|\le {\rm sinh}\bigg(2\,\Sc \bigg|\sin{c\over 2}\bigg|\bigg)
\end{equation}
with $\Sc=|\Gamma| R/v_0$.
Thus, angular differences $|\Delta|$ assumed at the exit point of an interaction
that are smaller than the critical value $\Delta_C$,
\begin{equation}
\Delta_C=2 {\rm atan}\bigg[
{\rm sinh}\bigg(2\,\Sc \bigg|\sin{c\over 2}\bigg|\bigg)
\bigg]
\end{equation}
cannot occur. Hence, $P_2=0$ for this state.
Since $\sin(c/2)$ does also depend on $\Delta$, it is better to substitute it be means of eq. (\ref{A_LINK_C2})
in the argument of the logarithm in eq. (\ref{TDUR1}) and to use the addition theorem
${\rm sinh}(a+b)={\rm sinh}(a)\,{\rm cosh}(b)+{\rm cosh}(a)\,{\rm sinh}(b)$ to 
obtain an alternative condition for the forbidden zone in terms
of $\Delta$ and $\phi$ as (note that we are always assuming that $\phi$ is in the {\em receding} interval $[\-\pi/2,\pi/2]$
where $\cos\phi\ge 0$)
\begin{equation}
\label{ALTERNATE_C}
\sin{|\Delta|\over 2}< {\rm tanh}(2\Sc \cos\phi)
\end{equation}
where the critical value $\Delta_C$ follows as
\begin{equation}
\label{DELTA_CRIT}
\Delta_C=2\,{\rm asin}({\rm tanh}(2\Sc \cos\phi))\text{.}
\end{equation}
The maximum possible critical value is obtained for $\cos\phi=1$ (or $\sin(c/2)=1$, according to eq. (\ref{A_LINK_C2})) 
with
\begin{equation}
\label{DELTA_CRIT_MAX}
\Delta_{C,\text{max}}=2\, {\rm atan}\bigg[
{\rm sinh}\bigg(2\,\Sc \bigg)
\bigg]=2\,{\rm asin}({\rm tanh}(2\Sc))\,.
\end{equation}
For large coupling $\Sc\gg 1$, $\Delta_{C,\text{max}}$ approaches $\pi$,
\begin{equation}
\Delta_{C,\text{max}}\approx \pi-4 {\rm e}^{-2\Sc}\text{.}
\end{equation}
This is the expected result, because it 
means that whatever the angular difference at the entrance of a collision, at strong coupling particles always depart
in almost opposite directions.
Thus, exit states where the particle angles do not differ that strongly, cannot occur.

\subsection[]{Boltzmann scattering theory for weak anti-alignment\texorpdfstring{, $\Sc\ll 1$}{}}

\label{sec:weak}

\subsubsection{Handling of the forbidden zone}

For negative $\Gamma$, a forbidden section exists in phase space, see Fig.~\ref{FIG_FORBIDDEN}.
This can be handled by an appropriate
reduction of the integration domain in the collision integral, which will be discussed here.
Condition eq. (\ref{DELTA_CRIT}) can be rewritten as a condition for a critical angle $\phi_C$ for a given
angular difference $\Delta$ as
\begin{equation}
\label{CRITICAL_COS_PHI}
\cos\,\phi_C={1\over 2 \Sc}{\rm atanh}\Big[
\sin{|\Delta|\over 2}
\Big]\text{.}
\end{equation}
The forbidden zone occurs now for angles $\phi$ with $\cos\phi> \cos\phi_C$. 
However, on one hand, for angular difference $\Delta$ that are larger than $\Delta_{\text{max}}$ given in eq. (\ref{DELTA_CRIT_MAX}), 
the expression
for $\cos\phi_C$ gives a value larger than one.
This means, no angles $\phi$ (at fixed $\Delta$) qualify for the definition of the  forbidden zone.
This simply means that once $|\Delta|$ is sufficiently large, all values of $\phi$ from the relevant interval $-\pi/2\le \phi \le \pi/2$ are possible, i.e. lead to non-vanishing values of $P_2$.
On the other hand, for angular differences $|\Delta|<\Delta_{\text{max}}$, the cosine $\cos\phi_C$ is smaller than one
and, for fixed $\Delta$ there is now a finite (forbidden) interval $[-\phi_C,\phi_C]$ where $P_2=0$.
Here, we define $\phi_C$ as
\begin{equation}
\label{PHI_C_FULL}
\phi_C={\rm acos}\left\{
\operatorname{min}\left[1, {1\over 2 \Sc}{\rm atanh}\left(
\sin{|\Delta|\over 2}
\right)
\right]
\right\}
\end{equation}
where only non-negative values of the ${\rm acos}$ function are used.

Expanding eq.~(\ref{DELTA_CRIT}) for small $\Sc$ gives the critical angular difference
$\Delta_C$,
\begin{equation}
\Delta_C=4 \Sc \cos\phi\Bigg[1-{2\over 3}\Sc^2\,\cos^2\phi\Bigg]+\mathcal{O}(\Sc^5)
\end{equation}
such that for $|\Delta|\le \Delta_C$ the two-particle probability density $P_2$ vanishes
for receding particles. 
In linear order in $\Sc$, this gives a simple definition of the forbidden zone in terms
of the angle $\phi$ for fixed $\Delta$:
\begin{equation}
\cos\phi\ge \cos\phi_C={|\Delta(t)|\over 4\Sc}+\mathcal{O}(\Sc^2)\text{.}
\end{equation}
If $|\Delta(t)|$ is larger than $4\Sc$, there is no forbidden zone 
and all angles of $\phi$ are possible.
If $|\Delta(t)|$ is smaller than that, there is a finite interval $[-\phi_C,\phi_C]$ 
in which $P_2$ does vanish.
This motivates the following splitting of the integration over $\theta_2$ in the 
collision integral (assuming periodicity of the integrand):
\begin{equation}
\int_{-\pi}^{\pi} \diff\theta_2\ldots=
\int_{-4\Sc+\theta}^{4\Sc+\theta} \diff\theta_2\ldots+
\int_{-\pi+\theta}^{-4\Sc+\theta} \diff\theta_2\ldots+\int_{4\Sc+\theta}^{\pi+\theta} \diff\theta_2\ldots \, \text{.}
\end{equation}
The first integral contains the forbidden zone, meaning that not all angles $\phi$ are allowed in it.
In the following two integrals, all angles of $\phi$ are possible.
This leads to the final splitting of the two-dimensional collision integral
for receding particles:
\begin{align}
\begin{split}
	  \int_{-\pi}^{\pi} \diff\theta_2 \int_{-\pi/2}^{\pi/2} \diff \phi\ldots=&
	 \int_{-4\Sc+\theta}^{4\Sc+\theta} \diff \theta_2 
	\Bigg\{
	\int_{-\pi/2}^{-\phi_C} \diff\phi \ldots
	+\int_{\phi_C}^{\pi/2} \diff\phi \ldots
	\Bigg\}
	\\
	\label{FINAL_INTEGRAL_SPLIT}
	 &+\int_{-\pi+\theta}^{-4\Sc+\theta} \diff\theta_2 \int_{-\pi/2}^{\pi/2} \diff\phi\ldots
	+\int_{4\Sc+\theta}^{\pi+\theta} \diff\theta_2 \int_{-\pi/2}^{\pi/2} \diff\phi \ldots
\end{split}
\end{align}
where we have already removed the forbidden part because the integrand is zero there.
As long as $\Sc$ is not too large, $\Sc<\pi/4$, this splitting works, at least formally. 
Of course, at larger $\Sc$ the expression for $\phi_C$ is not quantitatively correct anymore and  
nonlinearities of $\mathcal{O}(\Sc^2)$ and higher, according to eq. (\ref{PHI_C_FULL}), need to be taken into account.

\subsubsection{The duration of the encounter}

Expanding the duration time, eq. (\ref{TDUR1}), for small $\Sc$ gives
\begin{equation}
T_{\text{dur}}={R\over v_0\,\mu} \sin{c\over 2}
\Bigg[
\sqrt{1+\mu^2}-{\Gamma R\over v_0 \mu} \sin{c\over 2}
+{4 \Gamma^2 R^2\over 3 v_0^2 \mu^2}\sin^2{c\over 2}
\Bigg]
\end{equation}
where ${\rm cosh}({\rm asinh}\mu)=\sqrt{1+\mu^2}$ was used.
For zero coupling, $\Gamma=0$, particles just fly straight through the interaction circle
and the duration becomes
\begin{equation}
\label{TDURNULL}
T_{\text{dur},0}={R\over v_0} \sin\Big({c\over 2}\Big)
{\sqrt{1+\mu^2}\over \mu}
={R\over v_0} {\cos\phi\over 
|\sin{\Delta(t)\over 2}|}
\end{equation}
where eqs.(\ref{REL_DEL_MU}, \ref{A_LINK_C2}) were used.
eq. (\ref{TDURNULL}) confirms that $T_{\text{dur},0}$ is always positive for a receding collision where $-\pi/2\leq \phi \leq \pi/2$.
For small $\Gamma\neq 0$, one has 
\begin{equation}
\label{TDUR_LINEAR1}
T_{\text{dur}}=T_{\text{dur},0}-{\rm sgn}(\Gamma)\,\Sc {R\over v_0}\, 
\Bigg({ \cos\phi\over \tan{\Delta(t)\over 2}} \Bigg)^2+\mathcal{O}(\Sc^2)
\end{equation}
Anti-alignment, i.e. $\Gamma<0$, leads to a decrease of the angular difference during back-tracing,
which makes it harder for the particles to separate from each other.
In contrast, for positive $\Gamma$ the angular difference {\em increases} in the back-wards time evolution, leading
to a faster increase of their distance, and thus to a {\em shorter} interaction time.
Thus, in first order in $\Sc$, for $\Gamma<0$, 
the duration of the encounter is increased compared to the non-interacting case,
as quantified by eq. (\ref{TDUR_LINEAR1}).
At first sight, this result seems to be counter-intuitive as it is well-known that particles stay together {\em longer}
due to ferromagnetic alignment and shorter if anti-alignment is present.
This puzzle is resolved by noting that we do not consider the average duration of all encounters but that we
select only receding particles, that is particles that just underwent a binary collision, and analyze the effect of 
the interaction on the duration of their encounter by {\em backtracing} in time.
In the former case, one would place the condition of an approaching pair of particles on the evaluation, whereas in the latter case,
the condition of a receding pair is used, resulting in qualitatively different outcomes.

\subsubsection{The change of the flying angles during the interaction}
\label{sec:change}

To properly close the first BBGKY-equation by 
the extended super position principle  (\ref{FACTOR_REC2}),
explicit relations for the
entrance angles, $\theta_{\text{ent}}=\theta  (t_0)$ and $\theta_{2,\text{ent}}=\theta_2(t_0)$ as functions of the exit angles
and the connecting angle $\phi$ are needed.
Because $\theta+\theta_2$ stays constant during  a two-particle collision, see Appendix \ref{app:A}, one has 
\begin{align}
\begin{split}
	\theta  (t_0)=&{1\over 2}\big[\theta(t)+\theta_2(t)-\Delta(t_0)\big] \\
	\label{ENTRANCE_ANGLE1}
	\theta_2(t_0)=&{1\over 2}\big[\theta(t)+\theta_2(t)+\Delta(t_0)\big]
\end{split}
\end{align}
Substituting eq. (\ref{A_LINK_C2}) into (\ref{FINAL_DIFF1}), the difference of the angles at entrance time $t_0$
is obtained as
\begin{equation}
\label{A_FINAL_DIFF2}
\Delta(t_0)
=2 {\rm atan}\Bigg\{
{\rm sinh}\Bigg[{\rm asinh}\Big[\tan \bigg({\Delta(t)\over 2}\bigg)\Big]+
{2 \Gamma R\over v_0}\, \cos\phi\;{\rm sgn}\Bigg(\sin{\Delta(t)\over 2}\Bigg)
\Bigg]\Bigg\}
\end{equation}
For weak coupling $\Sc=|\Gamma| R/v_0\ll 1$ and sufficiently large initial differences $\Delta(t)$ such that
$|\mu|\gg 2\,\Sc |\cos{\phi}|$,
expression (\ref{A_FINAL_DIFF2}) can be expanded in terms of
$\epsilon\equiv 2\Gamma R \cos{\phi}\,{\rm sgn}(\sin{\Delta(t)/2})/v_0$ as
\begin{align}
\begin{split}
	\Delta(t_0)=&2\,{\rm atan}\Bigg[
		\mu+\epsilon \sqrt{1+\mu^2}+{\epsilon^2\over 2} \mu+ {\epsilon^3\over 6} \sqrt{1+\mu^2}+\ldots\Bigg]\\
		= & \Delta(t)+{2\epsilon\over \sqrt{1+\mu^2}}-{\epsilon^2 \mu\over 1+\mu^2}+\mathcal{O}(\Sc^3) \\
			= & \Delta(t)+2\epsilon\,\left|\cos{\Delta(t)\over 2}\right|-{\epsilon^2\over 2}\,\sin\Delta(t)+\mathcal{O}(\Sc^3) \\
		\label{EXPAND_DELTA1}
		= & \Delta(t)+2\,\sin\Delta(t)\Bigg[ \Gamma T_{\text{dur},0}-{\Gamma^2 R^2\over v_0^2}\cos^2\phi\Bigg]+\mathcal{O}(\Sc^3)
\end{split}
\end{align}
where the last equality is only valid for $-\pi\leq \Delta \leq \pi$.
Combining eqs. (\ref{ENTRANCE_ANGLE1},~\ref{EXPAND_DELTA1}) we find for weak coupling:
\begin{eqnarray}
\theta  (t_0)&=&\theta(t)- \sin\Delta(t)\Bigg[ \Gamma T_{\text{dur},0}-{\Gamma^2 R^2\over v_0^2}\cos^2\phi\Bigg]+\mathcal{O}(\Sc^3) \\
\theta_2(t_0)&=&\theta_2(t)+\sin\Delta(t)\Bigg[ \Gamma T_{\text{dur},0}-{\Gamma^2 R^2\over v_0^2}\cos^2\phi\Bigg]+\mathcal{O}(\Sc^3)
\end{eqnarray}

\subsubsection{The phase space compression integral}

The integral in the extended superposition equation, eq. (\ref{FACTOR_REC2}), which reflects phase space compression, is transformed by means
of a trigonometric identity, 
\begin{equation}
J_S=\int_{t_0}^t \diff \tilde{t}\, \cos\{\theta_2(\tilde{t})-\theta(\tilde{t})\}\Bigg]=
\int_{t_0}^t \diff \tilde{t}\,
{1-\tilde{\mu}^2\over 1+\tilde{\mu}^2} \label{eq:Jsdef}
\end{equation}
with $\tilde{\mu}\equiv \tan(\theta_2(\tilde{t})-\theta(\tilde{t})/2$.
Using eq.(\ref{ANG_DIFF_RES}), it is rewritten as 
\begin{equation}
J_S=\int_{t_0}^t \diff \tilde{t}\,{1-\mu^2 \mathrm{exp}(4\Gamma(t-\tilde{t})) \over
1+\mu^2 \mathrm{exp}(4\Gamma(t-\tilde{t}))}\,,
\end{equation}
and by means of the transformation $x=\mathrm{exp}(2\Gamma(t-\tilde{t}))$ it is solved exactly as
\begin{eqnarray}
\nonumber
J_S&=&{1\over 2 \Gamma} \int_{\mu}^{\mu\, \mathrm{exp}(2\Gamma (t-t_0))} {1-x^2\over x(1+x^2)}\,\diff x \\
\nonumber
   &=&{1\over 2 \Gamma} \int_{\mu}^{\mu\, \mathrm{exp}(2\Gamma (t-t_0))} 
\left({1\over x}-{2x\over 1+x^2}\right)\,\diff x\\
\label{J_S_FOR_EXACT}
   &=& {1\over 2\Gamma}\left[
{\rm ln}{\mu\, \mathrm{exp}(2\Gamma (t-t_0))\over 1+\mu^2\, \mathrm{exp}(4\Gamma (t-t_0))}
-{\rm ln}{\mu\over 1+\mu^2}\right] \\
\nonumber
   &=& t-t_0-{1\over 2\Gamma}{\rm ln} {1+\mu^2\, \mathrm{exp}(4\Gamma (t-t_0))
\over 1+\mu^2} \\
\label{J_S_EXPRESS}
   &=&\cos\Delta(t)\,T_{\text{dur},0}-\Sc {\rm sgn}(\Gamma) {R\over v_0}
\cos^2\phi\left[{1\over \tan^2(\Delta/2)}+2\cos^2(\Delta/2)\right]+\mathcal{O}(\Sc^2)
\end{eqnarray}
Note that substituting $P_2$ as given in eq. (\ref{FACTOR_REC2}) into (\ref{JCOLL4}) amounts to a
{\em non-local} closure of the first BBGKY-equation. This is because in this kinetic equation for $P_1$ at angle $\theta$, 
one inserts a functional
of $P_1$ at the {\em different} angles $\theta_{\text{ent}}\neq \theta$ and $\theta_{2,\text{ent}}\neq \theta$.
In principle, there is also a non-locality in position and time, something we neglected in the Boltzmann-style coarse-graining
of eq.~(\ref{FACTOR_REC1}).

\subsubsection{Evaluating the collision integral}
\label{sec:eval_coll}
\vspace{0.5ex}
 
~\\
\noindent
{\bf Calculation of the {\em approaching} part}
\vspace{0.5ex}
 
\noindent
To evaluate the collision integral in Fourier-space, we introduce the angular Fourier transform
of the one-particle probability, $\hat{P}_n$,
\begin{equation}
\hat{P}_n(\vec{r},t)=\int_0^{2\pi}{\diff \theta \over 2 \pi} P_1(\vec{r},\theta,t){\rm e}^{-\ii n \theta}
\end{equation}
and its inverse
\begin{equation}
P_1(\vec{r},\theta,t)=\sum_{n=-\infty}^{\infty}\hat{P}_n(\vec{r},t)
{\rm e}^{\ii n \theta}
\end{equation}
Multiplying the collision integral, eq. (\ref{JCOLL4}), by ${\rm e}^{-\ii m \theta}/(2\pi)$ and integrating over $\theta$ gives its Fourier component
\begin{equation}
\hat{J}^{(\text{coll})}_m=\hat{J}^{(\text{rec})}_m+\hat{J}^{(\text{app})}_m \label{eq:decompJcoll}
\end{equation}
where the contribution from approaching particles follows as 
\begin{equation}
\hat{J}^{(\text{app})}_m=-(N-1)R \sum_{n_1,n_2} \hat{P}_{n_1} \hat{P}_{n_2}\,\int_0^{2\pi}{\diff \theta\over 2 \pi} 
\int_0^{2\pi} \diff \theta_2\, v_{\text{rel}}\,\int_{-\pi/2}^{\pi/2}\diff \phi\,\cos\phi\;
{\rm e}^{\ii \theta(n_1-m)+\ii n_2\theta_2}
\end{equation}
Inserting $v_{\text{rel}}=2 v_0 |\sin(\Delta/2)|$ from eq. (\ref{VREL_VS_SIN}) and performing the integration over $\phi$
gives
\begin{equation}
\label{J_FOUR_EXPR1}
\hat{J}^{(\text{app})}_m=-4 (N-1)R v_0 \sum_{n_1,n_2} \hat{P}_{n_1} \hat{P}_{n_2}\,\int_0^{2\pi}{\diff \theta\over 2 \pi} 
{\rm e}^{\ii \theta(n_1-m)} \hat{K}_m(\theta)
\end{equation}
with the integral $\hat{K}_m$,
\begin{equation}
\label{K_DEF1}
\hat{K}_m(\theta)= 
\int_0^{2\pi} \diff \theta_2\, \left|\sin{\theta_2-\theta\over 2}\right| \,
{\rm e}^{\ii n_2\theta_2}
\end{equation}
Since the integrand is periodic, we can rewrite this integral as
\begin{align}
\begin{split}
	\hat{K}_m(\theta)=& 
	\int_{\theta}^{\pi+\theta} \diff \theta_2\, \sin{\theta_2-\theta\over 2} \,
	{\rm e}^{\ii n_2 \theta_2}
	-\int_{-\pi+\theta}^{\theta} \diff \theta_2\, \sin{\theta_2-\theta\over 2} \,
	{\rm e}^{\ii n_2 \theta_2}\\
	=&
	\int_{\theta}^{\pi+\theta} {\diff \theta_2\over 2 \ii}\, 
	\left[{\rm e}^{\ii(\theta_2-\theta)/2}-{\rm e}^{-\ii(\theta_2-\theta)/2}\right]\,
	{\rm e}^{\ii n_2 \theta_2} \\
	&- \int_{-\pi+\theta}^{\theta} 
	{\diff \theta_2\over 2 \ii}\, 
	\left[{\rm e}^{\ii(\theta_2-\theta)/2}-{\rm e}^{-\ii(\theta_2-\theta)/2}\right]\,
	{\rm e}^{\ii n_2 \theta_2}\\
	\label{K_INT_RES}
	 =&{1\over (1/4)-n_2^2}\, {\rm e}^{\ii n_2 \theta}
\end{split}
\end{align}
This expression is well-defined for all mode numbers $n_2$ and is periodic in $\theta$
as expected. Furthermore, for $n_2=0$ it is positive and equal to $4$ as can be verified easily by
a direct integration of eq. (\ref{K_DEF1}) for $n_2=0$.
Substituting $\hat{K}_m$ from eq. (\ref{K_INT_RES}) into (\ref{J_FOUR_EXPR1}) and integrating over $\theta$ yields
the final result:
 \begin{equation}
\label{J_APPROACH_FINAL}
\hat{J}^{(\text{app})}_m=-4 (N-1)\,R\, v_0\, \sum_{n=-\infty}^{\infty} {\hat{P}_n\, \hat{P}_{m-n} \over (1/4)-(m-n)^2} 
\end{equation}
Note that the largest contribution to the $m$-th mode of $\hat{J}^{(\text{app})}$ comes
from the same mode $n=m$ of $\hat{P}_n$, and that the contribution of mode products 
with larger differences $|m-n|\gg1$ of the mode numbers
decays inversely proportional to the square of that difference. 
Thus, we expect that a kinetic description using just a few modes should be sufficient.
\vspace{1.5ex}

\noindent
{\bf Calculation of the {\em receding} part in first order in $\Sc$}

\noindent
The contribution to the collision integral from {\em receding} particles, written
in Fourier-space, is 
\begin{equation}
\label{J_REC_1}
\hat{J}^{(\text{rec})}_m=(N-1)R \sum_{n_1,n_2} \hat{P}_{n_1} \hat{P}_{n_2}\left(
\hat{H}_1+
\hat{H}_2+
\hat{H}_3\right)
\end{equation}
where we used
the integral-splitting of eq. (\ref{FINAL_INTEGRAL_SPLIT}) to define
 \begin{subequations}\label{H123_DEF}
\begin{eqnarray}
\label{H1_DEF}
& &\hat{H}_1=
\int_0^{2\pi}{\diff \theta\over 2 \pi}
\int_{-4\Sc+\theta}^{4\Sc+\theta} \diff \theta_2\,
\Bigg\{
\int_{-\pi/2}^{-\phi_C} \diff \phi\,Q(\theta,\theta_2,\phi)
+\int_{\phi_C}^{\pi/2} \diff \phi \,Q(\theta,\theta_2,\phi)
\Bigg\} \\
& &\hat{H}_2=
\int_0^{2\pi}{\diff \theta\over 2 \pi}
\int_{-\pi+\theta}^{-4\Sc+\theta} \diff \theta_2\, \int_{-\pi/2}^{\pi/2} \diff \phi\, Q(\theta,\theta_2,\phi) \\
& & \hat{H}_3=
\int_0^{2\pi}{\diff \theta\over 2 \pi}
\int_{4\Sc+\theta}^{\pi+\theta} \diff \theta_2\, \int_{-\pi/2}^{\pi/2} \diff \phi\, Q(\theta,\theta_2,\phi)
\end{eqnarray}
with the kernel
\begin{equation}
Q(\theta,\theta_2,\phi)=v_{\text{rel}}\,{\rm e}^{2\Gamma\,J_S}\,\cos\phi\,\mathrm{exp}\left[-\ii\theta m+\ii n_1\theta_{\text{ent}}+\ii n_2 \theta_{2,\text{ent}}\right] \label{eq:kernel}
\end{equation}	
\end{subequations}
where $J_S$, given in eq. (\ref{J_S_EXPRESS}), describes phase space compression, and where
$v_{\text{rel}}=2v_0|\sin(\Delta/2)|$.

The goal is to first calculate the total collision integral in first order, $\mathcal{O}(\Sc)$, in the coupling strength
$\Sc$. Higher order contributions are straightforward but tedious to calculate, and are treated further below.
We therefore approximate the kernel $Q$ in the definitions of $\hat{H}_2$ and $\hat{H}_3$ as
\begin{align}
\begin{split}
	Q\approx & 2 v_0 \left|\sin{\Delta\over 2}\right|\,
	(1+2\Gamma T_{\text{dur},0}\,\cos\Delta)\,\cos\phi\,{\rm e}^{\ii\theta(n_1-m)+\ii n_2 \theta_2}\\
	&  \times\left(1-\ii n_1 \Gamma T_{\text{dur},0}\,\sin\Delta\right) 
	\left(1+\ii n_2 \Gamma T_{\text{dur},0}\,\sin\Delta\right) \\
	 =&2 v_0 \left|\sin{\Delta\over 2}\right|\,\cos\phi\,{\rm e}^{\ii\theta(n_1-m)+\ii n_2 \theta_2}\\
	\label{KERNEL_APPROX1}
	&\times \left(1+\Gamma\,T_{\text{dur},0}\left[2 \cos\Delta+\ii (n_2-n_1) \sin\Delta\right]\right)	
\end{split}
\end{align}
We cannot use this simple expansion in the integral $\hat{H}_1$ because the duration of an encounter $T_{\text{dur},0}$ diverges \label{pageH1}
for $\Delta\to 0$, and $|\Gamma\,T_{\text{dur}}|\ge 1$ inside the integral. In this case, an exact treatment of the
phase space compression factor is pursued. 
Defining $\beta$ as
\begin{equation}
\beta={\rm asinh}\mu+{2\Gamma R\over v_0}\sin{c\over 2}=
{\rm asinh}\mu+{2\Gamma R\over v_0}{\rm sgn}\left(\sin{\Delta/2}\right)\cos\phi
\end{equation}
we obtain 
\begin{equation}
\label{EXACT_T2}
\mu{\rm e}^{2\Gamma T_{\text{dur}}}={\rm sinh}\beta=\tan{\Delta(t_0)\over 2}
\end{equation}
where the first equality follows from eq. (\ref{TDUR1}), and the second equality from (\ref{FINAL_DIFF1}).
Inserting eq. (\ref{EXACT_T2}) into eq. (\ref{J_S_FOR_EXACT}), the phase space compression factor is
given by
\begin{align}
{\rm e}^{2\Gamma J_S}=&{ {\rm sinh}\beta \over 1+{\rm sinh}^2\beta} {1+\mu^2\over \mu}
={ \sin\Delta(t_0)\over \sin\Delta(t)} \\
=&1+\varepsilon\left[{\sqrt{1+\mu^2}\over \mu}-{2\mu\over \sqrt{1+\mu^2}}\right]
+\varepsilon^2{\mu^2-5\over 2(1+\mu^2)}+\mathcal{O}(\Sc^3) \\
\begin{split}
 =&1-2\,{\rm sgn}(\Gamma)\,\Sc \cos\phi\,{\rm sgn}\left(\tan{\Delta\over 2}\right)\,
\left\{ 2\sin{\Delta\over 2}-{1\over \sin{\Delta\over 2}} \right\}
\\
 &+2\,\Sc^2\,\cos^2\phi {\mu^2-5\over 1+\mu^2}+\mathcal{O}(\Sc^3)\end{split}
\end{align}
with $\varepsilon\equiv 2\,{\rm sgn}(\Gamma)\,\Sc {\rm sgn}\left(\sin{\Delta\over 2}\right)\,\cos\phi$.
The integral $\hat{H}_1$ takes the following form,
\begin{equation}
\hat{H}_1=2 v_0
\int_0^{2\pi}{\diff \theta\over 2 \pi}
\int_{-4\Sc+\theta}^{4\Sc+\theta} \diff \theta_2\,
\int_{\Omega} \diff \phi\, \cos\phi\, \left|\sin{\Delta\over 2}\right| 
{ {\rm sinh}\beta \over 1+{\rm sinh}^2\beta} {1+\mu^2\over \mu} {\rm e}^{\ii S}
\end{equation}
where $\Omega$ is the domain specified in eq. (\ref{H1_DEF}) for the integration over $\phi$, and $S(\theta,\theta_2,\phi)$ 
is some phase.
Transforming the $\theta_0$-integral to the new variable $x=\Delta/(4 \Sc)=(\theta_2-\theta)/(4 \Sc)$
one has
\begin{equation}
\hat{H}_1=8 \Sc v_0
\int_0^{2\pi}{\diff \theta\over 2 \pi}
\int_{-1}^{1} \!\diff x\!
\int_{\Omega} \diff \phi\,  \cos\phi\,\left|\sin(2\Sc x)\right| 
{ {\rm sinh}(2\Sc\tilde{\beta}) \over 1+{\rm sinh}^2(2\Sc \tilde{\beta})} {1+\tan^2(2 \Sc x)\over \tan(2 \Sc x)} 
{\rm e}^{\ii S}
\end{equation}
with $\tilde{\beta}={\rm asinh}[\tan(2\Sc x)]/(2 \Sc)+{\rm sgn}(\Gamma)\,{\rm sgn}(x)\cos\phi$.
Since $x$ is of order one, the ${\rm sinh}$, $\sin$,$\tan$ and ${\rm atan}$ functions can be expanded for small $\Sc$
with the result
\begin{equation}
\hat{H}_1=16\, \Sc^2\, v_0
\int_0^{2\pi}{\diff \theta\over 2 \pi}
\int_{-1}^{1} \diff x\,
\int_{\Omega} \diff \phi\, \cos\phi\, {\rm sgn}(x)\, 
 (x+{\rm sgn}(\Gamma)\,{\rm sgn}(x)\,\cos\phi)  
\,{\rm e}^{\ii S}+\mathcal{O}(\Sc^3)
\end{equation}
Thus, $\hat{H}_1$ is of order $\mathcal{O}(\Sc^2)$ and will be neglected in this first order approach.
Inserting the approximated kernel $Q$ from eq. (\ref{KERNEL_APPROX1}) into the expressions for $\hat{H}_2$ and $\hat{H}_3$
and integrating over $\phi$ gives
\begin{equation}
\hat{H}_j=\int_{0}^{2\pi}{\diff \theta\over 2 \pi}
\,{\rm e}^{\ii\theta(n_1-m)}\left\{4 v_0 B_{j1}+\Gamma R \pi 
\left[2 B_{j2}+\ii (n_2-n_1) B_{j3}\right]\right\}
\end{equation}

for $j=2,3$ and the auxiliary integrals
\begin{align}
\begin{split}
	B_{21}=&-\int_{-\pi+\theta}^{-4 \Sc+\theta}\diff \theta_2\, \sin{\theta_2-\theta\over 2}\,{\rm e}^{\ii n_2 \theta_2}\\
	B_{22}=&\int_{-\pi+\theta}^{-4 \Sc+\theta}\diff \theta_2\, \cos(\theta_2-\theta)\,{\rm e}^{\ii n_2 \theta_2}\\
	B_{23}=&\int_{-\pi+\theta}^{-4 \Sc+\theta}\diff \theta_2\, \sin(\theta_2-\theta)\,{\rm e}^{\ii n_2 \theta_2}\\
	B_{31}=&\int_{4 \Sc+\theta}^{\pi+\theta} \diff \theta_2\, \sin{\theta_2-\theta\over 2}{\rm e}^{\ii n_2 \theta_2}\\
	B_{32}=&\int_{4 \Sc+\theta}^{\pi+\theta} \diff \theta_2\, \cos(\theta_2-\theta)\,{\rm e}^{\ii n_2 \theta_2}\\
	B_{33}=&\int_{4 \Sc+\theta}^{\pi+\theta} \diff \theta_2\, \sin(\theta_2-\theta)\,{\rm e}^{\ii n_2 \theta_2}
\end{split}
\end{align}

One finds,
\begin{equation}
B_{31}={1\over  (1/2)-2n_2^2} {\rm e}^{\ii n_2 \theta}
\left[2 n_2 \ii (-1)^{n_2}+(1-4\Sc \ii \,n_2) {\rm e}^{4 \Sc \ii \,n_2}\right]+\mathcal{O}(\Sc^2)
\end{equation}
and the exact relation $B_{21}(n_2)={\rm e}^{2\ii n_2 \theta}\,B_{31}(-n_2)$.
Thus, the sum is
\begin{equation}
B_{31}+B_{21}=
{1\over  (1/4)-n_2^2} {\rm e}^{\ii n_2 \theta}
\left[ \cos(4 \Sc n_2)+4\Sc n_2\,\sin(4 \Sc n_2)\right]+\mathcal{O}(\Sc^2)
\end{equation}
This sum
has the expected limit for $\Sc\to 0$ but no contribution
in $\mathcal{O}(\Sc)$:
\begin{equation}
B_{31}+B_{21}=\hat{K}_M(\theta)+\mathcal{O}(\Sc^2).
\end{equation}
Since we only need the sum of $\hat{H}_2$ and $\hat{H}_3$ in linear order in $\Sc$,
it suffices to evaluate the following sums
\begin{eqnarray}
\left. B_{22}+B_{32}\right|_{\Sc=0}&=& \int_{-\pi+\theta}^{\pi+\theta} \cos(\theta_2-\theta)\,{\rm e}^{\ii n_2 \theta_2}\,\diff \theta_2=
\pi {\rm e}^{\ii n_2 \theta}\left[\delta_{n_2,-1}+\delta_{n_2,1} \right] \\
\left. B_{23}+B_{33}\right|_{\Sc=0}&=&\int_{-\pi+\theta}^{\pi+\theta} \sin(\theta_2-\theta)\,{\rm e}^{\ii n_2 \theta_2}\,\diff \theta_2=
-i\,\pi {\rm e}^{\ii n_2 \theta}\left[\delta_{n_2,-1}-\delta_{n_2,1} \right] 
\end{eqnarray}
Performing the sum 
\begin{equation}
\hat{H}_1+\hat{H}_2+\hat{H}_3=\delta_{m,n_1+n_2}
\left\{
\Gamma R \pi^2 
\left[ (1-n_1)\delta_{n_2,-1}+(1+n_1)\delta_{n_2,1}\right]+
4 v_0 {1\over (1/4)-n_2^2} \right\}
\end{equation}
and inserting into eq. (\ref{J_REC_1}) gives the collision contribution from the receding particles as
\begin{equation}
J^{(\text{rec})}_m=4 v_0 (N-1) R \sum_n {\hat{P}_n \hat{P}_{m-n}\over (1/4)-(m-n)^2}
+(N-1) \Gamma R^2 \pi^2 m 
\left[\hat{P}_{m-1}\hat{P}_1-\hat{P}_{m+1}\hat{P}_{-1}\right] \text{.}
\end{equation}
Adding both contributions from receding and approaching particles leads to the final result
for the collision integral in Fourier space at order $\mathcal{O}(\Sc)$,
\begin{equation}
\label{J_COLL_1st_SC}
J^{(\text{coll})}_m=J^{(\text{rec})}_m+J^{(\text{app})}_m=
(N-1) \Gamma R^2 \pi^2 m
\left[\hat{P}_{m-1}\hat{P}_1-\hat{P}_{m+1}\hat{P}_{-1}\right]+\mathcal{O}(\Sc^2)
\end{equation}
Remembering that $P=f/N$ and $\hat{P}_n/N=\hat{f}_n/N$, one realizes that the collision integral 
for large particle number, $N\gg1$, and in linear order in $\Sc$,
is {\em identical} to the one we had obtained by the much simpler mean-field approximation in section 
\ref{sec:vlasovkinet}.
This is rather surprising as we explicitly treated the (correlated) alignment interactions over the duration
of the collision encounter.
In the following section we will see that 
the difference between the simple factorization approximation and the one-sided molecular chaos assumption
shows up for the first time in second order in the coupling strength $\Sc$.
This is partly due to the fact that the effect of the ``forbidden zone''
did not enter the calculations in linear order in $\Sc$.
\vspace{1.5ex}

\noindent
{\bf Calculation of the {\em receding} part in second order in $\Sc$}

\noindent
We find 
\begin{align}
\begin{split}
\hat{H}_2+\hat{H}_3=&2v_0\delta_{n_1+n_2,m}\left\{ 2 B_1(\pi)-{\pi \over 2}\,\Sc 
\left[\ii \Delta n\,B_3(\pi)+2 B_2(\pi)\right]\right. \\
& +\left. {4\over 3}\,\Sc^2\left[-{(\Delta n)^2\over 2}A_1(\pi)+2I(\pi)+2\ii\,\Delta n\, A_2(\pi)-\ii\,\Delta n\,A_3(\pi)\right]\right. \\
\label{SECOND_ORD_1}
&  \left. -2 B_1(4\Sc)+{\pi\over 2}\Sc\left[\ii\,\Delta n\,B_3(4\Sc)+2 B_2(4\Sc)\right]+\mathcal{O}(\Sc^3)\right\}
\end{split}
\end{align}
with $\Delta n\equiv n_2-n_1$.
The quantities $A_i$ and $B_i$ are angular integrals which are defined and calculated in Appendix \ref{app:C}.
The auxiliary quantity $I$ in eq. (\ref{SECOND_ORD_1}) can be expressed in terms of integrals $A_i$ as
\begin{equation}
I(\pi)=\int_{-\pi}^{\pi} {\mu^2(x)-5 \over 1+\mu^2(x)}\,\left| \sin{x\over 2}\right|\,{\rm e}^{\ii n_2x}\,\diff x=
-A_1(\pi)-A_4(\pi)
\end{equation}
with $\mu(x)=\tan(x/2)$.
The part $\hat{H}_1$ is of second order in $\Sc$, and one obtains 
\begin{equation}
\hat{H}_1=16\, \Sc^2\,v_0 \delta_{n_1+n_2,m}\,J+\mathcal{O}(\Sc^3)
\end{equation}
where $J$ is the following double integral:
\begin{equation}
J=\int_{-1}^1\,\diff x\left\{ \int_{-\pi/2}^{-\phi_c}\,\diff \phi\, (|x|-\cos\phi)\cos\phi+
\int_{\phi_c}^{\pi/2}\,\diff \phi\, (|x|-\cos\phi)\cos\phi\right\}
\end{equation}
which can be written as $J=J_0-J_1$ with
\begin{eqnarray}
J_0&=&\int_{-1}^1\,\diff x\, \int_{-\pi/2}^{\pi/2}\,\diff \phi (|x|-\cos\phi)\cos\phi=2-\pi \\
J_1&=&\int_{-1}^1\,\diff x\, \int_{-\phi_c}^{\phi_c}\,\diff \phi (|x|-\cos\phi)\cos\phi
\end{eqnarray}
Since at this order in $\Sc$ we have $\cos\phi_c\approx |x|$, the integral can be solved by the transformation $x=\cos y$,
and one finds $J_1=-4/3$.
Thus, finally we obtain
\begin{equation}
\hat{H}_1=16\, \Sc^2\,v_0 \delta_{n_1+n_2,m}\left[ {10\over 3}-\pi\right]
\end{equation}
Adding $\hat{H}_1$ to $\hat{H}_2+\hat{H}_3$, inserting in eq. (\ref{J_REC_1}) and collecting only terms of order $\Sc^2$ one finds
the following addition to the collision integral $\hat{J}_m^{(\text{coll})}$ that goes beyond naive mean-field theory:
\begin{align}
\begin{split}
 &(N-1)R v_0\,\Sc^2\,\sum_{n_1,n_2}\hat{P}_{n_1}\hat{P}_{n_2} \delta_{n_1+n_2,m}\left\{ {8\over 3}
\left[
{1\over n_2^2-{1\over 4}} \left(1-3\Delta n\,n_2+{1\over 2}\Delta n^2\right) 
\right.\right. \\
& \left.\left.
+{1\over n_2^2-{9\over 4}}\left(9-5\Delta n\,n_2+{3\over 2}\Delta n^2\right)
\right]
\right\}
\end{split}
\end{align}
This new contribution can be written by means of a coupling matrix $g_{m,n}$,
\begin{equation}
\label{J_COLL_NEW_FIN}
\hat{J}_m^{(new)}=(N-1)R v_0\,\Sc^2\,\sum_{n=-\infty}^{\infty}\hat{P}_{n}\hat{P}_{m-n}
g_{m,n}
\end{equation}
with
\begin{equation}
\label{G_MN_DEF}
g_{m,n}={8\over 3} m\left[ {{3\over 2}m-n\over (m-n)^2-{1\over 4}}+{n+{1\over 2}m \over (m-n)^2-{9\over 4}}\right]
\end{equation}
The coupling matrix has a rather intricate form and it is useful to establish symmetry requirements to check its consistency.
Because of mass conservation, the mode $\hat{f}_0=\rho_0/(2\pi)$ should never change in a homogeneous system.
As a consequence, the collision integral should be zero for $m=0$. This amounts to a non-trivial cancellation of terms
in the contributions $\hat{H}_j$, $j=1,2,3$. Thus, the coupling matrix $g_{m,n}$ should have the property
$g_{0,n}=0$ for all $n$.
The distribution $f$ is a real function and therefore its complex Fourier coefficients obey the following relation:
$\hat{f}_{-k}=\hat{f}_k^*$. 
Considering only real Fourier coefficients (which amounts to solutions that are symmetric with respect to the x-axis) 
one then
expects the following symmetry relation for the coupling matrix
\begin{equation}
g_{-m,-n}=g_{m,n}
\end{equation}
The matrix $g_{m,n}$ in eq. (\ref{G_MN_DEF}) possesses both of the required properties.
The validity of the expression for the coupling matrix is further supported numerically in agent-based simulations 
in section \ref{sec:numer}, where the temporal relaxation of the Fourier modes $\hat{f}$ is measured and compared to the theoretical 
predictions of
the one-sided molecular chaos approximation.

Collecting all terms, we find that the Fourier representation Boltzmann equation is given by
\begin{equation}
	\begin{split}
\partial_t \hat{P}_m =& {v_0\over 2}\left[\nabla^*\hat{f}_{m-1}
+\nabla \hat{f}_{m+1}\right] \\&+ (N-1) \Gamma R^2 \pi^2 m [\hat{P}_{m-1} \hat{P}_1-\hat{P}_{m+1}\hat{P}_{-1}]\\&+(N-1)\frac{R^3}{v_0} \Gamma^2  \sum_{n=-\infty}^{\infty} \hat{P}_n\hat{P}_{m-n} g_{mn}
	\end{split}
\end{equation}
with the coupling matrix given by \cref{G_MN_DEF} and terms sorted by order in coupling. In anticipation of section~\ref{sec:disorder}, we can attest validity for any sign of $\Gamma$ to it, even though the binary collision approximation becomes invalid for aligning coupling. As a proper Boltzmann equation it is the correct kinetic equation to first order in density ($M$). Here we have evaluated the coupling matrix analytically and are, therefore, limited to second order in coupling. Evaluation of the integrals aside, the method itself is non-perturbative in the couplings and can be evaluated (likely numerically) in the strong coupling limit as well. 

\subsection{Self-diffusion and velocity autocorrelation: Boltzmann--Lorentz theory}
\label{sec:selfdiffusion}

In general, a Boltzmann equation describes the advection and binary collisions of particles 
in terms of their probability density $f$.
Therefore it contains information about how the particle velocities change during collisions, and thus about the
velocity autocorrelation function (vaf). Since one-sided molecular chaos is assumed in the derivation of the Boltzmann equation,
particles are expected to ``forget'' previous encounters before an interaction starts with a new partner. 
This means, subsequent collision events are uncorrelated and the vaf should decay 
exponentially, at least for time scales larger than $R/v_0$ and in situations where the Boltzmann equation is expected to be asymptotically exact, i.e. for 
$M\to 0$ and $\Sc\to 0$. 
A similar assumption was made for regular fluids before the numerical discovery of 
long-time tails in 1970 \cite{alder_70}, where it was observed that the vaf showed a power law decay $\sim t^{-d/2}$ at long times.
Here, $d$ denotes the spatial dimension.
It was shown that these tails are a consequence of the back-flow effect which relies on momentum-conservation 
\cite{dorfman_70,ernst_70,kawasaki_71,zwanzig_book}.
However, momentum-conservation does not hold in the ``artificial fluid'' of self-propelled particles considered here
and other sources of such tails seem to be absent in this system of point particles at low densities $M\ll 1$.\footnote{There is, however, another quantity that is conserved during interactions: the sum of the angles
of the involved particles, $\sum_j \theta_j$.~\cite{boltz2025,mihatsch2025}}
Hence, long-time tails are not plausible in the vaf of the current system; a purely exponential decay of the vaf is 
expected,  
\begin{equation}
\label{DEF_VAC1}
C(t)=\langle \vec{v}(0)^2 \rangle\, {\rm e}^{-t/\tau_C}\,,
\end{equation}
at least for times much larger than the duration of a particle collision, $T_{\text{dur}}$. 
According to the Green-Kubo relation,
\begin{equation}
D={1\over d}\int_0^{\infty}\langle \vec{v}(t)\cdot\vec{v}(0)\rangle\,\diff t
\end{equation}
which is valid in the stationary state of any fluid, 
the self-diffusion coefficient $D$ is given by the time integral over the vaf. 
Relying on the exponential behavior of the vaf, we can easily deduce its correlation time $\tau_C$ from
the self-diffusion coefficient as
\begin{equation}
\label{TAU_C_DEF}
\tau_C={2 D\over v_0^2}
\end{equation}
Therefore,  in order to find $\tau_C$ it suffices to calculate $D$.
This can be done by the so-called Boltzmann--Lorentz theory, see for example ref.~\citenum{hauge_70} and its applications 
to granular gases \cite{brey_99,garzo_03}.
The main idea is to suppose that several particles are tagged but otherwise all
particles are mechanically equivalent. Then, the system is formally considered as a binary system where a population
of tagged particles is immersed in a sea of untagged background particles. 
For our purposes, we tag only one particle, in particular particle $i=1$, and introduce the tagged particle density
$h$ as the ensemble average of the corresponding one-particle phase space density,
\begin{equation}
h(\vec{r},\theta)\equiv \langle \delta(\vec{r}-\vec{r}_1(t))\,\delta(\theta-\theta_1(t))\rangle
\end{equation}
The density of the remaining particles is given by 
\begin{equation}
\label{f_tilde_def}
\tilde{f}(\vec{r},\theta)\equiv \Big\langle \sum_{j=2}^N\delta(\vec{r}-\vec{r}_j(t))\,\delta(\theta-\theta_j(t) \Big\rangle
\end{equation}
In the thermodynamic limit (td) $N\to \infty$, it does not matter whether one particle is 
omitted in the summation 
of eq. (\ref{f_tilde_def}) or not, and $\tilde{f}$ agrees with the function $f$ defined previously, $f=NP_1$. 
In abstract notation, the collision term of the nonlinear Boltzmann equation is given as a functional of $f$ by $J[f,f]$.
Here, the first argument in $J$ denotes the function whose evolution is considered; for example $J[f,h]$ would occur
in an equation for the density $f$ and describes scatterings of particles from the f-population on members 
of the h-population. Hence, in general $J[f,h]\neq J[h,f]$.
The evolution equation for the tagged particle density contains a collision term of the {\em same} functional form
$J[h,f]$ because it describes the collision of particle 1 with the other mechanically identical particles. 
An additional term of type $J[h,h]$ would reflect collisions among tagged particles, which are impossible with only one tagged particle. In the case of a few tagged particles $N_S$, these collisions are negligible as 
their density $N_S/V$ goes to zero
in the thermodynamic limit $V\to \infty$ with $N/V=const$. 
This also means that the collision term in the  evolution equation for $h$, a particular example of the {\em Boltzmann--Lorentz equation}, is {\em linear} in $h$. 
In the thermodynamic-limit, the evolution equation for $\tilde{f}\approx f$ is decoupled from the tagged density $h$, 
because the collision term $J[\tilde{f},h]$ is smaller than $J[\tilde{f},\tilde{f}]$ by a factor of N.

As explained above, the collision term $J[h,f]$ of the Boltzmann--Lorentz equation and its Fourier-transformed version 
does not have to be rederived, 
it follows directly from eq. (\ref{J_COLL_1st_SC}) and (\ref{J_COLL_NEW_FIN}) by formally replacing $\hat{f}_k$ by $\hat{h}_k$ at the 
appropriate position. Then, the Boltzmann--Lorentz equation for the angular Fourier components
of the tagged particle density becomes: 
\begin{eqnarray}
\!\!\!\!\!\!\!\!\partial_t\hat{h}_m{+}{v_0\over 2}\left[\nabla^* \hat{h}_{m{-}1}{+}\nabla \hat{h}_{m+1}\right]\!\!
\!&=&\!\!\!\Gamma R^2 \pi^2m\left[\hat{h}_{m{-}1}\hat{f}_1{-}\hat{h}_{m+1}\hat{f}_{{-}1}\right]
   {+}R v_0 \Sc^2\!\!\!\sum_{n=-\infty}^{\infty}\!\!\!\hat{h}_{n}\hat{f}_{m{-}n}
g_{m,n}
\label{TAGGED_EVOLV}
\end{eqnarray}
where the coefficients $g_{m,n}$ are given in eq. (\ref{G_MN_DEF}), and the terms of order $\mathcal{O}(\Sc^0)$ follow from the 
hierarchy
of the Vlasov-like approach, eq. (\ref{VLASOV_HIERA}).
The first three members of the hierarchy (\ref{TAGGED_EVOLV}) read 
\begin{subequations}\label{HIERARCH_TAGG2}
\begin{align}
\partial_t\hat{h}_0+{v_0\over 2}\left[\nabla^*\hat{h}_{-1}+\nabla \hat{h}_{1}\right]=&0 \\
\begin{split}
\partial_t\hat{h}_1+{v_0\over 2}\left[\nabla^*\hat{h}_{0}+\nabla \hat{h}_{2}\right]=&
-A\,\pi \Gamma \left[
\hat{h}_{2}\hat{f}_{-1}-\hat{h}_0\hat{f}_1
\right]\\
&+Rv_0\, \Sc^2
\left[ \hat{h}_0\hat{f}_1 g_{1,0}+\hat{h}_1\hat{f}_0 g_{1,1}+\hat{h}_2\hat{f}_{-1} g_{1,-1}+\ldots\right]\end{split}\\
\begin{split}\partial_t\hat{h}_2+{v_0\over 2}\left[\nabla^*\hat{h}_{1}+\nabla \hat{h}_{3}\right]=&
-2 A\,\pi \Gamma \left[
\hat{h}_{3}\hat{f}_{-1}
-\hat{h}_{1}\hat{f}_1
\right]\\
&+Rv_0\, \Sc^2
\left[
\hat{h}_0\hat{f}_2 g_{2,0}+\hat{h}_1\hat{f}_1 g_{2,1}+\hat{h}_2\hat{f}_0 g_{2,2}+\ldots\right]\end{split}
\end{align}\end{subequations}
where $\nabla$ and $\nabla^*$ are the complex nabla operator and its conjugate, respectively, see eq. (\ref{COMPLEX_NABLA}).

Assuming a disordered and homogeneous background,
all modes of $\hat{f}_n$ vanish, except the $n=0$ mode:
\begin{equation}
\label{DEF_NULL_MODE}
\hat{f}_n=\delta_{n,0}{\rho_0\over 2 \pi}
\end{equation}
where $\rho_0$ is the average total particle density.
Then, the hierarchy, eq. (\ref{HIERARCH_TAGG2}), 
simplifies significantly. In particular, all terms related to the Vlasov-part of the collision integral vanish
and only terms proportional to $\Sc^2$ remain.
To obtain a diffusion equation for the mode $\hat{h}_0$ which is proportional to the density of the tagged particle $\rho_S$, 
we perform a Chapman-Enskog expansion \cite{chapman_52, hirschfelder_54, mcquarrie_76, ihle_16} and introduce a small ordering parameter $\epsilon$ which is set to one at the end of
the expansion. The spatial gradients are scaled as $\nabla\to \epsilon \nabla$ and the Fourier-modes are assumed 
to 
scale as $\hat{h}_n\sim\epsilon^{|n|}$ which can be verified a posteriori. 
In addition, we introduce the usual multiple time scale expansion of the time derivative,
\begin{equation}
\partial_t=\epsilon\partial_{t_0}+
\epsilon^2\partial_{t_1}+\ldots
\end{equation}
Inserting these expressions into eq. (\ref{HIERARCH_TAGG2}) and collecting terms of the same order, one obtains in linear
order, $\mathcal{O}(\epsilon)$, $\partial_{t_1}\hat{h}_0=0$, and
\begin{equation}
\label{FIRST_ORD}
\hat{h}_1={\nabla^*\hat{h}_0\over 2 R\,\Sc^2\,\hat{f}_0\, g_{1,1}}
\end{equation}
meaning that at this order, $\hat{h}_1$ is enslaved to the density mode $\hat{h}_0$.
In second order, $\mathcal{O}(\epsilon^2)$, one finds $\partial_{t_0}\hat{h}_1=0$, and
\begin{equation}
\label{SEC_ORD}
\partial_{t_1}\hat{h}_0=-{v\over 2}\left[ 
\nabla^* \hat{h}_{-1}+\nabla \hat{h}_1
\right]
\end{equation}
Inserting eq. (\ref{FIRST_ORD}) in (\ref{SEC_ORD}), substituting $g_{1,1}$ from eq. (\ref{G_MN_DEF}),
a diffusion equation is obtained,
\begin{eqnarray}
\label{DIFF_EQ1}
\partial_t\hat{h}_0&=&D\, \Delta \hat{h}_0 \\
\label{DIFF_COEFF1}
D&=&{9 \pi^2 R v_0\over 64 M\,\Sc^2}={9\pi^2\over 64}{v_0^3\over R \Gamma^2 M}
\end{eqnarray}
where $\hat{h}_{-1}=\hat{h}_1^*$, $\nabla\nabla^*=\partial_x^2+\partial_y^2\equiv \Delta$ and
$M=\pi R^2\rho_0=2 \pi^2 R^2\hat{f}_0$ from eq. (\ref{DEF_NULL_MODE}) was used, and $\epsilon$ was set to one.
The relaxation time $\tau_C$ of the velocity autocorrelation follows from the diffusion coefficient $D$ in 
eqs. (\ref{TAU_C_DEF}) and (\ref{DIFF_COEFF1}) as
\begin{equation}
\label{TAU_C_RESULT}
\tau_C={9\pi^2\over 32}{v_0\over R \Gamma^2 M}
\end{equation}
In Ref. \cite{ihle_23a} it is shown how the exponentially decaying vaf follows directly from Boltzmann--Lorentz theory without
the need for Chapman-Enskog expansions and Green-Kubo formulas, leading to the same result for $\tau_C$, eq. (\ref{TAU_C_RESULT}),
and supporting the arguments at the beginning of this section.

It is also interesting to note that there is a qualitative difference between the Vlasov-like approximation,
where the N-particle probability density is simply factorizing and the one-sided molecular chaos assumption
which takes two-particle correlations within the duration of binary encounters into account:
Only this approach can explain the finite relaxation time observed in simulations. The Vlasov-approach,
which can formally be considered by setting $\Sc=0$ or $\Gamma=0$ in eqs. (\ref{DIFF_COEFF1}, \ref{TAU_C_RESULT}), 
incorrectly 
predicts no relaxation at all, i.e. $\tau_C\to \infty$ which is equivalent to
an infinite diffusion coefficient. 
In VM models with explicit noise terms, the difference between the two approximations is not as drastic,
because the noise terms lead to a finite relaxation time, even within the Vlasov-like approach.
However, the theory with external noise that is large enough to significantly modify the particle trajectories during the collision time $\sim R/v_0$, is complicated to evaluate and will be left for the future.

The predicted diffusion coefficient, eq. (\ref{DIFF_COEFF1}), scales as $\sim 1/M$, i.e. goes to infinity for
vanishing density. This is expected since at low $M$, a particle very rarely encounters another one, and thus
moves ballistically for very long times.
eq. (\ref{DIFF_COEFF1}) also predicts scaling with the coupling constant as $D\sim \Sc^{-2}$.
This is plausible as a vanishing coupling leads to ballistic flights, which corresponds to
an infinite diffusion coefficient.
Detailed comparisons of the predicted behavior of the diffusion coefficient with agent-based simulations are 
presented in section \ref{sec:numer} and show excellent agreement at 
small density and small coupling strength.

\section{The effect of frozen disorder in the particle speeds}
\label{sec:disorder}

So far, the discussion assumed identical particles. Particularly, there is a single constant propulsion speed $v_0$ in eq.~\eqref{POS_EQ} for every particle. This is peculiar as it is very unlikely to be achievable in any possible realization of dry aligning active matter and, more fundamentally, it means that there are configurations that lead to infinite time spent within distance $R$ of each other even in the free problem for $\Sc\equiv 0$. While we address the consequences of this in the following section, we are for now considering ensembles of $N$ particles with the modified equations of motion (cp. eqs.~\eqref{POS_EQ} and  \eqref{ANGLE_EQ}) 
\begin{subequations}
	\begin{eqnarray}
	\label{POS_EQ_var}
	& & {\diff \vec{r}_i \over \diff t}=v_i\,\hat{\vec n}_i \\
	\label{ANGLE_EQ_var}
	& & {\diff\theta_i \over \diff t}={\Gamma\over N_i^{\beta}}\sum_{j\epsilon\Omega_i}
	\sin(\theta_j-\theta_i)\,. 
	\end{eqnarray}
	\end{subequations}
	Herein, $v_i$ are random speeds that are constant over time for a specific realization of this system, i.e. they constitute \emph{frozen} disorder (also called \emph{quenched} disorder).~\cite{binderkob} Their statistics are given by
	\begin{subequations} \label{eq:speeddisorder}
	\begin{eqnarray}
		v_i &=& v_0 + \omega \nu_i \\
		\nu_i &\sim& \mathcal{U}_{-1/2,1/2}\text{,}
	\end{eqnarray}	\end{subequations}
i.e. the speeds are uniform in $[v_0-\omega/2,v_0+\omega/2]$. The scale of fluctuations around the mean is given by $\omega\ll v_0$. While we specifically choose a uniform distribution here to avoid the possibility of negative speeds, the specific details should not be too important in the limit $\omega/v_0 \to 0$. This limit is in the spirit of the original model as $v_0$ can only be considered a typical speed in a meaningful way, if the fluctuations around it are small. As is usual in systems with frozen disorder, care is needed in regards to how averages are computed or, more generally, how a statistical theory is to be constructed. For a specific realization of the disorder, that is a given set $\{v_i\}$, we can compute a specific quantity, say $X(t)[\{v_i\}]$, which is dependent on the realization and then have to perform an additional quenched averaging over these frozen disorder parameters, i.e.
\begin{align}
	\overline{X} &= \int \ldots \int \operatorname{d}^N\!v_i\, P_v(\{v_i\}) X(t)[\{v_i\}]
\end{align}
with the distribution of speeds $P_v(\{v_i\})$. We use the notation with an overline, $\overline{\phantom{\lvert}\cdot\phantom{\rvert}}$, here to avoid any confusion with other kinds of averaging. 

In the Boltzmann--Lorentz theory of the prior section the degrees of freedom of the untagged particles were also treated as disorder, however they maintain their dynamics and so these parameters change on the time-scales of statistical computation making them \emph{annealed} disorder.

We are specifically interested in the Boltzmann-equation, see eq.~\eqref{eq:def_jcoll} in positional space. As discussed before, the Vlasov or mean-field contributions to the collision are independent from the particle dynamics and, thus, unchanged upon introducing speed disorder. In general, the structure of the kinetic theory will remain equivalent, such that we can for a spatially homogeneous system (without loss of generality) postulate the mode equations of the system with frozen disorder
\begin{align}
\partial_t \hat{f}_n[\{v_i\}] &= K_{mn}[\{v_i\}] \hat{f}_m[\{v_i\}] \hat{f}_{m-n}[\{v_i\}] \text{.}
\end{align}

We generally cannot directly infer a closed dynamics of the averaged distribution modes $\overline{\hat{f}_n}=\int \ldots \int \operatorname{d}^N\!v_i\, P_v(\{v_i\})\hat{f}_n[\{v_i\}]$ from this. However, the effective coupling $K_{mn}$ includes contributions from every pair and, thus, for large $N$ is expected to be \emph{self-averaging} and not explicitly depend on the specific realization, thus eliminating the quenched dependencies. While this rationale might seem intuitive and persuasive, it is also known to be false close to phase transitions: If there is a transition to global order with particles interacting for a long time in coherent groups, the pair structure is no longer explored equally and self-averaging has to fail. This is a fundamental issue and would severely put into question whether there is a reasonable definition of an averaged Boltzmann equation, but the kinetic strategy as pursued here breaks down at a possible transition anyway as the dynamics can no longer be modeled as a sequence of finite pair interactions. So any possible violation of self-averaging would be outside of the scope of this work anyhow and we will not dwell on this. As before, we focus on the case of $\beta\equiv 0$.

\subsection{Deriving the Boltzmann collision operator for positive alignment strengths \texorpdfstring{$\Gamma$}{}}

Before proceeding with any computations, we will give a less ad-hoc and more technical reasoning for considering quenched disorder in speeds. The preceding sections developed the scattering theory for arbitrary $\Sc$
 but relied on the finiteness of collision times, which is guaranteed for  $\Gamma<0$. For $\Gamma>0$, certain two-particle configurations lead to divergent encounter durations in the deterministic model with identical speeds. We now show that this divergence is an artifact of the idealized model and can be removed by a physical regularization. This is reasonable as the deterministic system with positive alignment strength would have very artificial dynamics in which almost every encounter is permanent and in the absence of self-mixing the initial conditions become dominant. However, the argument can be made that there is a parameter range for small parameter numbers $M$ (and the associated separation of timescales $T_{\text{dur}} \ll T_{\text{free}}$ discussed earlier) in which an additional orientational noise is small enough so as to not disturb the nature of the interactions, but strong enough to avoid the formation of global orientational order. Additionally, one can argue that in general the asymptotic series approach undertaken here should extrapolate from $\Gamma <0$ to $\Gamma >0$ or at least there does not seem to be an obvious reason to the contrary.

To address this, we consider what eqs.~\eqref{J_REC_1} and \eqref{H123_DEF} look like for positive $\Gamma$. Indeed most of the analysis still holds (in particular the two-particle scattering), but there is no forbidden zone for $\Gamma>0$. So the integral splitting introduced in \cref{FINAL_INTEGRAL_SPLIT} to account for the forbidden zone has to be undone. This means there will be new contributions and especially also new contributions to the dynamics of the zeroth mode. The zeroth mode is proportional to the density and therefore conserved in a homogeneous system. Inspecting the dynamics of $m=0$ without forbidden zone, we find that the density no longer is conserved and instead there is a contribution $\partial_t f_0 \propto - \Sc^2$ draining probability from the system and revealing it as an unphysical approximation.

The origin of this is the assumption that a small noise can be treated perturbatively within the current approach. We can make this more quantitative by considering the timescale for the duration of interactions, $T_\text{dur}$, and the timescale between interactions, $T_\text{free}\sim M^{-1}$. Already in the free, non-interacting ($\Gamma\equiv0$) system, we find $T_\text{dur} \sim \sin^{-1}{\frac{\Delta} 2}$ is diverging and thus will exceed $T_\text{free}$ for sufficiently small $\Delta$, in this sense three-particle (and higher) terms would have to be considered always relevant as they no longer require coincidence of three independent particles, but the triple (or n-tuple) can be built successively.

However, the expected number of particle pairs in the small $\Delta$-region, $\lvert \Delta\rvert \lesssim M$, for which this applies, is suppressed by an additional factor $M$ for analytic angle distributions and, therefore, not relevant to the dynamics considered here. For non-analytic distributions of the angle difference $\Delta$, in particular one displaying a singularity at $\Delta = 0$, this argument does not hold, and the Boltzmann equation has to be considered insufficient. This is the situation in a flocking state, which, in the deterministic model considered here, is the relevant steady state. Indeed and as discussed before, this inadequacy manifests itself in the fact that a naive evaluation of the collision integrals for positive $\Gamma$, i.e. without forbidden zones, does indeed lead to non-physical results. 

This is an inadequacy of the formalism and the restriction of the dynamics to binary collisions even in the deterministic model as three particles collisions will eventually occur and disrupt interacting dimers (particulary in the absence of established global order). We therefore have to conclude that this divergence in the interaction times is artificially introduced. On a coarser level, one can argue that there should be no non-analyticities present far from a transition and, thus, a direct perturbation expansion of any quantity would lead to the conclusion that the first non-mean-field contribution $\mathcal{O}(\Gamma^2)$ is invariant under $\Gamma \to - \Gamma$. To address this, some regularization is needed. We could introduce an ad-hoc forbidden zone around $\Delta=0$, but this is conceptually unsatisfying. There are two physical disorder-based approaches that address the issue on a more physical level. Firstly, we could account for orientational noise explicitly making the model non-deterministic. This is an avenue left for future work here as we are explicitly exploring deterministic equations of motion. Secondly, one can introduce a weak quenched disorder in the propulsion speeds, as introduced above. Thus, duration times are bounded above by the typical difference in velocity, $T_\omega \sim 1/\omega$. Within the Boltzmann equation approach, this leads to a regularization that avoids singularities as well as spurious terms in the expansion, and, thus, there cannot be a relevant $\mathcal{O}(1)$-contribution from the would-be forbidden zone. This approach therefore also suggests an equivalency of the first non-mean-field contribution for $\Gamma$ and $-\Gamma$, which for sufficiently small $\omega$ will coincide with the deterministic result.

In summation, we want to argue that the contributions to the collision integral from the zone of size $\mathcal{O}(\Sc)$ that is forbidden for $\Gamma <0$ should be discarded for $\Gamma > 0$ as well (in a theory to order $\mathcal{O}(M\Sc^2))$, because we can reach that model also in a regularized fashion via quenched speed disorder, $\omega\to 0, \omega >0$. For any finite $\omega$, the perturbative argument suggesting regularity holds on the global level and, on the microscopic level, there is an inaccessible region of small angle differences acting as an ersatz-``forbidden zone'' due to the enforced finiteness of  the interaction times. Which is also to say, that the naive extrapolation of simply changing the sign of $\Gamma$ in the results for $\Gamma <0$ does hold merit despite their derivation being no longer valid for $\Gamma>0$.

\subsection{Two-particle scattering with quenched speed disorder}
For simplicity, we consider the two-particle scattering of two particles with $v_1=v-\omega/2$ and $v_2=v+\omega/2$. Herein, we omitted $\nu\in [-1/2,1/2]$, a uniformly distributed random number, cp. \cref{eq:speeddisorder}, for the sake of notational simplicity. In principle both $\omega$ and $v$ are random quantities.  Trivially, the angular interactions in eq.~\eqref{ANGLE_EQ_var} are unchanged from the original model eq.~\eqref{ANGLE_EQ} and we still (cp. \cref{A_DIFF_SOL}) have (again using $\mu=\tan{\Delta(t)/2}$ and $\mu_0=\tan{\Delta(t_0)/2}$ as short-hands for the tangents after and prior to interaction)
\begin{align}
\mu_0 = \mathrm{e}^{2\Gamma T} \mu
\end{align}
wherein $T=T(\Gamma,\omega)=t-t_0$ is the interaction time that will change with disorder $\omega$ .

Employing the same geometric considerations as in Appendix \ref{app:B}, it is straightforward to find the absolute value of the relative velocity $\vec v_\text{rel}=\vec{v}_2-\vec{v}_1$
\begin{align}
v_\text{rel}&= \sqrt{4 v^2 \sin^2 \frac{\Delta}{2} + \omega^2 \cos^2\frac{\Delta}{2}} \text{.} \label{eq:vrel}
\end{align}
Using $\Delta \vec r=-R (\cos\beta,\sin\beta)^T$ as well as $\tilde c=\theta_1+\theta_2$ and $c=\tilde c -2 \beta$, we find
\begin{subequations}
\begin{align}
\cos\varphi &= \frac{\vec v_\text{rel}\cdot \Delta \vec r}{v_\text{rel} R} \\
&= v^{-1}_\text{rel} (2v \sin \frac{\Delta}{2} \sin \frac{c}{2}- \omega \cos\frac{\Delta}{2}\cos\frac{c}{2}) \label{eq:cphi}
\end{align}
\end{subequations}

The missing piece in terms of geometric quantities, is expressing $\sin\frac{c}{2}$ and $\cos\frac{c}{2}$ as functions of $\phi$ and $\Delta$. From squaring eq.~\eqref{eq:cphi}, we find directly the relations
\begin{subequations}
\begin{align}
    \sin \frac{c}{2} &= \frac{2 v \sin\frac{\Delta}{2}}{v_\text{rel}} \cos\varphi+ \frac{\omega \sin\varphi \cos\frac{\Delta}{2}}{v_\text{rel}} \\
    \cos\frac{c}{2} &=- \frac{\omega  \cos\frac{\Delta}{2}}{v_\text{rel}} \cos\varphi + \frac{2v \sin\varphi \sin\frac{\Delta}{2}}{v_\text{rel}} \text{.}
\end{align}
\end{subequations}

The collision condition, cp.~\eqref{ENTRANCE_COND} and Appendix \ref{app:A}, becomes
\begin{subequations}
    \label{eq:coll}
\begin{align}
0&= 2 R \left[2v \sin\frac{c}{2} G-\omega \cos\frac{c}{2} H\right] + \left[ 4v^2 G^2 + \omega^2 H^2\right]
\end{align}
with the new function
\begin{align}
H &= \int \mathrm{d}t' \cos\frac{\Delta}{2} = \left.\frac{1}{2\Gamma} \operatorname{arcotanh} \sqrt{1+\mu^2} \right\rvert_{\mu(t)}^{\mu(\tilde t)}
\end{align}
additional to the already used
\begin{align}
G &= \int\mathrm{d}t' \sin\frac{\Delta}{2} = -\left.\frac{1}{2\Gamma} \operatorname{arsinh}\mu\right\rvert_{\mu(t)}^{\mu(\tilde t)} \text{.}
\end{align}   
\end{subequations}

Any physical Boltzmann theory has to be mass-conserving (in the sense that the zeroth Fourier mode $\hat f_0$ is indeed conserved), this can be reduced to a necessary condition for the phase space compression $J_s$, cp.~eqs.~\eqref{eq:decompJcoll}, \eqref{J_FOUR_EXPR1} and \eqref{J_REC_1},
\begin{align}
    K&=\int_{-\frac{\pi}{2}}^{\frac{\pi}{2}}\mathrm{d}\varphi\cos\varphi\int_{-\pi}^\pi \mathrm{d}\Delta\, v_\text{rel} (\mathrm{e}^{2 \Gamma J_s}-1) \stackrel{!}{=} 0 . \label{eq:defK}
\end{align}
The meaning of this condition is that the contribution to the collision integral mode zero due to the coupling of this mode to itself has to vanish. We have already merged the contribution from the receding part (with a non-vanishing phase space compression factor $J_s$) and the approaching part (wherein $J_s \equiv 0$ due to the assumption of one-sided molecular chaos, the sign being a result of the $\pi$-shift of the angle $\phi$) into a single expression.

\begin{figure}
	\begin{center}
	\includegraphics[width=4.1in,angle=0]{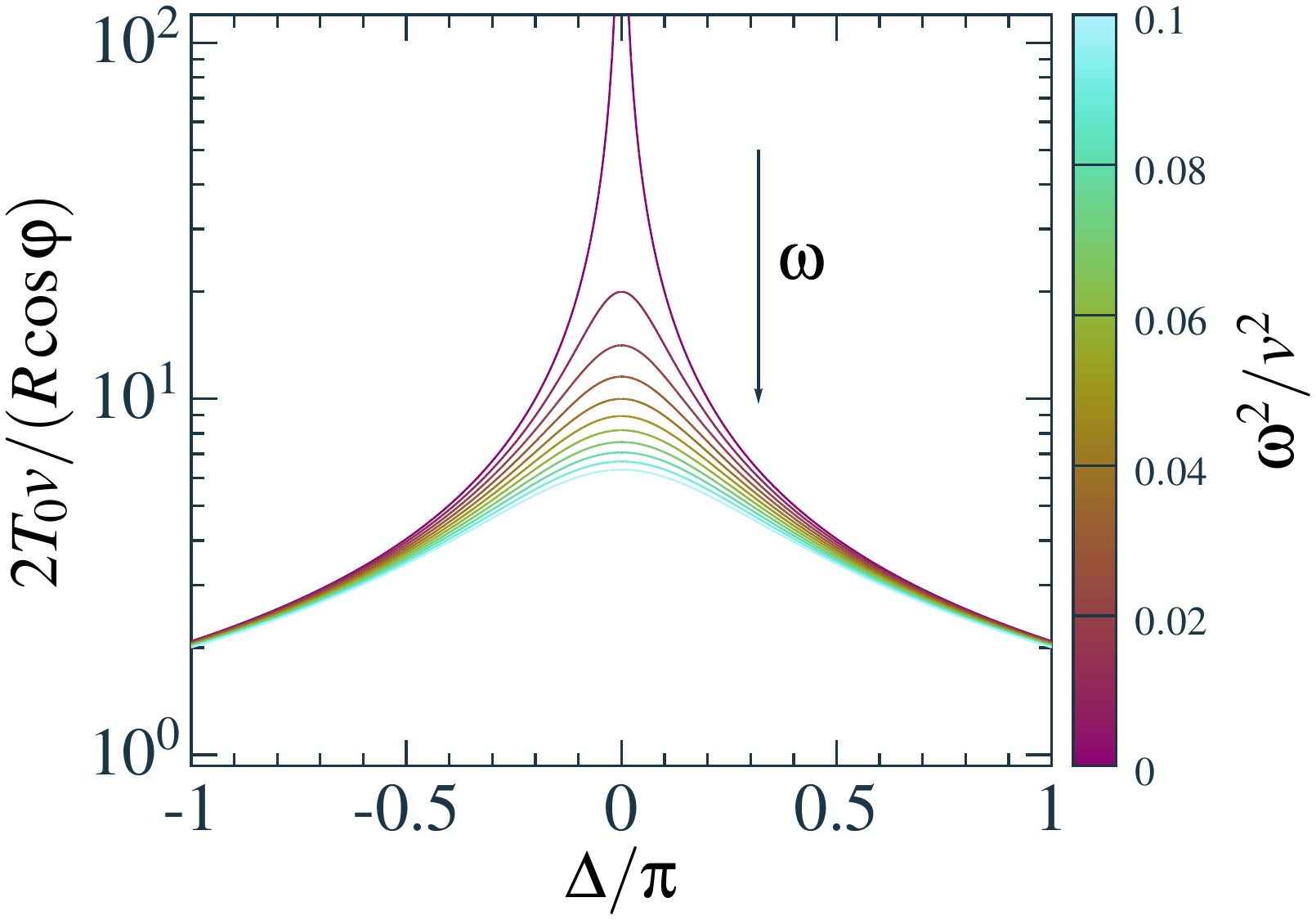}
	\caption{
	Plot of the free interaction time $T_0$ according to eq.~\eqref{eq:tfree} for varying values of the rescaled speed disorder $\omega/v$ (color-coded) as a function of the final angle difference $\Delta$. The regularization due to the quenched disorder in speeds is directly apparent. The consistency of the theory rests on $T_0<T_{\text{free}} \sim M^{-1}$.
	}
	\label{fig:vrel_comp}
	\end{center}
	\end{figure}

For the phase space compression $J_s$, we restate the earlier result that is solely dependent on the (unchanged) angular interactions from eqs.~\eqref{eq:Jsdef} and \eqref{J_S_FOR_EXACT}
\begin{align}
J_s &= \int \mathrm{d}t\,\cos(\theta_2-\theta_1)\\
&= \frac{1}{2\Gamma} \int\mathrm{d}\mu \frac{1-\mu^2}{\mu (1+\mu^2)} = \frac{1}{2\Gamma} \log(\frac{\mu_0}{\mu} \frac{1+\mu^2}{1+\mu_0^2}) \label{eq:Js}
\end{align}
and, hence,  find (cp. eq.~\eqref{EXACT_T2})
\begin{align}
\mathrm{e}^{2 \Gamma J_s} &= \frac{\mu_0}{\mu} \frac{1+\mu^2}{1+\mu^2_0} = \mathrm{e}^{2\Gamma T} \frac{1+\mu^2}{1+\mu^2 \mathrm{e}^{4\Gamma T}} \label{eq:expJ},
\end{align}
which highlights the peculiarities of a divergent interaction duration. In particular, this expression cannot be meaningfully evaluated perturbatively in $\Gamma$ in a region where $\Gamma T = \mathcal{O}(1)$. This was also discussed in the evaluation of the integral $\hat{H}_1$ on page~\pageref{pageH1}. However, here the quenched speed disorder guarantees finite interaction times and therefore we can proceed. Ultimately not successful but rather instructively along the way nonetheless, we make the following perturbative Ansatz for the duration time
\begin{subequations}
\begin{align}
T&=T_0 + \Gamma T_1 \label{eq:expandT}
\end{align}    
wherein $T_0$, see Fig.~\ref{fig:vrel_comp} denotes the flight time in the free problem
\begin{align}
T_0&=\frac{2R \cos\varphi}{v_\text{rel}} \text{.} \label{eq:tfree}
\end{align}
\end{subequations}
As seen before for negative $\Gamma$, this is a sufficient order as $T$ always enters with an additional factor $\Gamma$. Note that it is crucial to develop around the free problem (to have well-posed perturbation problem) and this is only possible for any value of $\Delta$ and $\phi$ because the speed disorder guarantees finite relative velocities.

Using this expression for the time, we can directly evaluate $G$ and $H$ yielding
\begin{subequations}
\begin{align}
G&= - \frac{\mu (T_0+\Gamma T_1)}{\sqrt{1+\mu^2}} - \frac{ \Gamma T_0^2 \mu}{\sqrt{1+\mu^2}^3}\\
&= G_0 (1+\Gamma \frac{T_1}{T_0}+\Gamma T_0 \frac{1}{1+\mu^2}) \notag
\end{align}
and
\begin{align}
H&= -\frac{1}{\sqrt{1+\mu^2}} (T_0+\Gamma T_1) +  \Gamma T_0^2 \frac{\mu^2}{\sqrt{1+\mu^2}^3}\\
&= H_0 (1+\Gamma \frac{T_1}{T_0}-\Gamma T_0 \frac{\mu^2}{1+\mu^2})  \notag
\end{align}
\end{subequations}
respectively. We introduced short-hands $G_0= - \mu T_0 / \sqrt{1+\mu^2}$ and $H_0 =G_0/\mu$ for the contributions to order $\mathcal{O}(\Gamma^0)$.

It is straightforward to check that the zeroth order of the collision condition eq.~\eqref{eq:coll} does indeed reproduce eq.~\eqref{eq:tfree}. To this end, it is helpful to rephrase that result using the expression for $\cos\varphi$ of \cref{eq:cphi} as
\begin{align}
T_0 &= \sqrt{1+\mu^2} \frac{2 R (2v \mu \sin\frac{c}{2}-\omega \cos\frac{c}{2})}{4v^2\mu^2+\omega^2}\text{.}
\end{align}

Using the zeroth order equation, the first order equation can be written as
\begin{align}
    \begin{split}
0=& [4 v^2 G_0^2 + \omega^2 H_0^2] \frac{T_1}{T_0} \\ & + T_0 \frac{1}{1+\mu^2} \left( 2 R \left[2v \sin\frac{c}{2} G_0 -\omega \cos\frac{c}{2} H_0 \mu^2 \right] + 2\left[ 4v^2 G_0^2 - \omega^2 H_0^2 \mu^2\right] \right)
    \end{split}
\end{align}
and, thus, we can identify
\begin{align}
T_1&= -\frac{T_0^2}{1+\mu^2}  \frac{\left( 2 R \left[2v \sin\frac{c}{2} G_0 -\omega \cos\frac{c}{2} H_0 \mu^2 \right] + 2\left[ 4v^2 G_0^2 - \omega^2 H_0^2 \mu^2\right] \right)}{4 v^2 G_0^2 + \omega^2 H_0^2 } \text{.}
\end{align}
For $\omega=0$, this does reproduce the correct result $T_1=-T_0^2(1+\mu^2)^{-1}$ found earlier,  cp. eqs. \eqref{TDURNULL} and \eqref{TDUR_LINEAR1}. On the flip-side, for $\omega\neq 0$ dominant, we find $T_1 \approx \frac{T_0^2\mu^2}{1+\mu^2} $. This is also the result found by formally setting $v=0$ from the beginning, i.e. in eq.~\eqref{eq:coll}. Still, we must not proceed with this perturbation in time, as the specific form of $T_1$ found here will always have regions in which $T_0+\Gamma T_1<0$ holds, which is unphysical. This reinforces the earlier notion that it is essential to handle the interaction times exactly as far as possible.

\begin{figure}
	\begin{center}
	\includegraphics[width=4.1in,angle=0]{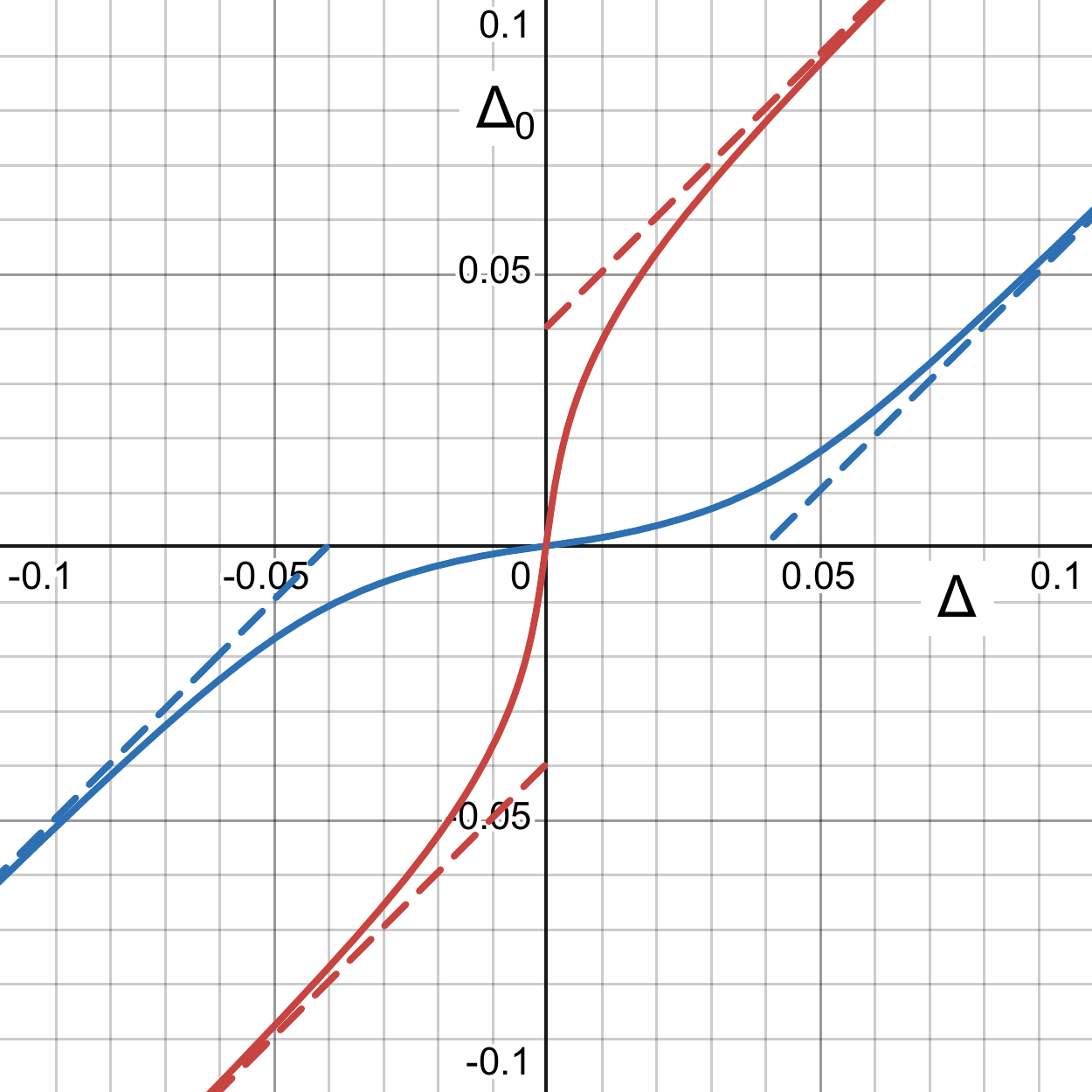}
	\caption{
	Plot  of the direct numerical solution of the collision condition, eq.~\eqref{eq:coll}. We restricted the numerical solution $\Delta_0(\Delta)$ to the physical solution branch ($T>0, \mu \mu_0 >0$). Shown are the solutions for $\Gamma=\pm 0.01$ (red: $\Gamma>0$, blue: $\Gamma<0$) each without (dashed lines) and with (solid lines) regularization by means of quenched speed disorder ($\omega^2/v^2=0.03$). The presence of the ``forbidden zone'' for $\Gamma<0$ is directly observable and the corresponding feature of the backtracing for positive coupling is a jump at $\Delta=0$. With any amount of finite quenched disorder, however, this function becomes analytic (also in $\Gamma$, because the interaction times are finite) and we can proceed with a perturbative calculation that does not account for these intricacies exactly (which was the approach for $\Gamma<0$). This is the graphical explanation for the existence of a regularized Boltzmann theory that consequently has to be adequate for positive and negative couplings. By extension, we argue that the exact asymptotic solution for $\Gamma<0$ presented earlier does represent the correct regularized description for $\Gamma>0$ as well.
	}
	\label{fig:desmos}
	\end{center}
	\end{figure}

Instead, we note that the regularization via quenched speed disorder has another direct consequence, see also Fig.~\ref{fig:desmos}: the relationship $\Delta_0(\Delta)$, i.e. the backtracing that is core to the actual implementation of \emph{one-sided molecular chaos}, is analytical and we can utilize it for a direct perturbation Ansatz
\begin{subequations}
	\begin{align}
		\Delta_0 \approx& \Delta + \epsilon \Gamma + \delta \Gamma^2\\
		\mu(\tilde t) =& \tan(\frac{\Delta_0}{2}) \approx \mu+\frac{\epsilon}{2} \Gamma (1+\mu^2)+ \frac{\epsilon^2\mu+2\delta}{4}\Gamma^2  (1+\mu^2) \text{.}
		\end{align}			
Thus, we can rewrite the expressions occurring in the collision condition, eq.~\eqref{eq:coll}, as
\begin{align}
G&= -\frac{\epsilon}{4} \sqrt{1+\mu^2} - \frac{\Gamma}{16} (4\delta+\epsilon^2 \mu) \sqrt{1+\mu^2}\\
H&= -\frac{\epsilon \sqrt{1+\mu^2}}{4\mu} + \Gamma{(\epsilon^2-4\delta\mu) \sqrt{1+\mu^2}}{16\mu^2}
\end{align}
and, using eq.~\eqref{eq:Js}, the phase space compression factor as
\begin{align}
\mathrm{e}^
{2\Gamma J_s} \approx 1- \frac{-1+\mu^2}{\mu} \varepsilon\Gamma + \frac{\delta (1-\mu^2) - \epsilon^2 \mu}{2\mu} \Gamma^2\text{.} \label{eq:Jspert}
\end{align}
\end{subequations}

The order-to-order solution of eq.~\eqref{eq:coll} is then
\begin{subequations}
\begin{align}
\epsilon =&\frac{8\mu R (2 v \sin{\frac{c}{2}} \mu -\omega \cos{\frac{c}{2}})}{\sqrt{1+\mu^2}(4v^2\mu^2+\omega^2)}\\
\delta =& -\frac{16\mu R^2(2v\sin\frac{c}{2} \mu-\cos\frac{c}{2}\omega)(-2\mu(2+\mu^2)\omega^2v\sin\frac{c}{2}+8\mu^5 v^3 \sin\frac{c}{2} + (\omega^3-4\mu^2(1+2\mu^2 \omega v^2))\cos\frac{c}{2})}{(1+\mu^2)(4v^2\mu^2+\omega^2)^3}
\end{align}
Using the explicit expressions for $\sin{\frac{c}{2}}$ and $\cos{\frac{c}{2}}$, we can simplify this to
\begin{align}
\epsilon =& \frac{8\mu}{1+\mu^2} \frac{R}{v_{\text{rel}}}\\
\delta =& - \frac{16\cos\phi \mu (-2\mu(1+\mu^2)\omega v \sin\phi-\cos\phi(\omega^2-4\mu^4 v^2))}{(1+\mu^2)^2 (4v^2\mu^2+\omega^2)}\frac{R^2}{v_{\text{rel}}^2}
\end{align}\end{subequations}
and, plugging this back into the zero-mode conservation integral from eq.~\eqref{eq:defK} via eq.~\eqref{eq:Jspert} while omitting odd terms in $\omega$, we find
\begin{subequations}
\begin{align}
K&\approx \int_{-\frac{\pi}{2}}^{\frac{\pi}{2}}\mathrm{d}\varphi\cos\varphi\int_{-\pi}^\pi \mathrm{d}\Delta\, (k_1 \Gamma + k_2\Gamma^2)\\
k_1&=\frac{2\cos\phi (-1+\mu^2) R}{1+\mu^2} \notag \\
k_2&=\frac{4 \cos\phi^2 R^2 ((1-5\mu^2)\omega^2+ 4\mu^2 (-5+\mu^2)v^2)}{(1+\mu^2)^3 v_{\text{rel}}^{3}} \text{.} \notag
\end{align}
Using trigonometric relations for the Weierstrass substitution terms (polynomials in $\mu=\tan\frac{\Delta}{2}$), we can rewrite the two integrands as
\begin{align}
k_1 &= 2\cos\phi R \cos\Delta\\
k_2 &= \frac{\cos^2\phi R^2 (5 (\omega^2 -4 v^2) \cos\Delta + 8 (\omega^2+4v^2) \cos(2\Delta) +3 (\omega^2-4v^2) \cos(3\Delta))}{\sqrt{2}(\omega^2+4v^2+(\omega^2-4v^2)\cos\Delta)^{3/2}}  
\end{align}	
\end{subequations}
and, finally, we do indeed find that the integration over $\Delta$ yields $K\equiv 0$ independent from $\Gamma$ and $\omega$.

Thus, we conclude with an important result for the applicability of the presented Boltzmann theory that we will summarize shortly, but, on a technical level, there is also a crucial lesson on the role of interaction times in kinetic theory to be gathered from this undertaking. The great power of Boltzmann theory is that it does account for interaction times as it includes (pairwise) interactions as finite events that have a beginning and end. This power comes at the equally great responsibility to pose a model that has physically sound interaction times (finite, positive) and this has to be manifestly established at any level.

\begin{figure}
	\begin{center}
	\includegraphics[width=4.1in,angle=0]{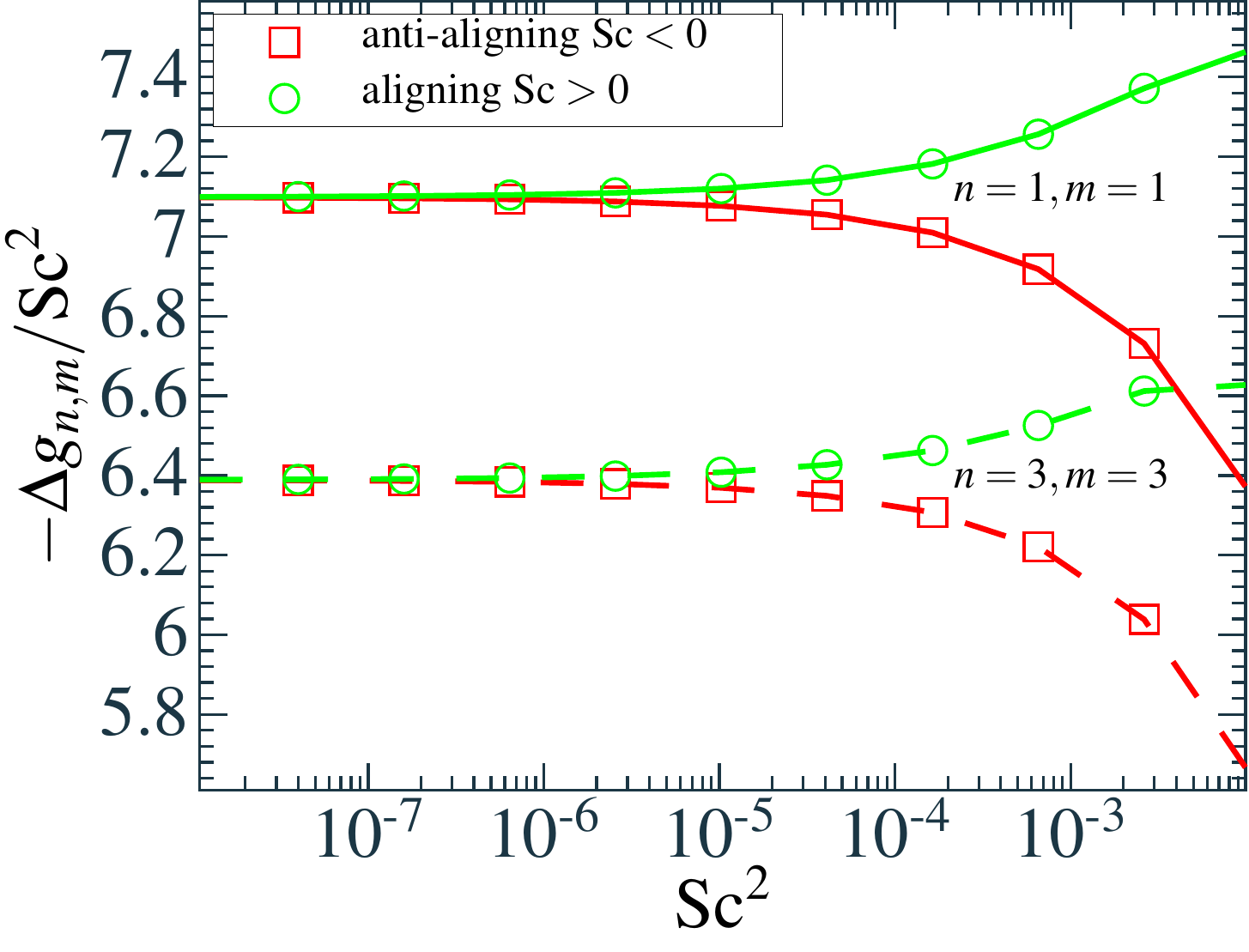}
	\caption{
	Numerical solution of the collision integral, as explained in the main text. We show two elements of the coupling matrix $\Delta g_{n,m}$ (explicitly having subtracted the mean-field contributions) which is expected to scale as $\Sc^2$ in leading order. We find that there is a good collapse at small values of $\Sc$ and that, in the regularized model with quenched speed disorder (here, we used $\omega/v=0.0256$), the kinetic theory is analytical in $\Sc$: The curves for positive and negative coupling agree for sufficiently small coupling strength and there respective deviations at larger values are consistent with the next relevant terms being $\mathcal{O}(\Sc^3)$.
	}
	\label{fig:vdis}
	\end{center}
	\end{figure}

As a sanity check and to make the perturbation theory argument that any analytic consistent kinetic theory should have equal applicability irrespective of the sign of the couplings more tangible, we show in Fig.~\ref{fig:vdis} results from a numerical evaluation of the collision integral. To this end, we computed the backtracing $\Delta(\Delta_0,\phi)$ by direct quadrature (employing an Euler scheme) of the equations of motion. As a byproduct, we can also compute $J_s$ along this integration. With both these results, the integration kernel (cp. eq.~\eqref{eq:kernel}) is known and the integrals over $\phi$,$\theta_1$ and $\theta_2$ can be performed numerically. For this step, we used a trapezoid rule with $2^{10}$ collocation points in every dimension. The results of Fig.~\ref{fig:vdis} corroborate the intuitive notion and the analytical computation of this very section that the introduction of quenched speed disorder into the model does regularize it to the point that the Boltzmann theory is sufficiently analytical and the asymptotic expansion considered here can be extended from its original realm of negative couplings to any coupling. We argue, therefore, that for $\Gamma>0$ in the original model one should really consider the $\omega^2=0^+$ limit of the disordered model (with other regularizations imaginable). Any finite amount of disorder serves as regularization that actually is done by means of noise (leading to stochastic equations of motion which are outside the scope of this work) and higher-order interactions. The applicability of the continuated theory is still limited: the Boltzmann closure, the idea of one-sided molecular chaos, rests on the assumption of essentially negligible correlations that are built up for a short time during interactions. For positive coupling in the deterministic setting, this limits the applicability to very short times after initialization. However, if the number of neighbors $M$ is sufficiently small, $M\ll 1$, there should be a regime where additional rotational white noise would not affect the dynamics on the shorter timescales $T_{\text{dur}}$ but be relevant enough on the long free-flight timescales $T_{\text{free}}$ to keep the system from globally flocking. As such, the continuated Boltzmann theory is a better description to describe the onset of flocking in noisy systems with a transition into a global flocking phase with polar order than the more frequently employed mean-field (or Vlasov) picture.

\section{Deriving an effective Langevin-equation}
\label{sec:langevin}

\subsection{Brownian motion of mobile rotators}
\label{sec:brownian}

The random motion of small particles such as pollen grains immersed in a fluid is known as
Brownian motion. 
Einstein's explanation of its nature can be regarded as the beginning of stochastic modeling of natural 
phenomena \cite{einstein_05}.
A simple, approximate way to treat the dynamics of the embedded particle is to model the kicks of the surrounding fluid molecules
by a random force in Newton's equation of motion, something we know now under the name of Langevin equation. 
The amount of simplification in this description is tremendous since there is no need to describe the details of the interactions 
among the surrounding molecules or between the molecules and the Brownian particle. All the information
about this many-body system is encoded in a friction term and the correlations of the noise, i.e. the random force.
In  this historic example, these correlations are simple, and the strength of the noise follows from a fluctuation-dissipation relation.
In more exotic cases, such as systems of self-driven particles with alignment interactions, 
it is not a priori clear whether a description
by an effective Langevin-equation always makes sense and how to determine the properties of the noise.

There are a few examples on how to analytically derive the properties of the noise by explicitly integrating over the irrelevant degrees of
freedom.
One of them is due to R.J. Rubin \cite{rubin_60} who considered a harmonic 
lattice in which one particle, the Brownian particle, is much heavier than the rest.~\footnote{Note as explained in Ref. \citenum{zwanzig_book}, these derivations lead to a so-called exact Langevin-equation 
that in general is non-linear
in contrast to the linear Langevin equations obtained by the well-known Mori-Zwanzig projector-operator approach.
This is because the latter is usually derived in the Hilbert space of dynamical variables where for example
the variables $x$ and $x^2$ are different vectors which are not necessarily parallel.}
For other, similar systems, see Ref. \cite{zwanzig_book}.
In a more recent example by van Meegen and Lindner \cite{vanmeegen_18}, a set of immobile rotators 
with fixed but randomly chosen frequencies and coupling constants were considered. 
Using a path integral formalism to average over the frozen disorder, it was shown how to derive the Langevin equation for 
the angular change $\dot{\theta}$ of a focal rotator and how to obtain the properties of the emerging 
dynamical noise
in this equation. In the case of fully connected networks, Steinhöfel et al.~\cite{steinhofel2026} used a mapping to the Schrödinger equation to solve the full $N$-body problem which also shed light on important fluctuations~\cite{spera2024}. 

In this section, we will pursue the main general idea of regular Brownian motion and 
assume that the effects of the surrounding 
particles
on a focal particle can be modeled by a typically colored but Gaussian noise term $\xi$, leading to an effective, one-particle
Langevin-equation
for the angular change,
\begin{equation}
\label{simpleLANG}
\dot{\theta}(t)=\xi(t)
\end{equation}
The major difference to Refs. \citenum{rubin_60} and \citenum{vanmeegen_18} is that our ``rotators'' are moving and have no
fixed set of interaction partners: the interactions in the VM take place on a time-varying 
network \cite{holme_15}
whose links depend on the outcome of the interactions.    

We assume here that the correlations of the dynamical noise $\xi=\xi_i$ for a given particle $i$ 
with the one 
of a different particle $j$, are negligible compared to the correlations of the noise of the same particle, 
that is $|\langle\xi_i(t)\,\xi_j(t')\rangle|\ll |\langle\xi_i(t)\,\xi_i(t')\rangle| $ for $i\neq j$.

In the following sections, we will show how the kinetic theory with one-sided molecular chaos from section 
\ref{sec:scatter} can be utilized
to calculate the properties of the noise $\xi$ with high precision in the limit of low density. 
Using a random-telegraph assumption, we also show how the noise can be determined in the opposite limit of large 
density.

\subsection{Noise calculation for low particle density}

The velocity autocorrelation function (vaf) for particles of constant speed $v_0$ can be written as
\begin{equation}
\label{AUTO_CN}
C(\tau)\equiv\langle \vec{v}(t+\tau)\cdot \vec{v}(t)\rangle
=v_0^2\langle \cos(\Delta \theta)\rangle={v_0^2\over 2}
\Big[\Big\langle {\rm e}^{\ii\Delta \theta}\Big\rangle+\Big\langle {\rm e}^{-\ii\Delta \theta}\Big\rangle
\Big]
\end{equation}
with
\begin{equation}
\label{DEF_LANG}
\Delta\theta\equiv \theta(t+\tau)-\theta(t)=\int_t^{t+\tau}  \diff t' \, \dot{\theta}(t')
\end{equation}
According to the assumed effective Langevin-equation, eq. (\ref{simpleLANG}),
the angular difference in eq. (\ref{DEF_LANG}) can be expressed as
the integral over the assumed Gaussian noise $\xi$, $\Delta \theta=\int_t^{t+\tau}  \diff t' \xi(t')$.
Thus, $\Delta \theta$ would also be a Gaussian noise.
For this kind of noise, the average of the exponentials in eq. (\ref{AUTO_CN}) can be expressed as
\begin{equation}
\label{GAUSS_EVAL0}
\Big\langle {\rm e}^{\pm i\Delta \theta}\Big\rangle=\mathrm{exp}\left[-{\langle (\Delta \theta)^2\rangle\over 2} \right]
\end{equation}
where one has
\begin{equation}
\label{DOUBLE_INT}
\langle (\Delta \theta)^2\rangle=\int_t^{t+\tau} \diff t' \int_t^{t+\tau} \diff t'' \langle \xi(t') \xi(t'')\rangle
\end{equation}
and
\begin{equation}
\label{GAUSSNOISE1}
C(\tau)=v_0^2 \mathrm{exp}\left[-{\langle (\Delta \theta)^2\rangle\over 2} \right]\,.
\end{equation}
Inserting the simplest kind of noise correlations,
\begin{equation}
\langle\xi(t)\xi(\tilde{t})\rangle=\sigma^2 \delta(t-\tilde{t})
\end{equation}
into eq. (\ref{DOUBLE_INT}) gives
\begin{equation}
\label{LINEAR_GROWTH1}
\langle (\Delta \theta)^2\rangle=\tau \sigma^2
\end{equation}
Plugging this result into (\ref{GAUSSNOISE1}) leads to an exponential decay of the vaf: 
\begin{equation}
\label{EXPON_VAC}
C(\tau)=v_0^2\,\mathrm{exp}\left(-{ \sigma^2 \tau \over  2}\right)
\end{equation}
i.e. exactly what is predicted by kinetic theory for times larger than $R/v_0$ and verified in agent-based simulations at low density $M\ll 1$ and 
small coupling strength $\Sc\ll 1$.
Moreover, comparing eq. (\ref{EXPON_VAC}) with (\ref{DEF_VAC1}) gives an explicit expression
for the noise strength of the effective Langevin equation for the angle of a particle:
\begin{equation}
\label{RELAT_ANG_SIGMA1}
\sigma^2={2\over \tau_C}={64 R \Gamma^2 M\over 9 \pi^2 v_0}
\end{equation}
where expression (\ref{TAU_C_RESULT}) from kinetic theory for $\tau_C$ was used.
One sees that the effective noise increases with increasing density and coupling strength. 
This is plausible, since increasing these parameters leads to a stronger scattering of a particle on others
which is reflected in a stronger noise.
The decrease of the noise  
with increasing particle velocity also makes sense since the mean free path of a particle increases with velocity.

In summary, by means of a quantitative kinetic theory we were able to construct an effective Langevin equation for
the time evolution of the particle's angle. 
In the limit of small density and small coupling, we observe that the noise
is delta-correlated, at least on the time-scale of the coarse-grained time of the underlying Boltzmann 
approach, $T_{\text{free}}$, and derive an explicit expression of the strength of this noise.
Note that the apparent white noise obtained by this kinetic approach does not exclude the possibility
that the noise is actually colored on very short time scales of order $T_{\text{dur}}$, where the coarse-grained
scattering approach is not applicable.

\subsection{Noise calculation for large densities: a random-telegraph approach}
\label{sec:large_dens}

\subsubsection{The variance \texorpdfstring{$\langle\dot{\theta}^2\rangle$}{}}

The calculation of the noise strength in the last section 
is based on the Boltzmann approach and thus fails at larger densities where $M$
is not very small anymore. Here, we go to the opposite limit, $M\gg 1$, but still assume small 
coupling,
$\Sc=|\Gamma| R/v_0\ll 1$. 
In this case, each particle has many neighbors on average, and the time scale of 
anti-alignment is much larger than the time 
scale at which particles loose contact to their neighbors.
The equation of motion for the orientation of a given particle reads
\begin{align}
	\dot{\theta}_i=-|\Gamma| \sum_{j\in \Omega_i} \sin(\theta_j-\theta_i)
=-|\Gamma| \sum_{j=1}^N a_{ij}\,\sin(\theta_j-\theta_i)\,,
	\label{eq:langevinangle}
\end{align}
where $a_{ij}=1$ if particles $j$ and $i$ are closer than the radius $R$, and is zero otherwise.
The modeling assumption of a Langevin-equation for $\dot{\theta}$, eq. (\ref{simpleLANG}), implies
that the right hand side of (\ref{eq:langevinangle}) should be interpreted as a random variable.
In the simplest mean-field approximation we consider all particles to be independent and equally distributed in space and orientation.
Then, we can estimate the order of the autocorrelation time of the right hand side of eq. (\ref{eq:langevinangle}) 
as $R/v_0$ because this is the inverse rate at which a given neighbor disappears and a new, independent neighbor appears. 
The reorientation of the neighbors occurs on much longer time scales, and therefore can be neglected.
Thus, for time scales much larger than $R/v_0$ we expect the fluctuating random quantity on the right hand side of \cref{eq:langevinangle} to be white noise asymptotically.

\begin{figure}
\begin{center}
\includegraphics[width=4.1in,angle=0]{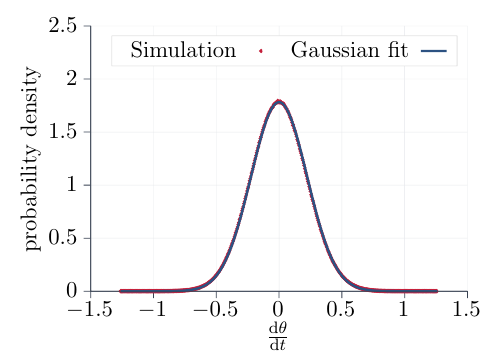}
\caption{
Histogram of $\dot{\theta}$ from agent-based simulations for $M=35.349$ and $\Gamma=-0.1$ (black curve) compared with fitted
Gaussian function (red curve) with variance $0.05$ whereas the measured value is $\langle \dot{\theta}^2\rangle=0.0506$. For visual clarity, only every tenth simulation data point is shown.
Parameters: $\Sc=0.15$, linear system size $L=40$, $N=2000$, $v_0=2$, $R=3$, $M\,\Sc^2=0.795211$, 
time step $\Delta t=0.025$, average over $40$ ensembles.
}
\label{GAUSS_M35_GAM0p1}
\end{center}
\end{figure}

\begin{figure}
\begin{center}
\includegraphics[width=4.1in,angle=0]{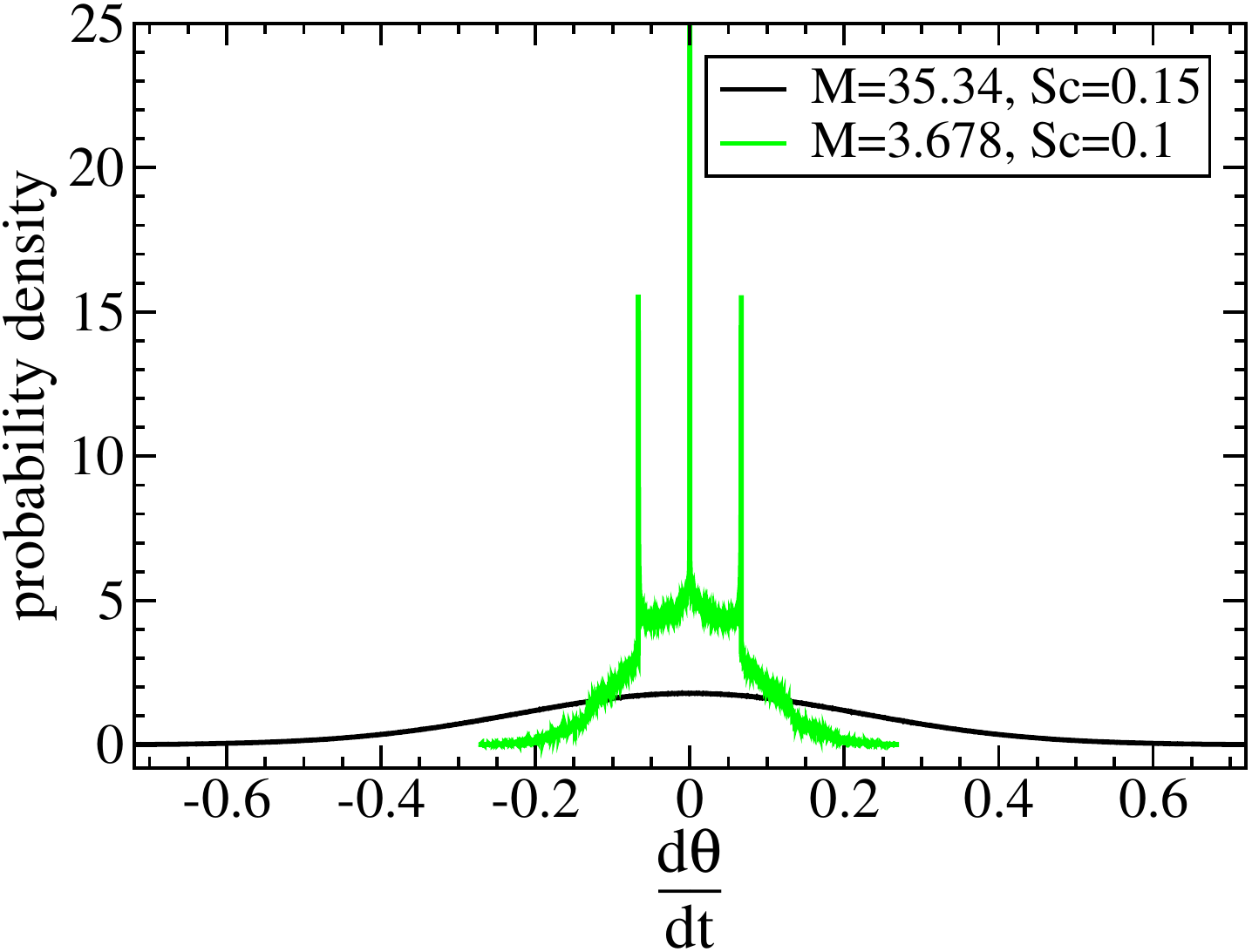}
\caption{
Histogram of $\dot{\theta}$ from agent-based simulations for $M=3.678$ and $\Gamma=-0.06666$ (green curve).
The central peak has a height of $212.3$. 
Parameters: $\Sc=0.1$, linear system size $L=40$, $N=2000$, $v_0=2$, $R=3$, $M\,\Sc^2=0.03677$, 
time step $\Delta t=0.025$, average over $40$ ensembles.
Comparison with 
the measured distribution for $M=35.34$ and $\Sc=0.15$ (black curve).
}
\label{NOGAUSS_M3p677_Sc0p1}
\end{center}
\end{figure}

For a given value of $\theta_i$, in our mean-field picture, the terms $\sin(\theta_i-\theta_j)$ are independent 
random variables.
For large $M$ there are typically many summands in \cref{eq:langevinangle}.
Hence, according to the central limit theorem, we can approximate the sum as Gaussian.
Fig.~\ref{GAUSS_M35_GAM0p1} shows that this is an excellent assumption already for $M=35.34$, whereas
for $M=3.68$ the distribution is still far from a Gaussian distribution and shows pronounced peaks, especially 
at $\dot{\theta}=0$ which corresponds to the case where the focal particle is alone in its collision circle, 
Fig.~\ref{NOGAUSS_M3p677_Sc0p1}. 
To proceed, we calculate the variance of $\dot{\theta}_i$ from the microscopic collision rule.
Within our simple mean-field assumption we consider particles even within the collision circle as uncorrelated.
This is in contrast to the more accurate one-sided molecular chaos approximation we employed at small density. 
Thus, we assume that
the two-particle distribution function $P(\theta_i,\theta_j)=P(\theta_i)P(\theta_j)=1/(2\pi)^2$,
factorizes
and obtain the variance as,
\begin{eqnarray}
\nonumber
	\langle \dot{\theta}_i^2\rangle &=& \langle (-|\Gamma| \sum_{j\in \Omega_i}\sin(\theta_j-\theta_i))^2\rangle
	\\
\nonumber
	&=& \Gamma^2 \sum_{n=0}^\infty \operatorname{Prob}(\# \text{neighbors} =n) \bigg[\sum_{j=2}^{n+1}\int_{0}^{2\pi} \sin^2(\theta_j-\theta_i)\frac{1}{2\pi}d \theta_j \\ \nonumber
&+&\sum_{j\neq k, =2}^{n+1}\int_{0}^{2\pi}\sin(\theta_j-\theta_i)\frac{1}{2\pi}d \theta_j\int_{0}^{2\pi}\sin(\theta_k-\theta_i)\frac{1}{2\pi}d \theta_k \bigg]
	\\
	&=&\Gamma^2 \sum_{n=0}^\infty \operatorname{Prob}(\#\text{neighbors} =n) \sum_{j=2}^{n+1}\int_{0}^{2\pi} \sin^2(\theta_j-\theta_i)\frac{1}{2\pi}d \theta_j 
	\nonumber
	\\
	&=&\Gamma^2 M \int_{0}^{2\pi} \sin^2(\theta_j-\theta_i)\frac{1}{2\pi}d \theta_j=\frac{\Gamma^2 M}{2}. 
	\label{eq:noisestrength}
\end{eqnarray}
In section \ref{sec:thetapunkt}, the validity of eq. (\ref{eq:noisestrength}) is checked by agent-based simulations.

\subsubsection{The relation between $\langle\dot{\theta}^2\rangle$ and the velocity autocorrelation function}

Differentiating the autocorrelation function, eq. (\ref{AUTO_CN}), with respect to time gives
\begin{equation}
{d C(\tau)\over d\tau}=\dot{C}=-v_0^2
\langle \dot{(\Delta \theta_i)}\,\sin(\Delta \theta_i)\rangle=
-v_0^2\langle \dot{\theta}_i(t+\tau)\,\sin(\Delta \theta_i)\rangle
\end{equation}
For vanishing time interval, $\tau\to 0$, $\sin(\Delta \theta_i)$ goes to zero and we find
\begin{equation}
\label{FIRST_DERIV1}
\dot{C}|_{\tau=0}=0
\end{equation}
Differentiating a second time results in
\begin{equation}
{d^2 C(\tau)\over d\tau^2}=\ddot{C}=-v_0^2
\Big[
\langle \ddot{\theta}_i(t+\tau)\,\sin(\Delta \theta_i)\rangle
+\langle \dot{\theta}^2_i(t+\tau)\,\cos(\Delta \theta_i)\rangle
\Big]
\end{equation}
and we obtain a relation to the variance as
\begin{equation}
\label{SECOND_DERIV_C1}
\ddot{C}|_{\tau=0}=-v_0^2\,\langle \dot{\theta}_i^2 \rangle
\end{equation}
For a Gaussian noise, the autocorrelation function has the form given by eq. (\ref{GAUSSNOISE1}).
Differentiating (\ref{GAUSSNOISE1}) and requiring consistency with (\ref{FIRST_DERIV1}) yields
\begin{equation}
\label{FIRST_DERIV2}
{d\over d \tau} \langle \Delta \theta_i^2\rangle|_{\tau=0}=0
\end{equation}
whereas differentiating eq. (\ref{GAUSSNOISE1}) a second time leads to
\begin{equation}
\label{SECOND_DERIV2}
\ddot{C}\big|_{\tau=0}=-{v_0^2\over 2}\,{d^2\over d\tau^2} \langle \theta_i^2\rangle\big|_{\tau=0}
\end{equation}
Comparison to (\ref{SECOND_DERIV_C1}) gives us a way to obtain the small time behavior of the angular displacement
from the microscopic collision rule,
\begin{equation}
\label{SMALL_TIME0}
{d^2\over d\tau^2} \langle \Delta\theta_i^2\rangle\big|_{\tau=0}=2 \langle \dot{\theta}_i^2 \rangle\big|_{\tau=0}
=\Gamma^2\,M\,,
\end{equation}
where eq. (\ref{eq:noisestrength}) was used.
Due to the requirements, (\ref{FIRST_DERIV2}) and (\ref{SMALL_TIME0}), a linear time dependence 
of $\langle \Delta\theta_i^2\rangle$ as in perfect angular diffusion
$\langle \Delta\theta_i^2 \rangle\sim \tau$ is ruled out for small time scales $\tau<R/v_0$ but is expected to appear 
at larger times. This just reflects the fact that in reality no noise is perfectly white. Here, 
the effective dynamical noise is colored with  
a finite autocorrelation time of order $R/v_0$ which describes the arrival and departure of collision partner in the 
collision circle of a given particle.
As we will see later, this small ``non-whiteness''
is essential for the long-time diffusive behavior of the displacement.
According to (\ref{FIRST_DERIV2}) and (\ref{SMALL_TIME0}), for small times, we expand the angular displacement as
\begin{equation}
\label{SMALL_TIME1}
\langle \Delta\theta_i^2 \rangle =\tau^2{\Gamma^2 M\over 2}+\alpha_3 |\tau|\tau^2+\alpha_4 \tau^4+\ldots
\end{equation}
with coefficients $\alpha_j$ given later in eq. (\ref{SMALL_TAU_EXP}).
\vspace{0.4cm}

\subsubsection{A random telegraph model}

The main difficulty in solving the microscopic evolution equations for the angles $\theta_i$, 
eqs. (\ref{eq:langevinangle}), is
that although the equations can be formally closed, there is a dependence on the {\em entire history} of all the angles.
This can be seen by integrating the position equations (\ref{POS_EQ}) and writing the indicator functions $a_{ij}$
as
\begin{equation}
a_{ij}(t)=a(|\vec{r}_i(t)-\vec{r}_j(t)|)=
a\left(v_0\left|\int_0^t [\hat{\vec n}(\theta_i(t'))-\hat{\vec n}(\theta_j(t'))]\, \diff t'\right|\right)
\end{equation}
For low densities, the difficulty could be resolved by expressing the dynamics in terms of isolated two-particle meetings 
within a Boltzmann-like theory.
However, to obtain a similar quantitative theory for moderate and high densities, one 
would have to abandon the one-sided molecular chaos assumption 
and to include a consistent treatment of two- or higher-particle correlation functions. 
In principle, this can be done by means of ring-kinetic theory \cite{chou_15,kuersten_21b} 
or the less accurate Landau kinetic theory \cite{landau1936,patelli_21,boltz_24} for active particles.
However, these theories are very complicated and often they can only be evaluated with a numerical effort of the 
same or higher order
as direct agent-based simulations.
Therefore, in this paper, we adopt a drastically simplified theoretical approach:
the quantities $a_{ij}$ in eq. (\ref{eq:langevinangle}) which can only take the values zero or one, are modeled
as independent random telegraph (RT) processes, see Appendix \ref{app:D},
\begin{equation}
\label{RT_CORR}
\langle a_{ij}(t+\tau)\,a_{ij}(t)\rangle=\delta_{jk}\,g(\tau)\;\;\;{\rm for}\; i\neq j,\, i\neq k
\end{equation}
with the correlation function $g(\tau)$ which is even in time, $g(\tau)=g(-\tau)$ and will be constructed
from the microscopic details of the active particle system.
Here, the focal particle $i$ is different from the particles $j$ and $k$ it ``collides'' with.
This simplification decouples the dynamics of the angles from the neighbor property $a_{ij}$.

To derive an equation for the angular displacement and thus for the vaf, we define the following complex numbers
at two different times $\tilde{t}$ and $t'$:
\begin{eqnarray}
\nonumber
z_j& \equiv &{\rm e}^{\ii\theta_j(t')} \\
\tilde{z}_j& \equiv &{\rm e}^{\ii\theta_j(\tilde{t})} 
\label{DEF_COMPLEX}
\end{eqnarray}
Because of eq. (\ref{DEF_LANG}) we can write the mean square angular displacement as
\begin{eqnarray}
\nonumber
\langle \Delta\theta_i^2 \rangle&=&
\int_t^{t+\tau} \diff \tilde{t}
\int_t^{t+\tau}  \diff t' 
\langle\dot{\theta}(\tilde{t})\,
\dot{\theta}(t')\rangle \\
&=&\Gamma^2
\int_t^{t+\tau} \diff \tilde{t}
\int_t^{t+\tau}  \diff t' 
\sum_{j=1}^N
\sum_{k=1}^N
\langle
a_{ij}(\tilde{t})\, a_{ik}(t')\,
\sin(\tilde{\theta}_j-\tilde{\theta}_i)\,
\sin(\theta_k-\theta_i)
\rangle
\label{DEF_SQUARE1}
\end{eqnarray}
with the abbreviations $\tilde{\theta}_j\equiv \theta_j(\tilde{t})$
and $\theta_j\equiv \theta_j(t')$.
Assuming that the $a_{ij}$ are independent of the angles with a second moment given by eq. (\ref{RT_CORR}), and using
the complex representation of the sine, one finds
\begin{equation}
\label{INTEGRAL_EQ0}
\langle \Delta\theta_i^2 \rangle=
-{\Gamma^2\over 4}
\sum_{j=1}^N
\int_t^{t+\tau} \diff \tilde{t}
\int_t^{t+\tau}  \diff t'
g(\tilde{t}-t')
\langle 
(\tilde{z}_j\tilde{z}_i^*-\tilde{z}_i\tilde{z}_j^*)
(z_j z_i^*-z_i z_j^*)
\rangle
\end{equation}
In evaluating the right hand side of (\ref{INTEGRAL_EQ0}) we assume that particles are uncorrelated (molecular chaos assumption),
that is, for example,
$\langle \tilde{z}_j z_i^*\rangle=
\langle \tilde{z}_j\rangle \langle z_i^*\rangle=0$ or $\langle \tilde{z}^*_j z_j \tilde{z}_i z_i^* \rangle=
\langle \tilde{z}^*_j z_j \rangle \langle \tilde{z}_i z_i^* \rangle $.
We also assume isotropy, i.e. that there is no preferred direction. This means that combinations of the $z_j's$ and $\tilde{z}_j's$
which are not rotationally invariant, such as $z_j\tilde{z}_j$ have a vanishing mean value, e.g. $\langle z_j\tilde{z}_j\rangle=0$. 
This can be seen by rotating the coordinate system by an arbitrary angle $\alpha$. The combination $z_j\tilde{z}_j$ would turn into 
$\mathrm{exp}(2i\alpha)z_j\tilde{z}_j$ i.e. would have an explicit dependence on the orientation of the coordinate system, and thus
is not rotationally invariant. In contrast, rotating the combination $z^*_j\tilde{z}_j$ would show no such dependence.
Note that the $N(N-1)$ terms proportional to terms $\langle a_{ij}\rangle\langle a_{ik}\rangle$ for $j\neq k$ in (\ref{DEF_SQUARE1})
have prefactors that vanish under the presumed molecular chaos assumption.

Finally, since all $j=1\ldots N$ particles have identical properties and since for $j=i$ there is no contribution to the right hand side,
we obtain
\begin{equation}
\label{INTEGRAL_EQ1}
\langle \Delta\theta_1^2 \rangle=
{\Gamma^2(N-1)\over 4}
\int_t^{t+\tau} \diff \tilde{t}
\int_t^{t+\tau}  \diff t'
g(\tilde{t}-t')
\Big[
\langle \tilde{z}_2 z_2^*\rangle \langle \tilde{z}^*_1 z_1\rangle
+\langle \tilde{z}^*_2 z_2\rangle \langle \tilde{z}_1 z^*_1\rangle
\Big]
\end{equation}
where we took particle $i=1$ as focal particle and particle $j=2$ as a representative neighbor of particle $1$.
Using the Gaussian assumption for the angular displacements we can express the products on the right hand side of (\ref{INTEGRAL_EQ1})
as given in eq. (\ref{GAUSS_EVAL0}). For example, one has
\begin{equation}
\langle \tilde{z}^*_1 z_1\rangle=\Big\langle {\rm e}^{\ii[\theta_1(t')-\theta_1(\tilde{t})]}\Big\rangle=
{\rm e}^{-{1\over 2}\langle \Delta \theta_1^2(\tilde{t}-t')\rangle }
=\langle \tilde{z}_1 z^*_1\rangle
\end{equation}
Requiring self-consistency, we drop the particle indices, since every particle's displacement should be the same 
on average. Assuming stationarity and by formally going to the thermodynamic limit, $N\to \infty$, $L\to \infty$ 
at constant
$\rho_0=N/L^2$, we obtain an integral equation for the mean angular displacement,
\begin{equation}
\label{INTEGRAL_PREFINAL}
\langle \Delta\theta^2(\tau) \rangle=
{\Gamma^2\over 2}
\int_0^{\tau} \diff \tilde{t}
\int_0^{\tau}  \diff t'\,
\hat{g}(\tilde{t}-t')\,
{\rm e}^{-\langle \Delta \theta^2(\tilde{t}-t')\rangle }
\end{equation}
with the scaled correlation function $\hat{g}$ of the random telegraph process,
\begin{equation}
\hat{g}(t)=\lim_{N\to \infty} (N-1)\, g(t)
\end{equation}
This correlation function is calculated
in Appendix \ref{app:D} with the result 
\begin{equation}
\label{FORMULA_CORR_RT}
\hat{g}(\tau)=M{\rm e}^{-w_{\text{off}}\,|\tau|}
\end{equation}
Here, $w_{\text{off}}$ is the rate by which the random variable $a_{ij}$ switches from one to zero.
Thus, $w_{\text{off}}$ parametrizes the statistical modeling of the effect that a collision partner of 
the focal particle leaves the collision circle at a particular time.
This connection to the microscopic dynamics will be made explicit in the following sections 
where $w_{\text{off}}$ will be determined self-consistently by means of kinetic theory.
Inserting eq. (\ref{FORMULA_CORR_RT}) into (\ref{INTEGRAL_PREFINAL}) leads to 
the final form of the integral equation for the mean square angular displacement 
\begin{equation}
\label{INTEGRAL_FINAL}
\langle \Delta\theta^2(\tau) \rangle=
{\Gamma^2M\over 2}
\int_0^{\tau} \diff \tilde{t}
\int_0^{\tau}  \diff t'\,
{\rm e}^{-w_{\text{off}}|\tilde{t}-t'|}\,
{\rm e}^{-\langle \Delta \theta^2(\tilde{t}-t')\rangle }\,.
\end{equation}
Solving this self-consistent equation for the function $\langle \Delta\theta^2(\tau) \rangle $ 
allows us to find the velocity auto-correlation function $C(\tau)$ at all times $\tau$ by means of
eq. (\ref{GAUSSNOISE1}).
A similar integral equation has been found earlier in Ref. \cite{vanmeegen_18} in the context of 
randomly coupled but fixed rotators.
As shown later, knowledge of the mean square angular displacement enables the calculation of the noise
correlations $\langle \xi(t)\xi(t') \rangle $ by inverting eq. (\ref{DOUBLE_INT}).

\subsubsection{Calculating the rate $w_{\text{off}}$}
\label{sec:woff}

The determination of the OFF-rate, $w_{\text{off}}$ is essential for a good model of the collision process by the random telegraph process. 
We consider the event that the variable $a_{ij}$ takes the value zero for the first time after having started 
from the value one at time $t=0$. The probability that this occurs at a time between $t$ and $t+ \diff t$ is denoted
by $W_{1\to 0}=P_{1\to 0}(t)\, \diff t$.
To find the probability density $P_{1\to 0}$ for the random telegraph process, we discretize time
$t=t_n=n\,\Delta t$ with a small time step $\Delta t$. 
The probability for the first ``success'' (corresponding to switching from 1 to 0 for the first time) at time $t_n$ is given
by the geometric distribution:
\begin{equation}
W_n=(1-w_{\text{off}}\Delta t)^{n-1}\,w_{\text{off}}\,\Delta t=\left(1-{w_{\text{off}}t_n\over n}\right)^{n-1}\,w_{\text{off}}\,\Delta t
\end{equation}

Setting $W_n=P_{1\to 0}(t)\,\Delta t$ and performing the continuous time limit $\Delta t\to 0$ at fixed
time $t=t_n$ and using the relation
\begin{equation}
\lim_{n\to\infty} \left(1-{x\over n}\right)^n={\rm e}^{-x}
\end{equation}
gives
\begin{equation}
P_{1\to 0}(t)=w_{\text{off}}{\rm e}^{-w_{\text{off}}\,t}
\end{equation}
This exponential probability density has a finite mean: The averaged first passage time for the flip from ON to OFF
follows as
\begin{equation}
\label{TFLIP_RT}
T_{flip}=\langle t \rangle=\int_{0}^{\infty} t\,P_{1\to 0}(t)\, \diff t={1\over w_{\text{off}}}
\end{equation}
because the density $P_{1\to 0}(t)$ is normalized to one.

To determine $w_{\text{off}}$ and to provide a link to the random telegraph process,
we first consider the actual contact process of a particle that has entered the collision circle of the focal particle
$i=1$ at time zero and leaves it at time $t$.
As a first approximation, we assume that $\Sc=|\Gamma| R/v_0\to 0$, which means that particles move 
ballistically in straight lines with constant speed
during their contact time.
Assuming furthermore that particles outside the collision circle are equally distributed in space and have no preferred 
direction, the average contact time turns out to be finite as in the RT-process and can be calculated exactly
by means of kinetic theory,
as shown in Appendix \ref{app:E}, with the result
\begin{equation}
\label{contact_time_moment}
\langle t \rangle ={\pi^2\over 8} {R\over v_0}
\end{equation}
Equating this moment with the one from the RT-process, eq. (\ref{TFLIP_RT}), leads to the OFF-rate,
\begin{equation}
\label{DEF_B}
w_{\text{off}}=B\, {v_0\over R}
\end{equation}
with the constant $B=8/\pi^2\approx 0.81$.
The behavior that $w_{\text{off}}\sim R/v_0$ with a proportionality constant of order one is expected
on dimensional and physical grounds since the time a particle of speed $v_0$ flies through an area of linear extension $R$ 
is of order $R/v_0$.
Modeling by the random telegraph process is not perfect as indicated by the fact that the distribution
of the exit time $t_{exit}$ is qualitatively different from the actual behavior of particles in the limit $\Sc\to 0$.
According to eq. (\ref{contact_dist_limit}) the distribution for the direct contact process has a power law tail, whereas in the RT-process
the distribution is exponential.
As a consequence, while the scaling behavior of $w_{\text{off}}$ is captured correctly, the prediction for the 
prefactor $B$ should only be taken as a first estimate. 
In section \ref{sec:matchRT} it is shown how $B$ can determined self-consistently by matching it to the Boltzmann kinetic theory
with the result $B=9\pi^2/64$.

\subsubsection{Solution of the integral equation}
\label{sec:sol_integral}

All parameters in the non-linear integral equation (\ref{INTEGRAL_FINAL}) for the mean square angular displacement
are defined and we proceed to solving it.
First, we discretize time with a small time step 
$\Delta t$
as $\tilde{t}=n\,\Delta t$, $t'=m\,\Delta t$, $\tau=p\,\Delta t$ with $n,m,p=0,1,2,\ldots$ and define
$\langle \Delta \theta(\tau)^2\rangle=\langle \Delta \theta(p\,\Delta t)^2\rangle\equiv x_p$.
The integral equation (\ref{INTEGRAL_FINAL}) is then discretized as
\begin{equation}
\label{DISCRETE_INTEGRAL1}
x_p={M\Gamma^2\over 2}\Delta t^2\sum_{n=1}^p \sum_{m=1}^p {\rm e}^{-w_{\text{off}}\,\Delta t|n-m|-x_{n-m}}\;\;\;\;
{\rm for}\;p\ge 1
\end{equation}
with initial value $x_0=0$ and
where we imply the discrete time-reversal symmetry $x_{n-m}=x_{m-n}$.
At the smallest non-zero time, $p=1$, eq. (\ref{DISCRETE_INTEGRAL1}) reproduces the quadratic small time behavior
required by eq. (\ref{SMALL_TIME1}),
$x_1={M\Gamma^2\over 2}\,\Delta t^2$.

\begin{figure}
\begin{center}
\vspace{0.2cm}
\includegraphics[width=5.1in,angle=0]{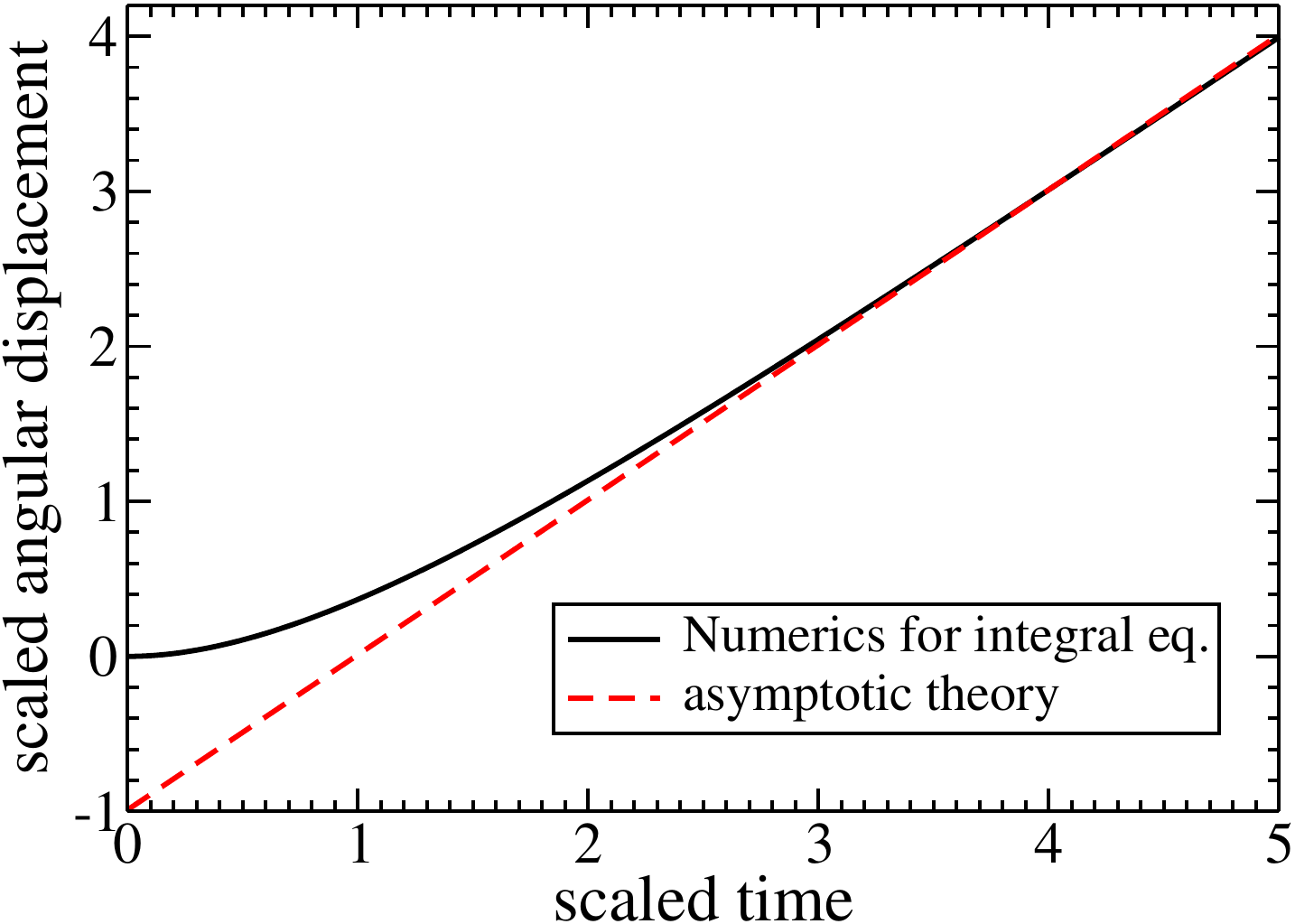}
\caption{
The normalized angular displacement $\langle\Delta \theta^2\rangle/\epsilon$ is plotted versus dimensionless
time $\tilde{t}=t\,w_{\text{off}}$. The black curve shows the numerical solution of the integral equation (\ref{RECURSIVE_INT}),
whereas the red line is the analytical linear result from eq. (\ref{PREDICT_LIN_SCAL1}), given here by
$-0.99+0.99995\,\tilde{t}$.
Parameters: $\epsilon=0.01$, $wt=\Delta t\,w_{\text{off}}=0.001$, $N_{tot}=5\times 10^{4}$ points, $M=1$, $|\Gamma|/w_{\text{off}}=0.1$. 
}
\label{FIG_INTEGRAL_EQ_eps0p01}
\end{center}
\end{figure}

By increasing $p$ step by step, we found a simpler form of the discretized integral equation
\begin{equation}
\label{RECURSIVE_INT}
x_p={M\Gamma^2 \Delta t^2\over 2}
\Big[ p+2\sum_{m=1}^{p-1} (p-m)\,{\rm e}^{-m\,w_{\text{off}}-x_m}
\Big]
\end{equation}
Because the right hand side of (\ref{RECURSIVE_INT}) contains only displacements $x_m$ at smaller times, i.e. with $m<p$,
it is a recurrence relation for the value of $x_p$ using the values $x_{p-1}, x_{p-2},\ldots x_0$.
Thus, by increasing the index $p$ in steps of one, storing all obtained values $x_p$ for use in the next sweep, 
the entire temporal behavior of the angular displacement can be found
numerically, see Fig. \ref{FIG_INTEGRAL_EQ_eps0p01}.
Numerical results obtained by this method in Fig. \ref{FIG_INTEGRAL_EQ_eps0p01} show that $x_p$ 
increases linear in time at large times, corresponding to a simple 
exponential decay of the vaf, see (\ref{GAUSSNOISE1}).
Performing the continuum limit of the recurrence relation (\ref{RECURSIVE_INT}) by
$\sum \Delta t\ldots \to \int\, \diff t\ldots$ for $\Delta t\to 0$
we arrive at a simpler integral equation, 
\begin{equation}
\label{SINGLE_INTEGRAL1}
\langle \Delta \theta(\tau)^2\rangle=M\Gamma^2 \int_0^{\tau}\,(\tau-t)\,{\rm e}^{-w_{\text{off}}\,t-\langle \Delta\theta(t)^2\rangle}
\, \diff t\,.
\end{equation}
Differentiating eq. (\ref{SINGLE_INTEGRAL1}) or (\ref{INTEGRAL_FINAL}) twice with respect to $\tau$ leads to a non-linear differential equation 
for $x(t)\equiv \langle \Delta\theta^2(t) \rangle$ with
explicit time dependence, 
\begin{equation}
\label{EQUIV_DIFF_EQ}
\ddot{x}(t)=M\Gamma^2 {\rm e}^{-w_{\text{off}}\, |t| -x(t) }\,.
\end{equation}
The explicit time dependence in the differential equation (\ref{EQUIV_DIFF_EQ})  
can be eliminated for $t\neq 0$ by the transformation
\begin{equation}
\label{PARA_X}
y=x+w_{\text{off}}\,|t|
\end{equation}
to yield
\begin{equation}
\label{TRANS_DIFF_EQ}
\ddot{y}(t)=M\Gamma^2 {\rm e}^{-y(t) }+2\,w_{\text{off}}\,\delta(t)
\end{equation}
with $\gamma\equiv M\Gamma^2$
because of $\dot{y}=\dot{x}+2\,w_{\text{off}}[\theta(t)-1/2]$ and $\ddot{y}=\ddot{x}+2 w_{\text{off}}\delta(t)$.
eq. (\ref{TRANS_DIFF_EQ}) can be interpreted as the equation of motion of a particle in an effective potential.
Multiplying by $\dot{y}$ and integrating in time one finds the corresponding 
conserved ``energy'' of this motion: 
\begin{equation}
\label{ENERGY_EQ}
E=\text{const}={\dot{y}^2\over 2}+\gamma{\rm e}^{-y}
\end{equation}
Because of the initial conditions $x(t=0)=0$ and $\dot{x}(0)=0$, one has $y(t=0)=0$ and $\dot{y}(t=0)=w_{\text{off}}$ 
which gives the value of the generalized energy, $E=\gamma+w_{\text{off}}^2/2$.
Solving eq.(\ref{ENERGY_EQ}) for the function $t(y)$ leads to
\begin{equation}
t=\int_0^{y} \left[ 2\gamma\left(1-{\rm e}^{-x}\right)+w_{\text{off}}^2 \right]^{-1/2}\,\diff x
\end{equation}
The transformation $\mathrm{exp}(-x)=z^2$ leads to a solvable standard integral,
\begin{equation}
t=-\sqrt{2\over \gamma} \int_1^{\mathrm{exp}(-y/2)} {\diff z\over z \sqrt{a^2-z^2} }
\end{equation}
with $a^2\equiv 1+w_{\text{off}}^2/(2\gamma)=1+1/(2\epsilon)$ and $\epsilon=M\Gamma^2/w_{\text{off}}^2$.
One obtains
\begin{equation}
\label{PARA_T}
t=\left. \sqrt{2\over \gamma a^2}\,{\rm ln}\left( {a+\sqrt{a^2-z^2}\over z}\right)\right|_1^{\mathrm{exp}(-y/2)}
=\sqrt{2\over \gamma a^2}\,{\rm ln}\left( {\left[a+\sqrt{a^2-\mathrm{exp}(-y)}\right]\mathrm{exp}(y/2)\over
a+\sqrt{a^2-1} }\right)
\end{equation}
Defining the variable $z\equiv\mathrm{exp}(-y/2)$ leads to a quadratic equation for $z$, which is, of course, solvable.
Thus, finally, it turns out that the integral equation (\ref{INTEGRAL_FINAL}) as well as the differential equation 
(\ref{EQUIV_DIFF_EQ}) are
exactly solvable, and the angular displacement follows as
\begin{equation}
\label{EXACT_SOL}
x(t)=2{\,\rm ln}\left\{
{c^2+{\rm e}^{-2\lambda\,|t|} \over
2b\,c\ {\rm e}^{-\lambda |t|}}\right\}
-w_{\text{off}}|t|
\end{equation}
with $\epsilon\equiv \gamma/w_{\text{off}}^2$, $b^2\equiv 1+1/2\epsilon$, $c\equiv b+\sqrt{b^2-1}$,
$\lambda=w_{\text{off}}\,b\sqrt{\epsilon/2}$.

The solution is rewritten in terms of the dimensionless time $\tilde{t}=t\,w_{\text{off}}$ and the scaled angular displacement
$\tilde{x}=x/\epsilon$.
Analysis for $\tilde{t}\gg 1$ allows neglecting exponentially small terms $\sim \mathrm{exp}(-\tilde{t})$
and gives the expected simple linear growth of the displacement at large times:
\begin{equation}
\label{PREDICT_LIN_SCAL1}
\tilde{x}={\sqrt{1+2\epsilon}-1\over \epsilon}\,\tilde{t}-{2\over \epsilon}\,{\rm ln}
\left[ {2\sqrt{1+2\epsilon} \over 1+\sqrt{1+2\epsilon}}  \right]
\end{equation}
In Fig.~\ref{FIG_INTEGRAL_EQ_eps0p01}
one sees excellent agreement of this asymptotic behavior with the full exact solution.
We checked that the analytical solution (\ref{EXACT_SOL}) agrees perfectly with the numerical solution of the 
integral equation (\ref{RECURSIVE_INT}).

For completeness, a perturbative solution of the differential equation (\ref{EQUIV_DIFF_EQ}) for small times 
will be given here in dimensionless form,
\begin{equation}
\label{SMALL_TAU_EXP}
\tilde{x}={1\over 2}\tilde{t}^2-{1\over 6} |\tilde{t}|^3
+{1\over 2}\left[{1\over 2}-\varepsilon\right]\tilde{t}^4+{1\over 40}\left[\varepsilon-{1\over 3}\right]|\tilde{t}|^5+
\mathcal{O}(\tilde{t}^6)\,
\end{equation}
where the first, quadratic, term agrees with the expectation, eq. (\ref{SMALL_TIME1}).

\subsubsection{Calculating the noise correlations}
\label{sec_calcnoise1}

Differentiating eq. (\ref{DOUBLE_INT}) twice with respect to $\tau$ gives the relation between the noise
correlations and the angular displacement $x(\tau)=\langle [\Delta \theta(\tau)]^2\rangle$,
\begin{equation}
\label{RELAT_CORREL_X1}
\ddot{x}(\tau)=2\,\langle \xi(t) \xi(0)\rangle
\end{equation}
From (\ref{EQUIV_DIFF_EQ}) we know how to express $\ddot{x}$ in terms of $x$ and $y$, 
\begin{equation}
\langle \xi(t) \xi(0)\rangle={\gamma\over 2}{\rm e}^{-y}
\end{equation}
where $y=x+w_{\text{off}}|t|$
and inserting the solution for $x$ from (\ref{EXACT_SOL})
we obtain an explicit expression for the noise correlations
\begin{equation}
\label{NOISE_EXACT}
\langle\xi(t)\xi(\tilde{t})\rangle
={\gamma\over 2}\left[
{2\,b\,c\, {\rm e}^{-\lambda|t-\tilde{t}|} \over
c^2+{\rm e}^{-2\lambda|t-\tilde{t}|}}
\right]^2
\end{equation}
As shown in Fig. \ref{FIG8_NOISE_CORR_THEORY},
the correlations become exponential for $|\tau|\gtrapprox R/v_0$.
When coarse-graining on time scales of order $R/v_0$ (as done in the Boltzmann kinetic theory), the colored network noise appears as an
effective white noise, $\langle \xi(\tau)\xi(0)\rangle\sim \sigma^2 \delta(\tau)$.
To calculate its strength $\sigma^2$ we integrate the noise correlations from zero to a very large time $T$.
For an assumed white noise this gives
\begin{equation}
\int_0^T \langle\xi(t)\xi(0)\rangle\, \diff t={\sigma^2\over 2}
\end{equation}
whereas from (\ref{RELAT_CORREL_X1}) it follows 
\begin{equation}
\int_0^T \langle\xi(t)\xi(0)\rangle\, \diff t=\int_0^T {\ddot{x}\over 2}\, \diff t={1\over 2} \dot{x}(T) \,.
\end{equation}
Equating the two results and performing the limit $T\to \infty$, we obtain the strength of the equivalent white noise
as
\begin{equation}
\label{SIGMA_SELF_CONS}
\sigma^2=\lim_{T\to \infty}\dot{x}(T)=\gamma\int_0^{\infty}{\rm e}^{-x(\tau)-w_{\text{off}}\tau}\,d\tau
=w_{\text{off}} \left[\sqrt{1+2\epsilon}-1\right]
\end{equation}
Because of $\tau_C=2/\sigma^2$, see eq. (\ref{RELAT_ANG_SIGMA1}), the RT-model predicts the auto-correlation 
time as
\begin{equation}
\label{RT_PREDICT_TAU}
\tau_C={2\over w_{\text{off}} \left[\sqrt{1+2\epsilon}-1\right]}
\end{equation}
Note that in the thermodynamic limit the ratio $M/N$ goes to zero, and the effective noise $\sigma^2$ only depends on the OFF-rate
$w_{\text{off}}$ of the random telegraph process.

\begin{figure}
\begin{center}
\vspace{0.2cm}
\includegraphics[width=5.1in,angle=0]{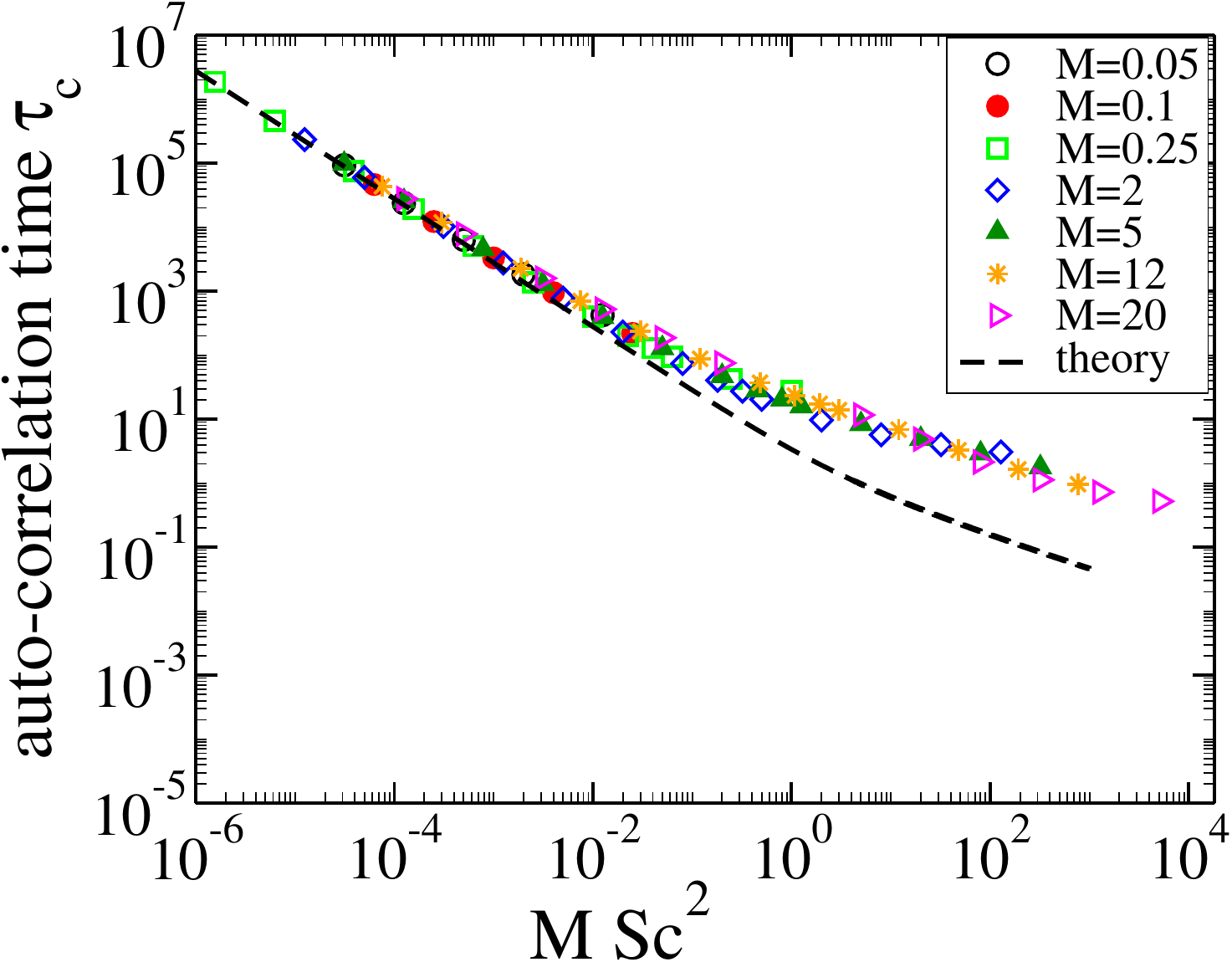}
\caption{
The scaled auto-correlation time $\tau_C\,v_0/R$ from agent-based simulations 
versus the scaling variable $\epsilon=M\,\Sc^2$ for different values of $M$,
compared to the theoretical prediction, eq. (\ref{FINAL_TAUC_PRED}).
}
\label{AUTOCORR_TIME_COMPARE1}
\end{center}
\end{figure}

\begin{figure}
\begin{center}
\vspace{0.2cm}
\includegraphics[width=5.1in,angle=0]{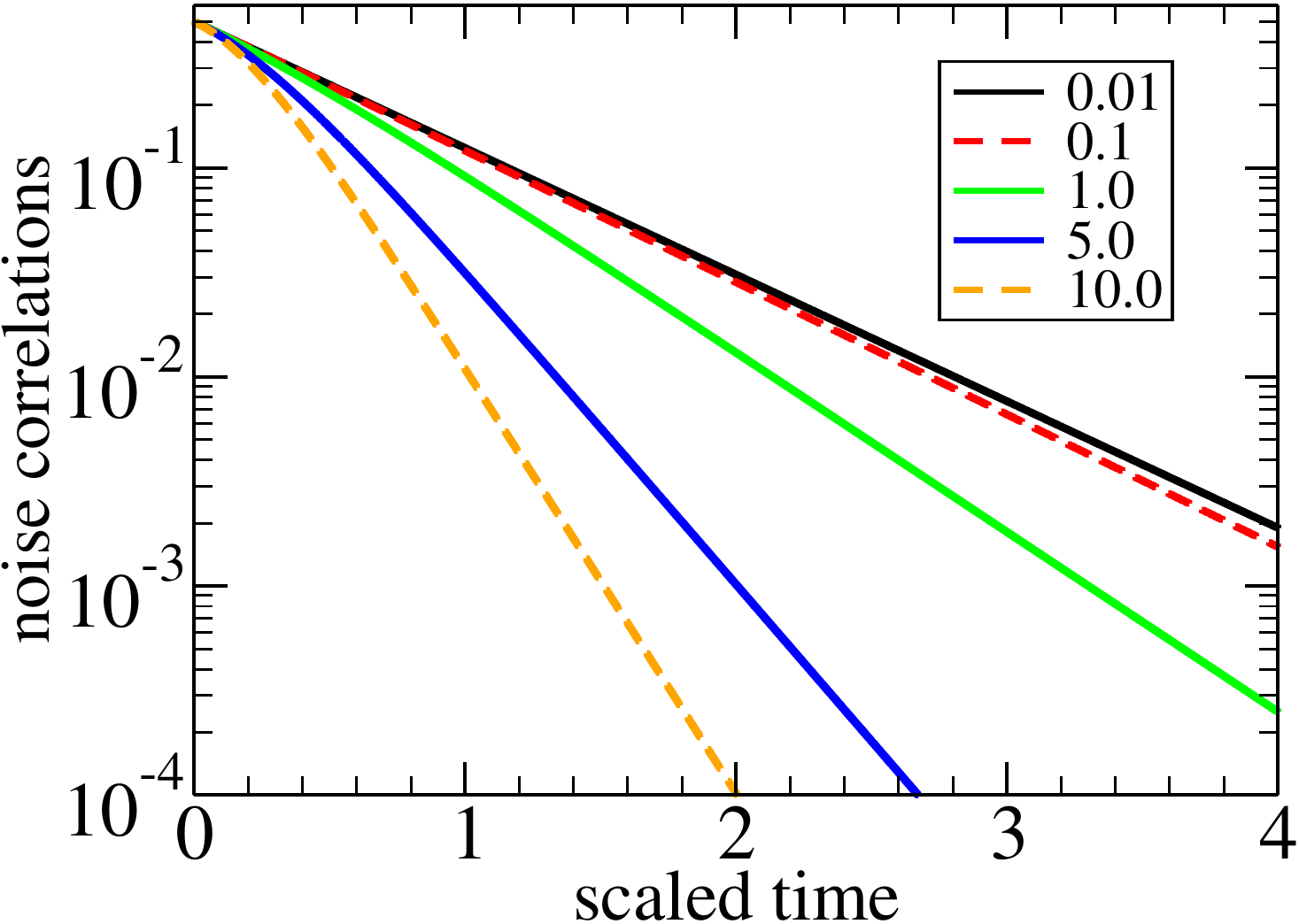}
\caption{
The theoretical prediction for the scaled noise correlations $\langle\xi(t)\xi(0)\rangle/M\Gamma^2$ according 
to eq. (\ref{NOISE_EXACT}) versus scaled time $\tilde{t}=tv_0/R$ for different values of the variable $M\,\Sc^2=0.01$ (black), $0.1$ (red)
till $M\,\Sc^2=10$ (orange). For $\tilde{t}\lessapprox 0.7$, the correlations are non-exponential.
}
\label{FIG8_NOISE_CORR_THEORY}
\end{center}
\end{figure}

\subsubsection{Matching the random-telegraph model and the Boltzmann theory}
\label{sec:matchRT}

eqs. (\ref{SIGMA_SELF_CONS}, \ref{RT_PREDICT_TAU}) suggest that
the noise strength $\sigma^2$ and the auto-correlation time $\tau_C$ are solely controlled  
by the composite variable $\epsilon\sim M\Sc^2$.
As shown in 
\cref{AUTOCORR_TIME_COMPARE1},
this is consistent with agent-based simulations: all data points for $\tau_C$ approximately 
lie on a Master curve when plotted versus $M\,\Sc^2$.
Under this assumption, expression (\ref{RELAT_ANG_SIGMA1}) from kinetic theory, 
valid for $M\ll 1$ and $\Sc\ll 1$, 
should match the small $\epsilon$ limit of (\ref{SIGMA_SELF_CONS}).
This is indeed the case and fixes the proportionality constant in the Ansatz $w_{\text{off}}=B\,v_0/R$ to the value
\begin{equation}
B={9\pi^2\over 64}\approx 1.38791 \,. 
\end{equation}
This leads to the final theoretical prediction for $\tau_C$ and the self-diffusion coefficient $D$,
\begin{eqnarray}
\label{FINAL_TAUC_PRED}
\tau_C&=&
{128 R\over 9 \pi^2 v_0}\,\left[ \sqrt{1+{8192\,M\,\Sc^2\over 81 \pi^4 }}-1 \right]^{-1} \\
\label{D_FINAL_EXPR1}
D&=&
{64 v_0 R\over 9 \pi^2}\,\left[ \sqrt{1+{8192\,M\,\Sc^2\over 81 \pi^4 }}-1 \right]^{-1}\,, 
\end{eqnarray}
supposedly valid for all densities $M$ but for small $\Sc\ll 1$.
Plotting this prediction 
in Fig. (\ref{AUTOCORR_TIME_COMPARE1})
shows excellent agreement with agent-based results at small $M\Sc^2$. At larger $M$ or $\Sc$, 
the theory underestimates the correlation time
$\tau_C$, probably because
the random-telegraph theory neglects correlations among particles inside the collision circle.
Since there are more such particles at larger $M$, these neglected contributions carry more weight in the final expression.
Furthermore, the determination of the coefficient $B$ was based on the Boltzmann kinetic theory which 
was only evaluated up to $\mathcal{O}(\Sc^2)$.
Thus, deviations for $\Sc>1$ are also not surprising.
However, modifications of the value of $B$ will not be sufficient to remedy the strong deviation between
theoretical predictions and simulations at large $M\,\Sc^2$ by about a factor of ten.
This is because in this limit, the expression (\ref{RT_PREDICT_TAU}) for $\tau_C$ becomes independent of $B$,
\begin{equation}
\label{TAU_C_LARGE_EPS}
\tau_C = \sqrt{2\over M}\,{1\over |\Gamma|}\;\;\;{\rm for}\,M\,\Sc^2 \gg 1
\end{equation}
Possible reasons for the discrepancies at larger $M\,\Sc^2$ will be investigated in section \ref{sec:thetapunkt} 
in more detail.


\section{Numerics}
\label{sec:numer}

\subsection{The relaxation of angular modes}
\label{sec:relax_f}

\begin{figure}
\begin{center}
\vspace{-0.3cm}
\includegraphics[width=5.1in,angle=0]{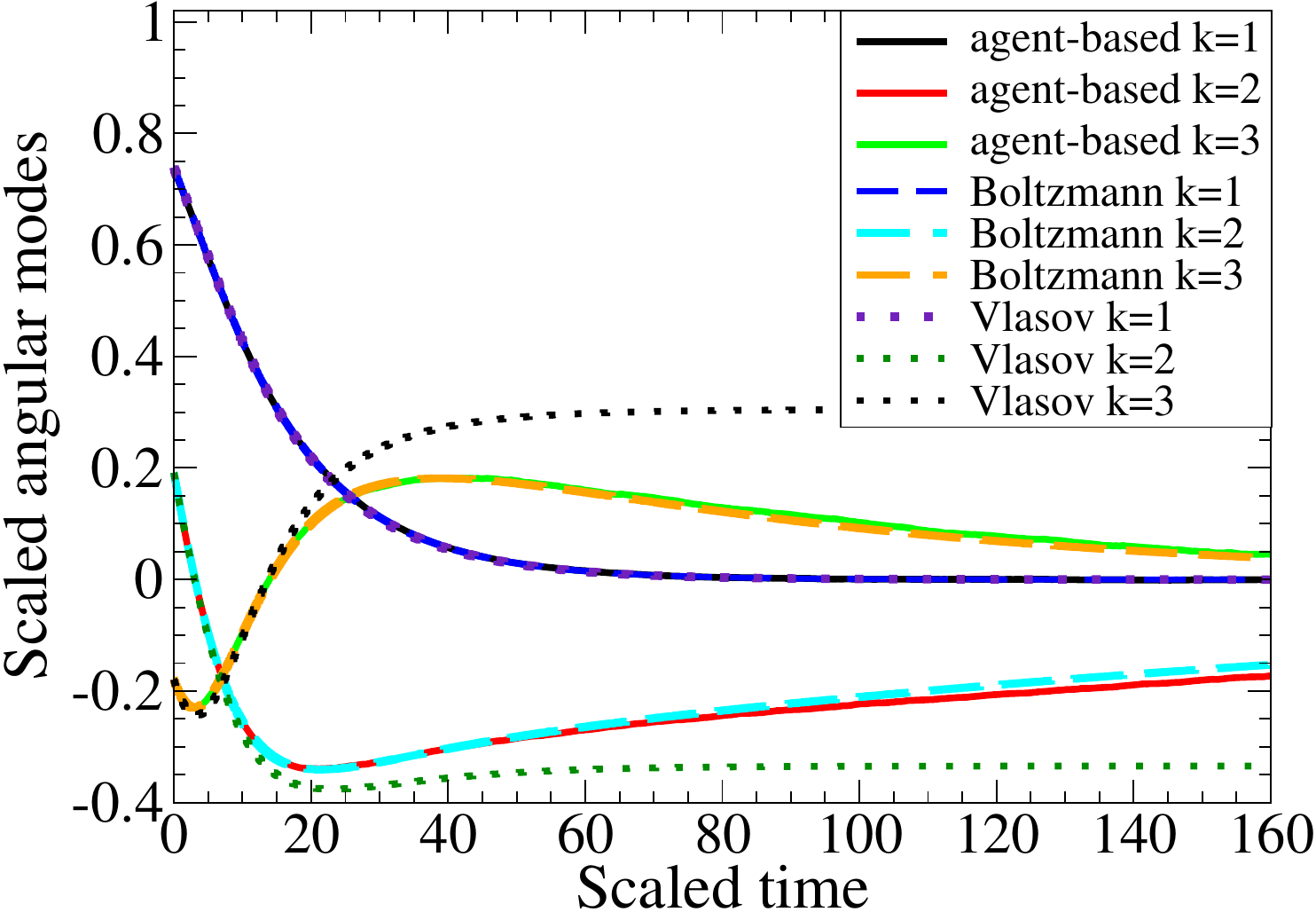}
\vspace{-0.3cm}
\caption{Relaxation of the first angular modes $\hat{f}_k$, divided by $\hat{f}_0$, as a function of dimensionless time
$\tilde{t}=t\,|\Gamma|$, obtained
by agent-based simulations (solid lines), Vlasov-like mean-field theory (dotted lines), 
and Boltzmann kinetic theory (dashed lines) for $\Sc=0.05$ and $M=0.1$.
Parameters for the agent-based simulations: $N=493$ particles, $\Gamma=-0.2$, 
integration time step $\Delta t=0.025$, $v_0=4$, $R=1$, linear system size $L=124$, 
initial opening angle $\alpha=150^\circ$,
number of replicas in the ensemble average, $n_{\text{ens}}=100$.
}
\label{FIG_RELAX1}
\vspace{-0.3cm}
\end{center}
\end{figure}

The usual hydrodynamics modes, particle density and momentum density, are encoded in the lowest angular modes $\hat{f}_0$ and $\hat{f}_1$.
The higher modes $\hat{f}_2,\hat{f}_3,\ldots$ relax faster and are thus considered kinetic modes. 
It is interesting to see if the Vlasov- and Boltzmann theories introduced above are able to correctly describe the time-evolution of the 
angular modes, even far away from stationary states.
Here, we will restrict ourselves to homogeneous systems and leave the evaluation of the full spatiotemporal behavior for future research.

We performed agent-based simulations by solving the time-discretized versions of eqs. (\ref{POS_EQ},\ref{ANGLE_EQ}) for $N$ 
particles using  
a small time step $\tau$ 
on a quadratic domain of linear size $L$ with 
periodic boundary conditions.
The system was initialized by randomly placing particles on the domain with flying directions which are 
equally distributed in the interval $[-\alpha/2,\alpha/2]$ with respect to the x-axis.  
The angular parameter $\alpha>0$ was chosen to be significantly smaller than $2\pi$. 
Thus, the initial state is spatially homogeneous but ordered where all angular Fourier modes are excited according to
\begin{equation}
\hat{f}_{k,init}={\rho_0\over 2\pi}\int_0^{2\pi} P(\theta){\rm e}^{-\ii k\theta} \,\diff \theta 
={\rho_0\over 2\pi \alpha} \int_{-\alpha/2}^{\alpha/2} {\rm e}^{-\ii k\theta} \,\diff \theta ={\rho_0\over 2\pi}\,{\rm sinc}\left({\alpha k\over 2}\right)
\end{equation}
which follows from the definition (\ref{DEF_ANG_FOUR}), and where $P(\theta)$ is the angular distribution of a single particle, and $\rho_0=N/L^2$.
The temporal evolution of the particles was calculated and the instantaneous angular Fourier modes were measured 
as 
\begin{equation}
\hat{f}_k(t)={\rho_0\over 2\pi} {1\over N} \sum_{j=1}^N{\rm e}^{-\ii k\theta_j(t)}
\end{equation}
These measurements were repeated on replica systems which were initialized in the same way but with different random seeds, and an ensemble average of the Fourier modes was performed. 
Fig.~\ref{FIG_RELAX1} shows the relaxation of the first modes, $\hat{f}_1$, $\hat{f}_2$ and $\hat{f}_3$.
As expected the higher modes relax faster. After a sufficiently long time, all modes had relaxed to zero, corresponding to a
disordered stationary state. The non-monotonic decay of the mode $\hat{f}_2$ which describes nematic order can be understood as follows:
Initially, the particles fly mostly in the x-direction, corresponding to relatively large positive
values of the polar and nematic order parameter, $\hat{f}_1$ and $\hat{f}_2$, respectively.
Due to the anti-alignment interaction, particles turn away from each other where about half of the particles tend to go towards 
the y-direction, whereas the other half goes more towards the negative y-direction.
This splitting corresponds to small polar order but significant nematic order with a negative value of $\hat{f}_2$, eventually reaching the minimum value of $\hat{f}_2$.
Later on, while attempting to turn anti-parallel to their respective neighbors, the particles get completely mixed and disoriented, and thus
all $\hat{f}_k$ with $k\ge 1$ go to zero.

To compare this behavior with the predictions from kinetic theory,
we also solved the hierarchy equations, eqs. (\ref{KINETIC-VLASOV1}) for the Vlasov hierarchy, and the Boltzmann hierarchy
where the additional collision term from eq. (\ref{J_COLL_NEW_FIN}) was added to the right hand side of (\ref{KINETIC-VLASOV1}).
The equations for the first 48 modes were solved explicitly where all higher modes were set to zero. 
We checked, that a truncation at a slightly different mode number does no significantly change the results.

Fig.~\ref{FIG_RELAX1} shows that at small times both kinetic theories agree very well with the agent-based result.
However, at larger times, the regular mean-field results start deviating from the direct simulations. Finally, at large times, 
a qualitatively different behavior is apparent: the angular modes do not converge to zero but non-zero asymptotic values
are reached, meaning that the stationary state is ordered.
In addition, by changing the initial opening angle $\alpha$, we observed that the actual asymptotic values 
depend on $\alpha$, i.e. on the initial conditions.

In contrast, the addition of the small correction due to non-mean field effects to the Vlasov-like collision operator
introduces a new long time scale, $\tau_{\text{long}}=(R/v_0)/M\,\Sc^2$, which describes the final relaxation of all modes to zero and
is very large at low densities and coupling strengths.
Furthermore, this small modification leads to an excellent, quantitative agreement between the Boltzmann kinetic theory and the agent-based simulations.
As expected, this agreement gets better, the smaller the scaled density $M$ and the smaller the coupling strength $\Sc$ is.
For large $M>1$, the assumptions of a Boltzmann theory are not valid. However, the qualitative picture stays the same:
the system eventually becomes disordered, which is confirmed by the improved kinetic theory but not by the Vlasov-like theory.

We conclude that in our noise-free system, 
the regular mean-field approximation of factorizing the N-particle probability distribution
leads to erroneous predictions, both at small and at large density, and should be abandoned. 
Instead, the one-sided molecular chaos approximation, as first used about 100 years ago in the standard Boltzmann-equation, leads to
quantitatively correct results, at least in the low density limit.

\subsection{Checking assumptions: measuring \texorpdfstring{$\langle \dot{\theta}^2\rangle$}{<theta^2>}}
\label{sec:thetapunkt}

In the derivation of the self-consistent integral equation (\ref{INTEGRAL_FINAL}) that leads to the determination of
the vaf in the limit of large $M$, a number of approximations had to be made.
One of them was a mean-field assumption in the calculation of $\langle \dot{\theta}^2\rangle$.
In agent-based simulations we checked whether the result, eq. (\ref{eq:noisestrength}), is actually correct.
We found that it is quite accurate at small $M$ and $\Sc$ but deviates increasingly for increasing $M$ 
which is not very encouraging as the result is needed in evaluating the large density case.
For example, for $M=35.34$, see Fig.~\ref{GAUSS_M35_GAM0p1} we observe
$\langle \dot{\theta}^2 \rangle/(M\Gamma^2/2)=0.28633$, i.e. the variance of the angular change is more than a factor of
three smaller than predicted by the molecular chaos assumption.
To rule out that this has to do with the average number of particles, we measured this independently with the result
that the average number of neighbors is only a factor of $0.9633$ smaller than $M$.
For a smaller density, $M=3.677$, see Fig.~\ref{NOGAUSS_M3p677_Sc0p1}, the difference is not so dramatic. Here, we 
find $\langle \dot{\theta}^2 \rangle/(M\Gamma^2/2)=0.68334$ and an average neighbor number of $0.91407\,M$.
Simulations at very large $M=123.7$, $\Sc=0.1$ and $N=7000$ resulted in
$\langle \dot{\theta}^2 \rangle/(M\Gamma^2/2)= 0.4341194$. For even larger $M=135.7$ and $\Sc=0.15$ ($M\,\Sc^2=3.054$) this ratio was measured at 
$\langle \dot{\theta}^2 \rangle/(M\Gamma^2/2)=0.4483190$. 

This lead us to the hypothesis that for larger $M$, different particles within the collision circle of particle $i$ 
are significantly anti-correlated. This causes relevant contributions of the $n(n-1)$ 
cross-terms $\sim \langle \sin(\theta_j-\theta_i)\,\sin(\theta_j-\theta_i)\rangle$ in eq. (\ref{eq:noisestrength}),
effectively lowering the impact of the $n$ diagonal contributions.
The variance $M\Gamma^2/2$ is the essential parameter in the analysis of the RT-approach. Lowering it means that 
for the predicted relaxation times $\tau_C$ one should assign values that correspond to much lower $M$ in the ideal theory (based on molecular chaos) which then leads to much larger predicted values of $\tau_C$.

It is interesting to note that these cross-correlations cannot occur in the small density limit, $M\ll 1$ because
in this case there is simply almost never a third particle involved in collisions.
We speculate that the increasing relevance of the anti-correlations at larger $M$ is due to the fact that there is $n-1$
times more cross-terms than diagonal terms, and larger $n$ become more likely at a larger average value $M=\langle n \rangle$.

Since a self-consistent theory merely for the autocorrelation of a particle, such as eq. (\ref{INTEGRAL_FINAL}), is 
apparently not sufficient 
for $M\,\Sc^2>0.01$, 
one could 
expand the theory by an equation for the cross-correlations or set up a ring-kinetic theory. 
However, this is beyond the scope of this paper and 
will be left for future work.

\section{Conclusions}
\label{sec:conclus}

In summary, we derive an asymptotically exact Boltzmann equation for the Kuramoto-Vicsek model of self-propelled particles without external noise. Starting at the N-particle Liouville equation and assuming low densities, we derive
an asymptotically exact scattering theory by means of a non-local closure of the first BBGKY-equation.
By means of this kinetic theory and a mapping to a random-telegraph process
we construct 
an effective Langevin equation for
the time evolution of a focal particle in a sea of host particles, which should be valid at arbitrary densities.
Analytical expressions for the temporal correlations of its effective noise term and the corresponding 
self-diffusion coefficient are provided. 
The mathematical details of all derivations are laid out and the underlying assumptions are thoroughly discussed.
Comparing to agent-based simulations we show that the theory accurately describes the time-evolution of the
hydrodynamic and kinetic modes
of the system, even far from stationary states. We demonstrate that the usual mean-field approach of 
molecular chaos which is based on a factorization of the N-particle probability density, fails in this
deterministic system and leads to unphysical predictions such as an infinite coefficient of self-diffusion.
The proposed theory opens a way to analytically treat other active systems beyond mean field, such as
mixtures of different SPPs and models
with non-reciprocal, chiral, and nematic interactions or interactions that couple dependent on the displacement vector (cohesion- or vision-like interactions).  The analytical development is carried out in detail for anti-aligning couplings, where collision times are finite and exact results are obtainable. We demonstrate that a physical regularization by quenched speed disorder establishes the validity of the same Boltzmann theory for aligning couplings.

The main technical results of this paper are (i) the coupling matrix $g_{mn}$, eq. (\ref{G_MN_DEF}), 
for the Fourier-modes which describe the
extension of the Vlasov-like theory beyond mean field, (ii) the integral- and corresponding differential equations for the angular displacement, 
eqs. (\ref{INTEGRAL_FINAL}, \ref{EQUIV_DIFF_EQ}), (iii) 
the noise correlations of the effective network noise, eq. (\ref{NOISE_EXACT}), and (iv) the expression for 
the self-diffusion coefficient, eq. (\ref{D_FINAL_EXPR1}).

\section*{Acknowledgments}
We thank J. Mihatsch for valuable discussions.   Parts of this work were carried out during a sabbatical stay of TI in the group of BL at Humboldt University Berlin, whose hospitality is gratefully acknowledged.

\appendix 

\section{Two-particle scattering}
\label{app:A}

The evolution equations for the angles of two interacting particles are
\begin{eqnarray}
\label{THET1_DGL}
\dot{\theta}_1&=&\Gamma\, \sin(\theta_2-\theta_1) \\
\label{THET2_DGL}
\dot{\theta}_2&=&\Gamma\, \sin(\theta_1-\theta_2) 
\end{eqnarray}
We define the auxiliary variables,
\begin{eqnarray}
\nonumber
\Delta &\equiv &\theta_2-\theta_1 \\
\tilde{c} & \equiv & \theta_1+\theta_2\,.
\end{eqnarray}
By adding and subtracting equations (\ref{THET1_DGL}, \ref{THET2_DGL}) one finds
\begin{eqnarray}
\label{A_SUM}
\dot{\tilde{c}}&=0 \\
\label{A_DIFF}
\dot{\Delta}   &=-2\Gamma\, \sin(\Delta)
\end{eqnarray}
Defining the angles $\alpha_1$ and $\alpha_2$ with respect to the 
vector $-\Delta \vec{r}=\vec{r}_1-\vec{r}_2$, see Fig.~\ref{FIG_ALPHADEF},
we have 
\begin{eqnarray}
\nonumber
\theta_1&=&\alpha_1+\beta \\
\label{alpha12_DEF}
\theta_2&=&\alpha_2+\beta 
\end{eqnarray} and thus,
$\Delta=\theta_2-\theta_1=\alpha_2-\alpha_1$ and 
\begin{equation}
\label{C-BETA-CONNECT}
c\equiv\alpha_1+\alpha_2=\tilde{c}-2\beta\,.
\end{equation}
Even though $\tilde{c}$ is conserved, the quantity $c$ depends on time.
This is because during a collision of finite duration, the positions of the particles change, resulting in
a change of the angle $\beta$ which is defined in Fig. ~\ref{FIG_ALPHADEF}.

The solution of the differential equation (\ref{A_DIFF}) is
\begin{equation}
\label{A_DIFF_SOL}
\tan\Big[{\Delta(\tilde{t})\over 2}\Big]=
\tan\Big[{\Delta(t  )\over 2}\Big]\,
\mathrm{exp}\big[2\Gamma (t-\tilde{t})\big]
\end{equation}

The time evolution of the connecting vector $\Delta \vec{r}=(\Delta r_x,\Delta r_y)$
follows from integrations of Eq.(\ref{POS_EQ}) over time as
\begin{eqnarray}
\nonumber
\Delta r_x(\tilde{t})-
\Delta r_x(t) &=& v_0\int_t^{\tilde{t}} \diff \hat{t}
\big[\cos(\theta_2(\hat{t}))-
     \cos(\theta_1(\hat{t}))\big] 
=-2v_0\int_t^{\tilde{t}} \diff \hat{t}\,\sin{\tilde{c}\over 2}\,\sin{\Delta(\hat{t})\over 2}
\\
\label{DELTA_R_EXP1}
\Delta r_y(\tilde{t})-
\Delta r_y(t) &=& v_0\int_t^{\tilde{t}} \diff \hat{t}
\big[\sin(\theta_2(\hat{t}))-
     \sin(\theta_1(\hat{t}))\big]   
=2v_0\int_t^{\tilde{t}} \diff \hat{t}\,\cos{\tilde{c}\over 2}\,\sin{\Delta(\hat{t})\over 2}\,.
\end{eqnarray}
With the abbreviation 
\begin{equation}
\label{G_FUN_DEF}
G\equiv \int_t^{\tilde{t}} \diff \hat{t}\,\sin{\Delta(\hat{t})\over 2}
\end{equation}
and using the invariance of $\tilde{c}=\theta_1+\theta_2$,
the evolution of the square of the distance between the particles $(\Delta r)^2=(\Delta r_x)^2+(\Delta r_y)^2$ follows
from Eq. (\ref{DELTA_R_EXP1}) as
\begin{equation}
\label{DELTA_R_EXP2}
(\Delta r(\tilde{t}))^2=
(\Delta r(t))^2+4v_0^2 G^2+4v_0G\Big\{\Delta r_y(t) \cos{\tilde{c}\over 2}-\Delta r_x(t) \sin{\tilde{c}\over 2}\Big\}
\end{equation}
Expressing the connecting vector in polar coordinates at the exit time, where $|\Delta\vec{r}(t)|=R$,
\begin{eqnarray}
\nonumber
\Delta r_x(t)=R\,\cos(\pi+\beta(t))=-R\,\cos\beta(t) \\
\label{POLAR_DEF}
\Delta r_y(t)=R\,\sin(\pi+\beta(t))=-R\,\sin\beta(t) 
\end{eqnarray}
where $\beta$ in defined in Fig. ~\ref{FIG_ALPHADEF} (and taken at the exit time), 
we insert Eq. (\ref{POLAR_DEF}) in (\ref{DELTA_R_EXP2}),
use Eq. (\ref{C-BETA-CONNECT}) and the trigonometric identity 
\begin{eqnarray}
\sin\left({\tilde{c}\over 2}-\beta(t)\right)=
\sin\left({c(t)\over 2}\right)=
\cos\beta(t)\, \sin\left({\tilde{c}\over 2}\right)
-\sin\beta(t)\, \cos\left({\tilde{c}\over 2}\right)
\end{eqnarray}
to find
\begin{equation}
\Delta r(\tilde{t})^2=
\Delta r(t)^2+4v_0^2 G^2+4v_0\,G\, R\, \sin{c(t)\over 2}
\end{equation}
which after taking the square root yields Eq. (\ref{DR_EVOLVE}) of the main text.
To evaluate the integral in the definition of $G$, Eq. (\ref{G_FUN_DEF}), we use the identity
\begin{equation}
\label{REL_DEL_MU}
\sin{\Delta\over 2}={\tan{\Delta\over 2}\over \sqrt{1+\tan^2{\Delta\over 2}}}=
{\mu\over \sqrt{1+\mu^2}}
\end{equation}
and the known temporal behavior of the tangent of $\Delta/2$ from Eq. (\ref{A_DIFF_SOL}) to obtain
\begin{equation}
G= \int_t^{\tilde{t}} \diff \hat{t}\,
{\mu\,\mathrm{exp}(2\Gamma(t-\hat{t}))\over
\sqrt{1+\mu^2\,\mathrm{exp}(4\Gamma(t-\hat{t}))}}
\end{equation}
Switching to the new variable $x=\mathrm{exp}(2\Gamma(t-\hat{t}))$ with $\diff \hat{t}=-\diff x/(2\Gamma x)$, the integral becomes
\begin{equation}
G={1\over 2\Gamma}\int_{\mathrm{exp}(2\Gamma(t-\tilde{t}))}^1 {\mu\,\diff x \over \sqrt{1+\mu^2 x^2}}
\end{equation}
The additional transformation $x\mu={\rm sinh}y$ results in a trivially solvable integral and leads
to Eq. (\ref{GDEF}) in the main text.

\section{Incorporating microscopic scattering into the collision integral}
\label{app:B}

The collision integral contains the relative velocity 
$\vec{v}_{\text{rel}}=\vec{v}_2-\vec{v}_1=v_0[\hat{\vec n}(\theta_2)-\hat{\vec n}(\theta)]$ and involves integrations over the angles
$\phi$ and $\theta_2$, where we take particle $1$ as focal particle with $\theta=\theta_1$.
However, the results of the two-particle scattering are expressed in terms of the angles $\alpha_1$ and 
$\alpha_2$, and in particular $c=\alpha_1+\alpha_2$.
Hence, a connection between these different sets of variables is needed.
From Figs. \ref{FIG_COLLCIRCLE1}, \ref{FIG_ALPHADEF} and 
the definition of the relative velocity we see that the scalar multiplication of $\vec{v}_{\text{rel}}$
with the unit vector $\hat{r}=\Delta \vec{r}/\Delta r$ at the exit time $t$ gives 
\begin{equation}
\label{SCAL_PROD_VREL}
v_{\text{rel}}\,\cos\phi=v_0(\cos\alpha_1-\cos\alpha_2)=2v_0
\sin{c\over 2}\,\sin{\Delta\over 2}
\end{equation}
where the last statement comes from the trigonometric identity for a sum of cosine functions.
The quantity $c$ can now be expressed in terms of the variables $\phi,\theta,\theta_2$ as
\begin{equation}
\label{A_LINK_C1}
\sin{c\over 2}={v_{\text{rel}}\over v_0} { \cos \phi \over 2\, \sin{\Delta\over 2}}
\end{equation}
with $\Delta=\theta_2-\theta$.
Furthermore, by means of other trigonometric identities, one finds that
\begin{equation}
\label{VREL_VS_SIN}
{v_{\text{rel}}\over v_0}=2\Big|\sin{\Delta\over 2}\Big|
\end{equation}
Inserting this into expression (\ref{SCAL_PROD_VREL}) leads to the simple relation
\begin{equation}
\label{A_LINK_C2}
\sin{c\over 2}={\rm sgn}\Big[ \sin{\Delta\over 2}\Big]\,\cos\phi
\end{equation}
where $sgn$ is the signum function.
Note that in Eqs. (\ref{SCAL_PROD_VREL} --\ref{A_LINK_C2}) all quantities are taken at the fixed exit time $t$.
Another useful identity involving the angle $\Delta$ is given by
\begin{equation}
{\rm sgn} \left( \sin {\Delta\over 2} \right)
\left[ {1\over \sin{\Delta\over 2}}-2\sin{\Delta\over 2}\right]=
{ \cos\Delta \over \left|\sin{\Delta\over 2}\right|}
\end{equation}

\section{Angular collision integrals}
\label{app:C}

In the perturbative evaluation of the collision integral in powers of $\Sc$ in section \ref{sec:eval_coll} the following
angular integrals appear:
\begin{eqnarray}
\nonumber
& & A_1(a)=\int_{-a}^a {\rm e}^{\ii\,n\,\theta}\, { \sin^2(\theta)\over \left| \sin{\theta\over 2}\right|}\,
\diff \theta=-{3\over n^2-\left({3\over 2}\right)^2}
        -{1\over n^2-\left({1\over 2}\right)^2} 
        +{ \cos\left(a\left[n-{3\over 2}\right]\right) \over n-{3\over 2}} \\
& &     -{ \cos\left(a\left[n+{1\over 2}\right]\right) \over n+{1\over 2}}
        -{ \cos\left(a\left[n+{3\over 2}\right]\right) \over n+{3\over 2}}
        +{ \cos\left(a\left[n-{1\over 2}\right]\right) \over n-{1\over 2}}
\\
\nonumber
& &A_2(a)=\int_{-a}^a {\rm e}^{\ii\,n\,\theta} { \sin(\theta)\, \cos(\theta)\over \left| \sin{\theta\over 2}\right|}\,
\diff \theta=2n\,i\left\{
{1\over n^2-\left({3\over 2}\right)^2}
        +{1\over n^2-\left({1\over 2}\right)^2}
\right\} \\
& &-i\left[
     { \cos\left(a\left[n-{1\over 2}\right]\right) \over n-{1\over 2}}
    +{ \cos\left(a\left[n+{1\over 2}\right]\right) \over n+{1\over 2}}
    +{ \cos\left(a\left[n+{3\over 2}\right]\right) \over n+{3\over 2}}
    +{ \cos\left(a\left[n-{3\over 2}\right]\right) \over n-{3\over 2}}
\right] \\
\nonumber
& &A_3(a)=\int_{-a}^a {\rm e}^{\ii\,n\,\theta} \sin(\theta)\, \left| \sin{\theta\over 2}\right|\,
\diff \theta=n\,i\left\{
{1\over n^2-\left({1\over 2}\right)^2}
        -{1\over n^2-\left({3\over 2}\right)^2}
\right\} \\
& &+{1\over 2i}\left[
    -{ \cos\left(a\left[n+{3\over 2}\right]\right) \over n+{3\over 2}}
    +{ \cos\left(a\left[n-{1\over 2}\right]\right) \over n-{1\over 2}}
    +{ \cos\left(a\left[n+{1\over 2}\right]\right) \over n+{1\over 2}}
    -{ \cos\left(a\left[n-{3\over 2}\right]\right) \over n-{3\over 2}}
\right] \\
\nonumber
& &A_4(a)=\int_{-a}^a {\rm e}^{\ii\,n\,\theta} \cos(\theta)\, \left| \sin{\theta\over 2}\right|\,
\diff \theta={1\over 2}\left\{
-{3\over n^2-\left({3\over 2}\right)^2}
        +{1\over n^2-\left({1\over 2}\right)^2}
\right\} \\
& &+{1\over 2}\left[
    -{ \cos\left(a\left[n+{3\over 2}\right]\right) \over n+{3\over 2}}
    -{ \cos\left(a\left[n-{1\over 2}\right]\right) \over n-{1\over 2}}
    +{ \cos\left(a\left[n+{1\over 2}\right]\right) \over n+{1\over 2}}
    +{ \cos\left(a\left[n-{3\over 2}\right]\right) \over n-{3\over 2}}
\right] 
\end{eqnarray}
For the special case $a=\pi$ one obtains,
\begin{eqnarray}
& & A_1(\pi)=
-{3\over n^2-\left({3\over 2}\right)^2}
        -{1\over n^2-\left({1\over 2}\right)^2}
\\
& &A_2(\pi)=
2n\,i\left\{
{1\over n^2-\left({3\over 2}\right)^2}
        +{1\over n^2-\left({1\over 2}\right)^2}
\right\} \\
& &A_3(\pi)=
n\,i\left\{
{1\over n^2-\left({1\over 2}\right)^2}
        -{1\over n^2-\left({3\over 2}\right)^2}
\right\} \\
& &A_4(\pi)=
{1\over 2}\left\{
-{3\over n^2-\left({3\over 2}\right)^2}
        +{1\over n^2-\left({1\over 2}\right)^2}
\right\} 
\end{eqnarray}
Another simpler set of angular integrals 
is
\begin{eqnarray}
& &B_1(a)=\int_{-a}^a {\rm e}^{\ii\,n\,\theta} \, \left| \sin{\theta\over 2}\right|\,
\diff \theta=
-{1\over n^2-\left({1\over 2}\right)^2}
    -{ \cos\left(a\left[n+{1\over 2}\right]\right) \over n+{1\over 2}}
    +{ \cos\left(a\left[n-{1\over 2}\right]\right) \over n-{1\over 2}}
\\
& &B_2(a)=\int_{-a}^a {\rm e}^{\ii\,n\,\theta} \, \cos(\theta)\,
\diff \theta=
a\left\{ {\rm sinc}\left(a\left[n-1\right]\right)
+{\rm sinc}\left(a\left[n+1\right]\right)
\right\} \\
& &B_3(a)=\int_{-a}^a {\rm e}^{\ii\,n\,\theta} \, \sin(\theta)\,
\diff \theta=
i\,a\left\{ {\rm sinc}\left(a\left[n-1\right]\right)
-{\rm sinc}\left(a\left[n+1\right]\right)
\right\} 
\end{eqnarray}
with the special case $a=\pi$:
\begin{eqnarray}
& &B_1(\pi)=
-{1\over n^2-\left({1\over 2}\right)^2}
\\
& &B_2(\pi)=\pi\left(\delta_{n,1}+\delta_{n,-1}\right)
\\
& &B_3(\pi)=
          \ii \,\pi\left(\delta_{n,1}-\delta_{n,-1}\right)
\end{eqnarray}
We also need the B-integrals for the small argument $a=4\,\Sc$:
\begin{eqnarray}
& &B_1(4\Sc)=8 \Sc^2+\mathcal{O}(\Sc^4) \\
& &B_2(4\Sc)= 8 \Sc+\mathcal{O}(\Sc^3) \\
& &B_3(4\Sc)= {128\over 3}n\,\ii \,\Sc^3 +\mathcal{O}(\Sc^5) 
\end{eqnarray}

\section{The correlation function of the Random Telegraph Process}
\label{app:D}

\noindent
Since the RT-process has only two allowed states, $a_{ij}=1$ and $a_{ij}=0$, we define the probability
to be in the ON-state and the OFF-state as
\begin{eqnarray}
p_+(t)\equiv {\rm prob}(a_{ij}=1\;{\rm at}\;{\rm time}\;t) \\
p_-(t)\equiv {\rm prob}(a_{ij}=0\;{\rm at}\;{\rm time}\;t) 
\end{eqnarray}
These probabilities obey two coupled Master equations
\begin{eqnarray}
\dot{p}_+=w_{\text{on}}\, p_- -w_{\text{off}}\,p_+ \\
\dot{p}_-=w_{\text{off}}\, p_+ -w_{\text{on}}\,p_- 
\end{eqnarray}
with the ON and OFF rates $w_{\text{on}}$, $w_{\text{off}}$, respectively.
Because of $p_-=1-p_+$ this system can be solved easily with the result
\begin{equation}
\label{SOLVE_MASTER1}
p_+(t)=p_+(0)\, {\rm e}^{-\lambda\,t}+{w_{\text{on}} \over \lambda}\left(1-{\rm e}^{-\lambda\, t}\right)
\end{equation}
with abbreviation $\lambda\equiv w_{\text{on}}+w_{\text{off}}$.
In the stationary state, for $t\gg 1/\lambda$, one has
\begin{equation}
p_{+,\infty}=\lim_{t\to \infty} p_+(t)= {w_{\text{on}} \over \lambda}
\end{equation}
The average of the random variable $a_{ij}$ in the stationary state is given by
\begin{equation}
\label{AVERAGE_A}
\langle a_{ij}\rangle =1\times p_{+,\infty}+0\times p_{-,\infty}={w_{\text{on}} \over \lambda}
\end{equation}
This average is identical to the probability to find particle $j\neq i$ in the collision circle of particle $i$,
which is given by the ratio of the area of the collision circle to the total area of the system,
\begin{equation}
\label{RT_stationary1}
p_{+,\infty}={w_{\text{on}} \over \lambda}={\pi R^2\over L^2}={M\over N}
\end{equation}
because of $M=\pi R^2 N/L^2$
This gives us a connection of the ON/OFF-rates of the random telegraph process to the actual particle dynamics.

To determine the auto-correlation in the stationary state
\begin{equation}
g(\tau)=\langle a_{ij}(t+\tau)\,a_{ij}(t)\rangle
=\langle a_{ij}(\tau)\,a_{ij}(0)\rangle
\end{equation}
with $\tau>0$ we need the two-time probability
\begin{equation}
p_{++}(\tilde{t},t)={\rm prob}(a_{ij}=1\;{\rm at}\;{\rm time}\;\tilde{t}\;{\rm AND}\; a_{ij}=1\;{\rm at}\;{\rm time}\;t) 
\end{equation}
because
\begin{eqnarray}
\nonumber
&&\langle a_{ij}(\tau)\,a_{ij}(0)\rangle =\\\nonumber 
&&
1\times 1\times p_{++}(\tau,0)+
1\times 0\times p_{+-}(\tau,0)+
0\times 1\times p_{-+}(\tau,0)+
0\times 0\times p_{--}(\tau,0) \\
&&=
p_{++}(\tau,0)
\end{eqnarray}
We introduce the conditional probability $p(+,\tau|+,0)$ which is the probability to find the random variable switched on
at time $\tau$ under the condition that it was switched on also at the earlier time $0$.
We write
$p_{++}(\tau,0)=p(+,\tau|+,0)\,p_+(0)
$
and find
\begin{equation}
g(\tau)=p(+,\tau|+,0)\,p_0
\end{equation}
where we abbreviated the probability to be in the ON-state at time 0 as $p_0$.
Finally, the conditional probability is just the solution, Eq. (\ref{SOLVE_MASTER1}), of the Master equation with the specific 
initial condition $p_+(0)=1$. This leads to 
\begin{equation}
g(\tau)=p_0\Big[
{\rm e}^{-\lambda\,\tau}+{w_{\text{on}} \over \lambda}\left(1-{\rm e}^{-\lambda\, \tau}\right)\Big]\;\;\;\tau\geq 0
\end{equation}
The value of the correlation at time $\tau=0$ is given by $p_0$,
$\langle a_{ij}^2 \rangle=g(0)=p_0$.
Since $a_{ij}$ can only be one ore zero, we have $\langle a_{ij}^2 \rangle=\langle a_{ij} \rangle$ which is given in Eq. (\ref{AVERAGE_A}).
Thus 
$p_0={\pi R^2/L^2}$ and because of time-reversal symmetry we obtain
\begin{equation}
\label{G_EQ3}
g(\tau)={\pi R^2\over L^2}\Big[
{\rm e}^{-\lambda\,|\tau|}+{w_{\text{on}} \over \lambda}\left(1-{\rm e}^{-\lambda\, |\tau|}\right)\Big]
\end{equation}
The OFF-rate of the process is related to how long a particle travels once it enters the collision circle of
the focal particle. Thus, this rate should be of the order of $v_0/R$ and stays finite in the thermodynamic limit.
Because of this, we see from Eq. (\ref{RT_stationary1}) that the ON-rate becomes very small for large particle number $N\gg 1$.
Thus, $w_{\text{on}}\ll w_{\text{off}}$ in this limit and we find
\begin{equation}
w_{\text{on}}\sim w_{\text{off}} {M\over N}
\end{equation}
This means, that the decay rate $\lambda=w_{\text{on}}+w_{\text{off}}$ is dominated by the OFF-rate.
Performing the thermodynamic limit $N\to \infty$ in $\hat{g}=(N-1)\,g(\tau)$ by using expression (\ref{G_EQ3}) with
$p_0=M/N=\pi R^2/L^2$ leads to the expression for the exponential auto-correlation function $\hat{g}(\tau)$
of the main text, Eq. (\ref{FORMULA_CORR_RT}).

\section{Calculating the contact time distribution}
\label{app:E}

We assume that the distance between particles one and two is larger than $R$ for $t<0$ and that this distance is equal to $R$ at $t=0$.
That means that particles one and two start interacting at time $t=0$.
We are now interested in the contact time that is the maximum time for which the distance between particles one and two is still less or equal to $R$ neglecting interactions with all other particles.

Without loss of generality we assume that particle one moves into the $x$-direction at $t=0$, that means that $\theta_1=0$.
The contact time depends on the position and direction of motion of particle two at $t=0$.
We denote the angle between $x$-axis and the line connecting particles one and two by $\Phi$ and the direction of motion of particle two by $\theta_2$.

The velocity component of particle two within the rest frame of particle one that points towards particle one is given by
\begin{align}
	v_{\perp}=v_0(-\cos \theta_2\cos \Phi + \cos \Phi -\sin \theta_2 \sin \Phi).
	\label{eq:vperp}
\end{align}
Clearly, the contact time is positive only if $v_{\perp}>0$.
\vspace{0.4cm}

\noindent
{\bf Case $\Sc=0$:}\\
For simplicity we first assume zero coupling, $\Sc=0$.
In that case, simple geometric considerations lead to the contact time
\begin{align}
	t(\theta_2, \Phi)=\frac{-\cos \theta_2\cos \Phi + \cos \Phi -\sin \theta_2 \sin \Phi}{1-\cos \theta_2}.
	\label{eq:contact_time1}
\end{align}

The distribution of contact times depends on the rates of contacts that appear with orientation parameters $\theta_2, \Phi$ per time, $r(\theta_2, \phi)$ as
\begin{align}
	p(\hat{t})=\frac{1}{Z} \int_{0}^{2\pi} d \theta_2 \int_{0}^{2\pi} d \Phi r(\theta_2, \Phi)\delta(\hat{t}- t(\theta_2, \Phi)),
	\label{eq:contact_time_distribution1}
\end{align}
where $Z$ is a normalization constant.
The rate of contacts is proportional to the velocity component of particle two within the rest frame of particle one that is perpendicular to the surface of the interaction region, that is the circle of radius $R$ around particle one.
Thus, the rate is given by
\begin{align}
	r(\theta_2, \Phi)= v_{\perp} \Theta(v_{\perp}),
	\label{eq:contact_rate}
\end{align}
where the Heaviside function $\Theta(v_{\perp})$ ensures that only approaching particles are considered.
Determining $Z$ via normalization of $p(t)$ we find with Eqs. \eqref{eq:vperp}, \eqref{eq:contact_time1}, \eqref{eq:contact_time_distribution1} and \eqref{eq:contact_rate} the contact time distribution
\begin{align}
	p(t)=&\frac{1}{16} \int_{0}^{2\pi} d \theta_2 \int_{0}^{2\pi} d \Phi \delta\bigg(t + \frac{R}{v_0}\frac{\cos \theta_2\cos \Phi - \cos \Phi +\sin \theta_2 \sin \Phi}{1-\cos \theta_2} \bigg)
	\notag
	\\
	&\times \Theta(-\cos \theta_2\cos \Phi + \cos \Phi -\sin \theta_2 \sin \Phi)
	\notag
	\\
	&\times(-\cos \theta_2\cos \Phi + \cos \Phi -\sin \theta_2 \sin \Phi).
	\label{eq:contact_time_distribution2}
\end{align}
For large time, we find
\begin{equation}
\label{contact_dist_limit}
p(t)={2 R^2\over 3 v_0^2}\,t^{-3}
\end{equation}

The average contact time, that is, the first moment of $p(t)$ is finite
and can be calculated from the distribution, leading to the result, $(\pi^2/8)\,R/v_0$, 
Eq. (\ref{contact_time_moment}),
in the main text. 
\vspace{0.4cm}

\noindent
{\bf Case $\Sc\to \infty$:}\\
In the limit of strong coupling $\Sc\to \infty$ we assume that particles immediately anti-align at the beginning of the interaction.
We make the same assumptions as above: $\theta_1=0$ at the start of the interaction between particles one and two at $t=0$.
After the immediate anti-alignment the orientations of particles one and two are
\begin{align}
	\theta_1'&= \Delta,
	\notag
	\\
	\theta_2'&=\theta_2 - \Delta,
	\label{eq:anti_align1}
\end{align}
where $\Delta$ needs to satisfy the anti-alignment condition
\begin{align}
	\theta_1'-\theta_2'= \pi + 2k \pi,
	\label{eq:anti_align2}
\end{align}
which can be rewritten as
\begin{align}
	\Delta= \frac{\theta_2}{2} + \frac{\pi}{2} + k\pi,
	\label{eq:anti_align3}
\end{align}
where $k$ is an integer.
Rotating the coordinate system about $\Delta$ around particle one we arrive at
\begin{align}
	\theta_1''&=0,
	\notag
	\\
	\theta_2''&=-\pi,
	\notag
	\\
	\Phi''&=\Phi - \frac{\theta_2}{2} - \frac{\pi}{2} - k\pi.
	\label{eq:anti_align4}
\end{align}
Inserting these orientations into the contact time \eqref{eq:contact_time1} we obtain
\begin{align}
	t=\cos(\Phi- \frac{\theta_2}{2} - \frac{\pi}{2} - k\pi),
	\label{eq:contact_time2}
\end{align}
where the integer $k$ has to be chosen such that the contact time is positive for approaching particles.
Thus, the contact time can be expressed as
\begin{align}
	t=|\cos(\Phi- \frac{\theta_2}{2} - \frac{\pi}{2})|.
	\label{eq:contact_time3}
\end{align}
In analogy to Eq. \eqref{eq:contact_time_distribution2} we obtain
\begin{align}
	p(t)=&\frac{1}{16} \int_{0}^{2\pi} d \theta_2 \int_{0}^{2\pi} d \Phi \delta\bigg(t - \frac{R}{v_0} |\cos(\Phi- \frac{\theta_2}{2} - \frac{\pi}{2})| \bigg)
	\notag
	\\
	&\times \Theta(-\cos \theta_2\cos \Phi + \cos \Phi -\sin \theta_2 \sin \Phi)
	\notag
	\\
	&\times(-\cos \theta_2\cos \Phi + \cos \Phi -\sin \theta_2 \sin \Phi).
	\label{eq:contact_time_distribution3}
\end{align}
Eventually, we calculate the first moment of the distribution as
\begin{align}
	\langle t \rangle =& \int_0^{\infty} \diff t\, p(t) t,
	\notag
	\\
	=&\frac{1}{16}\int_0^{2\pi} \diff \Phi \int_0^{2\pi} \diff \theta_2 \frac{R}{v_0} |\cos(\Phi- \frac{\theta_2}{2} - \frac{\pi}{2})| 
	\notag
	\\
	&\times \Theta(-\cos \theta_2\cos \Phi + \cos \Phi -\sin \theta_2 \sin \Phi)
	\notag
	\\
	&\times (-\cos \theta_2\cos \Phi + \cos \Phi -\sin \theta_2 \sin \Phi)
	\notag
	\\
	=&\frac{\pi}{4}\,\frac{R}{v_0}\,\;\;\;{\rm for}\; \Sc\to \infty
	\label{eq:first_moment_contact_time_infinite_coupling}
\end{align}
Thus, we find that the collisional paradigm holds for anti-aligning couplings meaningfully even in the limit of strong couplings. 

\bibliography{lit_boltzmann_long.bib}
\end{document}